# The impulse cutoff an entropy functional measure on trajectories of Markov diffusion process integrating in information path functional

**Vladimir S. Lerner, USA, lernervs@gmail.com**

**Abstract**

Integrating discrete information, composed of information Bits extracted from observing random process, solves the impulse cutting off entropy *functional* (EF) measure on trajectories Markov diffusion *process* whose information integrates path functional (IPF). Each cut brings memory of the entropy being cut, which provides both reduction of the process entropy and discrete unit of the cutting entropy – a Bit. Consequently, *information is memorized entropy cutting in random observations which process interactions*. The origin of information associates with an "anatomy of creation of impulse" enables both cut and stipulate random process while generating information under the cut. Memory stands as the impulse' cut time interval.

Defining the EF via the process additive functional with functions drift and diffusion allows reducing this functional on trajectories to a regular integral functional.

Compared to conventional Shannon entropy measure of a random state, cutting the process on separated states decreases quantity of process information by the amount, concealed in correlation connections between these states, which hold hidden process information.

The EF-IPF measure integrates information covered in both continuous process and discrete impulses which generate information and transmit it between the impulses.

The $n$-dimensional process cutoff generates a finite information measure, integrated in the IPF whose information approaches the EF measure at $n \to \infty$, restricting maximal information of the Markov diffusion process. Studied impulse delta-function cutoff and the discrete impulse deliver equivalent information at each cutoff. The constructed finite restriction limits the impulses' discrete stepwise actions applied for cutting the regular integral on the functional increments between the cutoffs.

Finite impulse step-up action transfers EF increment to following impulse whose step-down action cuts off information and step-up action starts imaginary (virtual) impulse carrying entropy increment to next real cut. Step-down cut generates maximal information while the step-up action delivers minimal information from impulse cut to next impulse step-down action.

A virtual impulse transfers conjugated entropy increments during a microprocess ending with adjoining increment within actual step-down action at cutoff. Extracting maximum of minimal impulse information and transferring minimal entropy between impulses implement maxmin-minimax principle of converting process entropy to information. Each cutoff sequentially and automatically converts entropy to information, holding information Bit from random process, which connects the Bits sequences in the IPF and predicts next cut. The built information macroprocess, as the EF minimax extremal, integrates both imaginary entropy of microprocess and cutoff information of real impulses in IPF information physical process. Each IPF dimensional cut measures Feller kernel information. Cutting entropy memorizes the cutting time interval which freezes the probability of events with related dynamics of information micro-macroprocess. Estimation extracting information confirms nonadditivity of the IPF measured fractions.

Keywords: *integral information measure; cutting off the diffusion process; impulse action, Feller kernel, information path functional, finite process information, information micro- macroprocesses.*

**Introduction**

Conventional information science generally considers an information *process*, but traditionally uses the probability measure for random *states* and Shannon's entropy measure as the uncertainty *function* of the states (C.E.Shannon [1], E.T. Jaynes [2], A.N. Kolmogorov [3], others).

Such a measure evaluates information for a sequence of each process' states and does not measure *inner connections* between these states *through all process*.

The problem is to find an *integral* information measure, evaluating the inner connection and dependencies of the *random* process' states to measure the process' total information.

Such measure should integrate information enclosed in both continuous and discrete processes.

Using definition of conditional information entropy and applying it to transitional probability density' measure, defined along the Markov diffusion process, we introduce the process' integral information measure on the process' trajectories.

Such an entropy functional (EF) measure has an analogy with R. P. Feynman's path functional [4], which M. Kac [5] has expanded on trajectories of Brownian processes.

The difference is transformation of the transition probability's density through the process' multiplicative and additive functionals whose mathematical expectation of a random trajectory defines the EF functional measure.

To estimate inner connections and dependencies of the diffusion process' states with EF, we use a jump-wise control action, cutting off the diffusion process through its de-correlation.

The cutting actions, while measuring the EF along the process trajectory, implement multiple sequence of the transitive transformations, which instantaneously convert probability of Markovian process to probability of Brownian process and vice versa.

Theory of additive functional is developed by E.B.Dynkin [6,7], I. V. Girsanov [8], A.D. Ventsel [9], R.M. Blumenthal, R.K.Getoor [10], K. Ito, N. Ikeda, S. Watanabe [11,12], Y.V. Prokhorov, Y.A. Rozanov [13], I.I.Gikhman, A.V.Scorochod [14], other authors.

The information measures of a random process were introduced by Fisher [15], S.Kulback, R.A, Leibler [16], A.Rényi [17], R.L. Stratonovich [18] and some others.

Fisher's logarithmic measure calculates the covariance matrices associated with maximum-likelihood estimates in statistics. Rényi's entropy is a generalization of Shannon entropy by introducing probability of a discrete random variable with various parameter of the probability's power. Kullback–Leibler's divergence between the probabilities of a process' states distributions is measured by a relative Shannon's entropy.

R.L.Stratonovich specified Radon-Nikodym's density for a probability density measure, applying it to Shannon's entropy of a random process.

Controllable Markov processes with the optimal stopping and estimation methods are studied by E.B. Dynkin [19], I.I.Gihman, A.V. Scorochod [20], N.V. Krylov [21], J. M. Harrison *et al* [22], F.B.Handson [23], A. Bensoussan, J.L. Lions [24], A. L. Bronstein *et al* [25], and many others.

W. Feller [26, 27] introduced a diffusion operator of "kernel", which absorbs the diffusion motion on its limited time interval, covering Markov property.

H.P. McKean and H. Tanaka [28] connected the Feller's kernel with additive functionals of Brownian path and related these functionals to both an instant of the path' passage boundary time, limited the kernel, and to the kernel finite energy.
M.Fukushima *et al* [29] estimated an energy, covering the kernel, and the minimal boundary time.
A.Borodin and V.Gorin [30] studied a correlation function of a discrete time Markov process with the transitional probability measure of Feller's kernel that links the kernel's measure to the correlations. The cutting off kernel information measure is still *unspecified*.
The paper cutoff impulse control "intervenes" in the diffusion process during the minimal boundary instant time of a kernel's jumping mechanism [29]. This leads to extraction of information hidden in a kernel, which is evaluated through the additive functional at such cut off.
A deterministic "intervention" in a random process, considered by A.A.Yushkevich [31], M.A.H.Dempster and J.J.Ye [32], others, studied mostly a diffusion process with a drift function.
Using the jump transition probability under the impulse control, which is described through the additive functional, defined by both drift and diffusion, allows us studying a more general case.
The paper shows that cutting off the entropy integral information measure on the separated states' measures decreases the quantity of process information by the amount, which was concealed in the correlation connections between the separating states. This integral functional measure accumulates more process information than the sum of the information measures of its separated states and integrates information covered in both continuous process and discrete impulses cutting information of the process.
Study discrete impulses shows how each generates information and transmit it between the impulses.
Each impulse intervention interacts with the process in different forms: the cutoff-reaction, observation, measurement, which brings both irreversibility and a memory of cutting entropy.
While each cutoff reveals information hidden in the process before the cutoff, the paper information path functional (IPF) integrates all cutoffs impulse information.
The $n$-dimensional process cutoffs generates a finite information measure**,** integrated by information path functional (IPF) whose total information approaches the EF *measure* at $n \to \infty$**,** restricting maximal information of the Markov diffusion process.
Studied impulse delta-function cutoff and the discrete impulse deliver equivalent information at each cutoff. The constructed finite restriction limits the impulses' discrete step-wise actions applied for cutting the regular integral on the entropy increments between the cutoffs. Finite impulse step-up action transfers EF increment to following impulse whose step-down action cuts off information and step-up action starts imaginary (virtual) impulse carrying entropy increment to next real cut.
Step-down cut generates maximal information while the step-up action delivers minimal information from impulse cut to next impulse step-down action.
A virtual impulse transfers conjugated entropy increments during a microprocess ending with adjoining increment within actual step-down action at cutoff. Extracting maximum of minimal impulse information and transferring minimal entropy between impulses implement maxmin-minimax principle of converting process entropy to information.
Each cutoff consequentially and automatically converts entropy to information, holding information Bit of random process, which connects the Bits sequences in the IPF.

Macroprocess, as the EF minimax extremal, integrates imaginary entropy of microprocess and cutoff information of real impulses in the IPF information physical process.

Each EF dimensional cut measures Feller kernel information. Estimation extracting information confirms non-additivity of the EF measured process fractions.

*The paper organizes in two Parts:*

Part 1 includes three Sections, evaluating integral information measure cutting from entropy *functional* on trajectories of Markov diffusion process and collected by information path functional.

Part 2 includes ten Sections, evaluating integral information, cutting from entropy *regular* integral, the EF fractions between the cutoffs, the IPF discrete integration, and the information micro-and macroprocess as the integral variation solutions, along with estimations, validations and applications.

*Section* 1.1 introduces the EF, defined on Markov diffusion process through the process' additive functional of transitive probabilities, which is determined by the functions drift and diffusion of Ito's stochastic differential equation.

The relational EF probability measure is conditional to Feller kernel and evaluates optimal transition to the kernel, minimizing the EF, which originates the mathematical problem solving in the paper.

*Section* 1.2 provides the information evaluation of the process' cut off action by applying Dirac's impulse control, estimates the controls' extracting information, which includes the Feller kernel, evaluates its maximum; along with extracting maximum of minimal impulse information and transferring minimal entropy between impulses.

These implement maxmin-minimax principle of converting process entropy to information.

*Section* 1.3 considers properties of the EF cutoff functional, confirms non-additivity of the EF measured process' fractions and defines information path functional (IPF) with its properties.

*In Section* 2.1, the entropy functional is expressed via the process' dispersion (correlation) and the averaged controllable drift functions. This allows both *identifying* the EF on observed Markov process by measuring the correlations at applying control functions and *reducing* the EF functional to a regular integral of non-random functions with the integrant defined by an averaged additive functional.

*Section* 2.2 applies variation condition Sec. 1.1. to the regular entropy integral which minimizes the entropy integrant and decreases time intervals between the cutoff at $t \to T$.

*Section* 2.3 evaluates cutting the regular entropy integral under Dirac' and Kronecker' delta-functions.

*Section* 2.4. introduces a finite restriction on the cutting function that limits the impulses discrete and determines an extremal increment of entropy functional between the discrete cutoff for the regular integral. The discrete impulse' step-up action between the cutoff transfers the EF increment to the following impulse' step-down cut, whose cutoff information delivers information to this impulse step-up action. The step-up action starts a virtual impulse, carrying the entropy increment between the impulses, until its step-down real cut extracts the information contribution. The step-down action cuts maximal information, while the impulse step-up action delivers minimal information from the impulse cut to next impulse's step-down action; both extract maximum of minimal information of this impulse. The sequential impulse cutoffs and the minimal entropy transfer between the impulses implement maxmin-minimax principle of converting the process entropy to information.

The cutting off random process brings information Bit that includes: information delivered by capturing external entropy during transition to the cut; information cut from the random process, and information transferring to the nearest impulse, which keeps persistence continuation of the impulse sequence and the Bits increasing density.

*Section* 2.5 analyzes structure of Information Path functional (IPF) in $n$-dimensional Markov process under $n$-cutoff impulses and evaluates summary of the cutoff information contributions. The decreasing distance between the finite cutoff intervals with growing process' dimension at $n \to \infty$ limits total time of getting summary information contributions, which IPF collects, approaching the process EF that restricts the process finite IPF.

The IPF maximum, integrating unlimited number of the Bits' units with finite decreasing distances, limits total information carrying by the process' Bits.

The EF functional time course is a source of the entropy increment between impulses, moving the nearest impulses closer. The initial EF measures a potential information functional of the Markov process whose entropy increment of cutting random states delivers information, hidden between these states, which allows prediction of a next cutoff information contributions. Estimating the times at the cutoffs allows finding the discrete increments of correlations at each cutting information contributions. The nearest opposite action on entropy integral models the control interaction with random process, which provides external influx entropy, that the control captures and carries to the cut.

*Section 2*.6 describes microprocess within the impulse, the conjugated entropy increments under step-functions, the emergent transitional impulse transferring the initial time-located vector to equivalent space-vector, entanglement of the increments of microprocess, and appearance information at ending impulse time interval with injection of an energy at step-up control transition; the probabilities functions of the microprocess. Examples confirm that entropy increments between the impulses are imaginary.

*Section 2*.7 determines information dynamic processes resulting from EF and IPF functionals: real and imaginary microrocesses, and macroprocess that holds these microprocesses.

The macroprocess averages solution of Ito Eq. under the optimal controls' multiple cutoffs.

*Section* 2.8 considers solution of variation problem for the EF, which describes information macroprocess, integrating the imaginary entropy between impulses with the imaginary microprocesses and the cutoff information of real impulses, builds information physical information process of the collected entropy-information. The EF extremal trajectories describe the IPF information macroprocess which averages all real and imaginary microprocesses. The identified initial conditions for the entropy functional and its extremals allowing build both micro- and macroprocess from observation.

*Section* 2.9, discusses the specific of theoretical results, focusing on very practical outcomes, such as structure of information Bit, cutting from random process, connecting the Bits, transformation entropy to information, integrating the Bits in the finite information measure, finiteness of information for infinite process' dimension, process time course and its entropy, prediction of information, others.

Since information processes chose and connect all observations in certain- most probable sequence of events, which include hidden information between interacting events, these allow the information processes correctly describe all worlds' natural processes.

*Section* 2.10 estimates the extracting information, hidden by the interstates' connection of the diffusion process, and provides applications of the cutoff method, revealing the generation of phenomena with cutting process' information.

**Part1. Evaluation of integral information measure cutting from entropy functional on trajectories of Markov diffusion process**

**1. Entropy functional on the trajectories**

Let us have Markov diffusion process $\tilde{x}_t$ with transition probabilities $P(s,\tilde{x},t,B)$, $\sigma$-algebra $\Psi(s,t)$ created by events { $\tilde{x}(\tau)\in B$ } at $s\le\tau\le t$, and conditional probability distributions $P_{s,x}$ on $\Psi(s,t)$.

A family of the real or complex random values $\varphi_s^t=\varphi_s^t(\omega)$ depending on $s\le t$ defines an *additive functional* of process $\tilde{x}_t=\tilde{x}(t)$ [27, 28], if each $\varphi_s^t=\varphi_s^t(\omega)$ is measured regarding the related $\sigma$-algebra $\Psi(s,t)$ at any $s\le\tau\le t$ with probability 1 at $\varphi_s^t=\varphi_s^\tau+\varphi_\tau^t$, and $E_{s,x}[exp(-\varphi_s^t(\omega))]<\infty$, where $E_{s,x}[\bullet]$ are the related mathematical expectations.

Transformation of probability measures [7, p.48]:

$$\tilde{P}_{s,x}(d\omega)=p(\omega)P_{s,x}(d\omega), \qquad (1.1)$$

on trajectories of Markov process $(\tilde{x}_t,P_{s,x})$ holds distributions $\tilde{P}_{s,x}=\tilde{P}_{s,x}(A)$ on extensive $\sigma$-algebra $\Psi(s,\infty)$ with density measure:

$$p(\omega)=\frac{\tilde{P}_{s,x}(d\omega)}{P_{s,x}(d\omega)}=\exp\{-\varphi_s^t(\omega)\}, \qquad (1.2)$$

and transitional probabilities of transformed diffusion process $\varsigma_t$ with the additive functional:

$$\tilde{P}(s,\varsigma,t,B)=\int_{\tilde{x}(t)\in B}\exp\{-\varphi_s^t(\omega)\}P_{s,x}(d\omega) \qquad (1.3)$$

Applying the definition of conditional entropy [17] to mathematical expectation of logarithmic probability *functional density measure* (1.2) for process $\tilde{x}_t$ regarding process $\varsigma_t$ we have

$$S(\tilde{x}_t/\varsigma_t)=E\{-\ln[p(\omega)]\}=\int_{\tilde{x}(t)\in B}-\ln[p(\omega)]\tilde{P}_{s,x}(d\omega), \qquad (1.4)$$

where $E=E_{x,s,\tilde{x}_t}$ is conditional mathematical expectation, taken along the process trajectories $\tilde{x}_t$ at a varied $(\tilde{x},s)$ (by analogy with M.Kac [5, 194-218]).

From (1.2) and (1.3) we get conditional entropy functional expressed via the additive functional on the trajectories of considered diffusion processes:

$$S[\tilde{x}_t/\varsigma_t]=E[\varphi_s^t(\omega)] , \qquad (1.5)$$

which is entropy measure of a distance between distributions $\tilde{P}_{s,x},P_{s,x}$ .

Minimum of this functional, depending on the additive functional measures closeness above distributions:

$$\min_{\varphi_s^t}S[\tilde{x}_t/\varsigma_t]=S^o . \qquad (1.5a)$$

Let diffusion process $\tilde{x}_t$ be a solution of stochastic Ito $n$-dimensional differential Eqs [12]:

$$d\tilde{x}_t=a(t,\tilde{x}_t)dt+\sigma(t,\tilde{x}_t)d\xi_t,\tilde{x}_s=\eta,t\in[s,T]=\Delta,s\in[0,T]\subset R_+^1, \qquad (1.6)$$

with standard limitations [19] on drift function $a = a(t,x)$, function of diffusion $\sigma = \sigma(t,x)$, and Wiener process $\xi_t = \xi(t,\omega)$, defined on a probability space of elementary random events $\omega \in \Omega$ for variables located in $R^n$.
The process additive functional, according Girsanov's Theorem [8], satisfies the form [13,p.355-360], [14,Theorem11] with its upper limit $T$:

$$\varphi_s^T = 1/2\int_s^T a(t,\tilde{x})^T (2b(t,\tilde{x}))^{-1} a(t,\tilde{x}_t)dt + \int_s^T \sigma(t,\tilde{x})^{-1} a(t,\tilde{x}) d\xi(t),\ 2b(t,\tilde{x}) = \sigma(t,\tilde{x})\sigma^T(t,\tilde{x}) > 0. \quad (1.7)$$

Let the transformed process be

$$\varsigma_t = \int_s^t \sigma(v,\xi_v)d\xi_v \qquad (1.8)$$

having the same diffusion as the initial process, but the zero drift.
Measures $\tilde{P}_{s,x} = \tilde{P}_{s,x}(A)$, defined for diffusion process $\varsigma_t$ (1.8) (with transitional probability (1.2)) holds dispersion $b(t,\tilde{x})$ of $\tilde{x}_t$.
Process $\varsigma_t$ is a transformed version of process $\tilde{x}_t$ whose transition probability satisfies (1.1), and transformed probability $\tilde{P}_{s,x}$ for this process evaluates the Feller kernel measure [27, 29].
Since transformed process $\varsigma_t$ (1.8) has the same diffusion matrix but zero drift, the right part of additive functional in (1.5) satisfies

$$E[\int_s^T (\sigma(t,\tilde{x})^{-1} a(t,\tilde{x})d\xi(t)] = 0. \qquad (1.9)$$

We come to entropy functional expressed via parameters of the Ito stochastic equation in form:

$$S[\tilde{x}_t / \varsigma_t] = 1/2E[\int_s^T a(t,\tilde{x})^T (2b(t,\tilde{x}))^{-1} a(t,\tilde{x})dt] \ . \qquad (1.10)$$

Formulas (1.2-1.5), (1.6), (1.7), (1.8) and (1.10) are in [33] with related citations and references.
The entropy functional in forms (1.4,1.5,1.10) is an *information indicator* of distinction between the probability measures of processes $\tilde{x}_t$ and $\varsigma_t$; it measures a *quantity of information* of process $\tilde{x}_t$ regarding process $\varsigma_t$. For the process' equivalent measures, this quantity is zero, and it is positive for the process' equivalent measures.
Since mathematical expectation on the *process' trajectories* (1.5) is conditional to probability measure of Feller kernel, it is invariant at Markovian transformations defined through Radon-Nikodym's probability density measure (1.3) where both $P_{s,x}$ and $\tilde{P}_{s,x}$ are defined.
Thus, integral (1.10) is the entropy measure of Markov process $\tilde{x}_t$ while being conditional to the kernel probability measure. Entropy measure (1.4) is conditional to any transformed Markov diffusion process, not necessary satisfying (1.10).
Measuring conditional entropy (1.4), (1.10) relatively to diffusion process $\varsigma_t$, which models standard perturbations in controllable systems, is practically usable.

*Comments.* Differential entropy defined on set of random variable $X$ (with probability density $p(\upsilon)$) as a Shannon form of conditional entropy:

$$H_{\upsilon}=E_X[-\ln p(\upsilon)]=-\int_X p(\upsilon)\ln p(\upsilon)d\upsilon\,, \qquad (1.11)$$

is not equal to (1.4), and is not invariant under change of variables.
Therefore, it cannot be an entropy measure for an arbitrary continuous process, which is unlimited by special requirements [35, pp.224-238].
For the same reason, Shannon entropy (type $H_{\upsilon}$) with its density $p(\upsilon)$ is not founded for continuous process and is not sufficient and applicable for considered random process. •
The problem consists of evaluation information, that curries entropy functional (EF) at transforming above distributions while moving probability measure of processes $\tilde{x}_t$ to probability measure of $\varsigma_t$.
The transformation or movement involves an action on the EF through changing the drift function of it additive functional. Measuring EF along trajectory of process $\tilde{x}_t$ requires a multiple sequence of such transformations, which instantaneously convert probability of Markovian process to probability of Brownian process and vice versa:

$$P_{s,x}\to\tilde{P}_{s,x},\tilde{P}_{s,x}\to P_{s,x}\quad . \qquad (1.12)$$

Such nonstop transformations bring instant conditional probability's Brownian measure to the EF, implementing direct and correct measuring the random process entropy through this Radon-Nikodym's probability density measure.
The instant action on the functional drifts requires finding function which immediately executes that. The requirement satisfies Dirac's delta-function and its discrete form represented through Heaviside's step-up and step-down functions cutting the process on small intervals. Concurrent cutting each process dimension will instantaneously implement these transformations for $n$-dimensional Markov process under the manifold cutoff impulses.
Cutting the process by applying impulse control extracts concealed hidden information, which evaluates the EF process cutoff fractions.

**2. The information evaluation of the process' cut off operation by an impulse control**

Let us define control on the space $KC(\Delta,U)$ of a piece-wise continuous step functions $u_t$ at $t\in\Delta$:

$$u_-\overset{def}{=}\lim_{t\to\tau_k-o}u(t,\tilde{x}_{\tau_k}),\; u_+\overset{def}{=}\lim_{t\to\tau_k+o}u(t,\tilde{x}_{\tau_k})\,, \qquad (2.1)$$

which are differentiable on the set

$$\Delta^o=\Delta\setminus\{\tau_k\}_{k=1}^m, k=1,...,m, \qquad (2.1a)$$

and applied on diffusion process $\tilde{x}_t$ from moment $\tau_{k-o}$ to $\tau_k$, and then from moment $\tau_k$ to $\tau_{k+o}$, implementing the process' transformations $\tilde{x}_t(\tau_{k-o})\to\varsigma_t(\tau_k)\to\tilde{x}_t(\tau_{k+o})$; $n$-dimensional process holds $m$ such transformations.

At a vicinity of moment $\tau_k$, between the jump of control $u_-$ and the jump of control $u_+$, we consider a control *impulse*

$$\delta u_{\pm}(\tau_k)=u_-(\tau_{k-o})+u_+(\tau_{k+o}). \qquad (2.2)$$

The following statement evaluates the EF information contributions at such transformations.

<u>*Proposition 1.*</u>

Entropy functional (1.10) at the switching moments $t=\tau_k$ of control (2.2) takes values

$$\Delta S[\tilde{x}_t(\delta u_{\pm}(\tau_k)]=1/2, \qquad (2.3)$$

and at locality of $t=\tau_k$: at $\tau_{k-o}\to\tau_k$ and $\tau_k\to\tau_{k+o}$, produced by each of the impulse control's step functions in (2.1), is estimated by

$$\Delta S[\tilde{x}_t(u_-(\tau_k)]=1/4,\; u_-=u_-(\tau_k),\; \tau_{k-o}\to\tau_k \qquad (2.3a)$$

and

$$\Delta S[\tilde{x}_t(u_+(\tau_k)]=1/4,\; u_+=u_+(\tau_k),\; \tau_k\to\tau_{k+o}. \qquad (2.3b)$$

*Proof.* The jump of the control function $u_-$ in (2.1) from moment $\tau_{k-o}$ to $\tau_k$, acting on the diffusion process, might cut off this process after moment $\tau_{k-o}\to\tau_k$. The cut off diffusion process has the same drift vector and the diffusion matrix as the initial diffusion process.

The additive functional for this cut off has the form [13]:

$$\varphi_s^{t-}=\begin{cases}0, t\le\tau_{k-o};\\ \infty, t>\tau_k\end{cases}, \qquad (2.4)$$

The jump of the control function $u_+$ (2.1) from $\tau_k$ to $\tau_{k+o}$ might cut off the diffusion process *after* moment $\tau_k\to\tau_{k+o}$ with the related additive functional

$$\varphi_s^{t+}=\begin{cases}\infty, t>\tau_k\\ 0, t\le\tau_{k+o}.\end{cases}. \qquad (2.5)$$

For the control impulse (2.2), the additive functional at a vicinity of $t=\tau_k$ acquires the form of an *impulse function*

$$\varphi_s^{t-}+\varphi_s^{t+}=\delta\varphi_s^{\mp}, \qquad (2.6)$$

which summarizes (2.3) and (2.4).

Entropy functional (1.10) following from (2.4-2.5) takes values

$$\Delta S[\tilde{x}_t(u_-(t\le\tau_{k-o};t>\tau_k))]=E[\varphi_s^{t-}]=\begin{cases}0, t\le\tau_{k-o}\\ \infty, t>\tau_k\end{cases}, \qquad (2.7a)$$

$$\Delta S[\tilde{x}_t(u_+(t>\tau_k;t\le\tau_{k+o}))]=E[\varphi_s^{t+}]=\begin{cases}\infty, t>\tau_k\\ 0, t\le\tau_{k+o}\end{cases}, \qquad (2.7b)$$

changing from 0 to $\infty$ and back from $\infty$ to 0, acquiring *absolute maximum* at $t>\tau_k$, between $\tau_{k-o}$ and $\tau_{k+o}$.

The multiplicative functional [11], related to (2.4-2.5), are:

$$p_s^{t-} = \begin{cases} 0, t \le \tau_{k-o} \\ 1, t > \tau_k \end{cases}, \quad p_s^{t+} = \begin{cases} 1, t > \tau_k \\ 0, t \le \tau_{k+o} \end{cases}. \qquad (2.8)$$

Control impulse (2.2) provides an impulse probability density in form of multiplicative functional

$$\delta p_s^{\mp} = p_s^{t-} p_s^{t+} \ , \qquad (2.9)$$

where $\delta p_s^{\mp}$ holds $\delta[\tau_k]$-function, which determines probabilities $\tilde{P}_{s,x}(d\omega) = 0$ at $t \le \tau_{k-o}, t \le \tau_{k+o}$ and $\tilde{P}_{s,x}(d\omega) = P_{s,x}(d\omega)$ at $t > \tau_k$.

For the cutoff diffusion process, transitional probability (at $t \le \tau_{k-o}$ and $t \le \tau_{k+o}$) turns to zero, and the states $\tilde{x}(\tau_k - o), \tilde{x}(\tau_k + o)$ become independent, while their mutual time correlations *are dissolved*:

$$r_{\tau_k - o, \tau_k + o} = E[\tilde{x}(\tau_k - o), \tilde{x}(\tau_k + o)] \to 0. \qquad (2.10)$$

Entropy increment $\Delta S[\tilde{x}_t(\delta u_{\pm}(\tau_k))]$of additive functional $\delta \varphi_s^{\mp}$ (2.5), produced within or at a border of control impulse (2.2), defines equality

$$E[\varphi_s^{t-} + \varphi_s^{t+}] = E[\delta \varphi_s^{\mp}] = \int_{\tau_{k-o}}^{\tau_{k+o}} \delta \varphi_s^{\mp}(\omega) P_{\delta}(d\omega), \qquad (2.11)$$

where $P_{\delta}(d\omega)$ is a probability evaluation of the impulse $\delta \varphi_s^{\mp}$.

Integral of symmetric $\delta$-function $\delta \varphi_s^{\mp}$ between the above time intervals on the border is

$$E[\delta \varphi_s^{\mp}] = 1/2 P_{\delta}(\tau_k) \text{ at } \tau_k = \tau_{k-o}, \text{ or } \tau_k = \tau_{k+o}. \qquad (2.12)$$

The impulse, produced by deterministic controls (2.2) for each process dimension, is random with probability at $\tau_k$-locality

$$P_{\delta c}(\tau_k) = 1, k = 1, ..., m. \qquad (2.13)$$

This probability holds a jump-diffusion transition probability in (2.12) (according to [19]), which is conserved during the jump.

From (2.11)-(2.13) follows estimation of the entropy functional's increment under impulse control (2.2) applying at $t = \tau_k$ in form

$$\Delta S[\tilde{x}_t(\delta u_{\pm}(\tau_k))] = E[\delta \varphi_s^{\mp}] = 1/2, \qquad (2.14)$$

which proves (2.3), while delta impulse $\delta \varphi_s^{\mp} \to \infty$ brings absolute maximum to (2.14) within each $k$ cutoff impulse.

Symmetrical entropy contributions (2.6) at a vicinity of $t = \tau_k$:

$$E[\varphi_s^{t-}] = \Delta S[\tilde{x}_t(u_-(t \le \tau_{k-o}; t > \tau_k))] \text{ and } E[\varphi_s^{t+}] = \Delta S[\tilde{x}_t(u_+(t > \tau_k; t \le \tau_{k+o}))] \qquad (2.15)$$

estimate relations

$$\Delta S[\tilde{x}_t(u_-(t \le \tau_{k-o}; t > \tau_k))] = 1/4, \ u_- = u_-(\tau_k), \ \tau_{k-o} \to \tau_k; \qquad (2.16a)$$

$$\Delta S[\tilde{x}_t(u_+(t > \tau_k; t \le \tau_{k+o}))] = 1/4, \ u_+ = u_+(\tau_k), \ \tau_k \to \tau_{k+o}, \qquad (2.16b)$$

which proves (2.3a,b).

Entropy functional (1.10), defined through Radon-Nikodym's probability density measure (1.3), holds all properties of the considered cutoff controllable process, where both $P_{s,x}$ and $\tilde{P}_{s,x}$ are defined. Thus, cutting correlations (2.10) extracts entropy of hidden process information which directly measures each $\delta -$cutoff:

$$\Delta I_k[\tilde{x}_t(\delta u(\tau_k))] = \Delta S[\tilde{x}_t(\delta u_{\pm}(\tau_k))] = 1/2 \qquad (2.17)$$

Known information measures do not provide such measuring.

According the definition of entropy functional (1.5), it is measured in natural $\ln$ where each its Nat equals $\log_2 e \cong 1.44 bits$; therefore, it does not using Shannon entropy measure. •

*Corollaries*

From the Proposition it follows that:

(a)-Stepwise control function $u_- = u_-(\tau_k)$, implementing transformation $\tilde{x}_t(\tau_{k-o}) \to \varsigma_t(\tau_k)$, converts the EF from its minimum at $t \le \tau_{k-o}$(2.16a) to maximum at $\tau_{k-o} \to \tau_k$ (2.17);

(b)-Stepwise control function $u_+ = u_+(\tau_k)$, implementing transformation $\varsigma_t(\tau_k) \to \tilde{x}_t(\tau_{k+o})$, converts the EF from its maximum at $\tau_{k-o} \to \tau_k$ (2.17) to the minimum at $\tau_k \to \tau_{k+o}$(2.16b);

(c)-Impulse control function $\delta u_{\tau_k}^{\mp}$, implementing transformations $\tilde{x}_t(\tau_{k-o}) \to \varsigma_t(\tau_k) \to \tilde{x}_t(\tau_{k+o})$, switches the EF from its minimum to maximum and back from maximum to minimum, while the maximum of the entropy functional at a vicinity of $t = \tau_k$ allows the impulse control to deliver *maximal amount* of information (2.17) from these transformations;

(d)-Dissolving the correlation between the process' cutoff points (2.10) cuts *functional connections* at these discrete points, which border the Feller kernel measure [27, 29];

(e)-The relation of that measure to additive functional [29] in form (1.6) allows evaluating the *kernel's information* by the EF (1.5). The jump action (2.2) on Markov process, associated with "killing its drift", selects the Feller measure of the kernel [12,29,34], while the cutoff EF *provides information measure* of the Feller kernel (2.17);

(g)-Stepwise control $u_- = u_-(\tau_k)$, transferring the EF from $\tau_{k-o} \to \tau_k$, maximizes by moment $\tau_k$ the minimal information increment (brought at $t \to \tau_{k-o}$), implementing condition

$$\max_{\tau_k} \min_{\tau_{k-o}} \Delta I_k[\tilde{x}_t(\delta u(\tau_k))] \;; \qquad (2.17a)$$

(f)-Stepwise control $u_+ = u_+(\tau_k)$, transferring the EF from $\tau_k \to \tau_{k+o}$, kills the additive functional at stopping moment $\tau_{k+o}$ minimizing the maximal information increment by the end of this transformation, implementing condition

$$\min_{\tau_{k+o}} \max_{\tau_k} \Delta I_k[\tilde{x}_t(\delta u(\tau_k))]. \qquad (2.17b)$$

Such transformation associates with killing the Markovian process at the rate of increment of related additive functional $d\varphi_s^{ti} / \varphi_s^{ti}$ for each single dimension $i$ [11]. Control $u_+ = u_+(\tau_k)$ transfers the rate of killed Markov process to a process with probability (2.13), conserved during the jump, and starts a certain (non-random) process with the eigenvalue of diffusion operator [40], creating the process that balances the killing at the same rate [39].

The step-wise controls, acting on the multi-dimensional diffusion process dimensions, sequentially stops and starts the process, evaluating its multiple functional information.

The dissolved element of the correlation matrix at these moments provides independence of the cutting off fractions, leading to orthogonality of their correlation matrix $\bullet$.

### 3. Properties of the cutoff functional

Let us consider a sum of increments (2.17) under impulse control $\delta u(\tau_k)$, cutting the process $x_t$ at moments $\tau_k, k=1,2,...,m$ along the process' trajectory on intervals

$$s > \tau_1 > t_1 > \tau_2 > t_2,...,t_{k-1} > \tau_k > t_k,...,t_{m-1} > \tau_m > t_m = T, \quad (3.1)$$

where borders of these interval $t_{k-1}, t_k$ belong to the EF fractions before the cuts:

$$\Delta S_k[\tilde{x}_t(\mathrm{t} \to t_k)], \Delta S_{k+1}[\tilde{x}_t(\mathrm{t} \to t_{k+1})]. \quad (3.2)$$

Inside the border, at $t_{k-1} \le \tau_{k-o}, \tau_k, \tau_{k+o} \le t_k$, arise cutting information contributions $\Delta I_k[\tilde{x}_t(\delta u(\tau_k))]$.

1. Applying additive principle for the process' information, measured at the moments of dissolving correlation, we get summary information

$$I_{mo} = \sum_{k=1}^{m} \Delta I_k[\tilde{x}_t(\delta u(\tau_k))] = \Delta I_1[\tilde{x}_t(\delta u(\tau_1))] + \Delta I_2[\tilde{x}_t(\delta u(\tau_2))]|,...,+\Delta I_m[\tilde{x}_t(\delta u(\tau_m))]. \quad (3.3)$$

where impulses $\delta u(\tau_k)$ implement transitional transformations (1.2-1.3), initiating the Feller kernels along the process and extracting total kernel information for $n$-dimensional process with $m$ cuts off.

2. Sum $S_{mo}|_s^T$ of the additive fractions of EF on the finite time intervals

$s, t_1; t_1 + o_1, t_2; .., t_{k-1} + o_k, t_k; ..., t_{m-1} + o_m, t_m = T$; at

$$S_{mo}|_s^T = \Delta S_{1o}[\tilde{x}_t / \varsigma_t]|_s^{t_1} + \Delta S_{2o}[\tilde{x}_t / \varsigma_t]|_{t_1+o_1}^{t_2},...,+\Delta S_{mo}[\tilde{x}_t / \varsigma_t]|_{t_{m-1}+o_m}^{t_m} \quad (3.4)$$

approaches $I_{mo}$ (3.3) at $o_k \to \tau_k, k=1,2,...,m$:

$$\lim_{o_k \to \tau_k} S_{mo}|_s^T = I_{mo}, \quad (3.5)$$

while at any $o(\tau_k) > \tau_k$, $I_{mo} < S_{mo}$. (3.5a)

3. Sum (3.4) is less than $S[\tilde{x}_t / \varsigma_t]_s^T$, which is defined by the additive functional (1.7).

As a result, additive principle for a process' information, measured by the EF, is *violated*:

$$S_m|_s^T < S[\tilde{x}_t / \varsigma_t]_s^T. \quad (3.6)$$

4. *Information path functional* (IPF) defines distributed actions multi-dimensional delta-function on entropy functional (1.10) through the additive functional for all dimensions:

$$I_{pm} = \delta_m\{S[\tilde{x}_t / \varsigma_t]|\} = 1/2E\{\int_s^T \delta_m[a(t,\tilde{x})^T (2b(t,\tilde{x}))^{-1} a(t,\tilde{x})dt)]\} \quad (3.7)$$

which determines sum (3.2) of information measures $\Delta I_k[\tilde{x}_t(\delta u(\tau_k))]$ along the path on the cutting process intervals (3.1). In a limit it leads to

$$I_p = \lim_{m\to\infty} \sum_{k=1}^{m} \Delta I_k[\tilde{x}_t(\delta u(\tau_k))].\qquad (3.7a)$$

Formal definition (3.7) allows the IPF representation by Furies integral [36] leading to frequency analysis with Furies series.

5. IPF is the sum of *extracted* information which approaches theoretical measure (1.10):

$$I_p = \lim_{m\to\infty} I_{mo}|_s^T = \lim_{m\to\infty} S_{mo}|_s^T \to S[\tilde{x}_t / \varsigma_t]_s^T,\qquad (3.8)$$

if all finite time intervals $t_1 - s = o_1, t_2 - t_1 = o_2, ..., t_{k-1} - t_k = o_k, ..., t_m - t_{m-1} = o_m$, at $t_m = T$ satisfy condition

$$(T-s) = \lim_{m\to\infty} \sum_{t=s,m}^{t=T} o_m(t).\qquad (3.8a)$$

Then, the initially undefined (in (1.7)) upper time $T$ of EF integral (3.6) become limited.

At infinite sequence of the time intervals, this sequence has limit

$$\lim_{m\to\infty} o(t_m) \to 0.\qquad (3.8b)$$

Therefore, sum of such sequence is finite [36, p.130, 4.8-1].)

Realization (3.7),(3.8),(3.8a,b) requires applying the impulse controls at each instant $(\tilde{x}, s), (\tilde{x}, s+o(s)),...$ along the process trajectories of conditional math expectation (1.10).

However for any *finite* number $m$ of these instants, the integral process information, composed from the information, measured for the process' fractions, is not complete.

***The*** $I_p$ ***properties***:

1. The IPF measures information of the cutting process's interstate connections hidden by the states correlations, which are not covered by traditional Shannon information measure.
2. Since each cutting $\Delta S_k[\tilde{x}_t(\delta u(\tau_k))] = \Delta I_k[\tilde{x}_t(\delta u(\tau_k))]$ maximizes the cutting information, $I_p$ measures a total (integral) maximal information on the path.

The cutting control provides equal maxmin-minimax information contributions

$$\max_{\tau_k} \min_{\tau_{k-o}} \Delta I_k[\tilde{x}_t(\delta u(\tau_k))] = \min_{\tau_{k+1}} \max_{\tau_k} \Delta I_k[\tilde{x}_t(\delta u(\tau_k))]\qquad (3.9)$$

on each path $t_{k-1} \to (\tau_{k-o} \to \tau_k \to \tau_{k+o}) \to t_k$ from cutting $t_{k-1}$ to following cut $t_k$ (Coroll.a-c).

3. If each $k$-cutoff "kills" its process dimension at moment $\tau_{k+o}$, then $m = n$, and (3.7), (3.8),(3.8a,b) require infinite process dimensions.
4. At $m = n \to \infty, o_k = t_k - t_{k-1} \to \tau_k$ , the process time

$$(T-s) = \lim_{n\to\infty} \sum_{t=s,m}^{t=T} \tau_k(t)\qquad (3.10)$$

approaches the summary of the discrete intervals cutting all correlations.

5. Sequential cuts transform the IPF information contributions from each maximum through minimum to next maximal information contributions

$$\max_{\tau_k} \Delta I_k[\tilde{x}_t(\delta u(\tau_k))] \to \min_{\tau_{k+o}} \Delta I_k[\tilde{x}_t(\delta u(\tau_k))] \to \max_{\tau_k+1} \Delta I_k[\tilde{x}_t(\delta u(\tau_{k+1}))] \ , \qquad (3.11)$$

where each next maximum decreases at the cutoff moments

$$\max_{\tau_k+1} \Delta I_k[\tilde{x}_t(\delta u(\tau_{k+1}))] < \max_{\tau_k} \Delta I_k[\tilde{x}_t(\delta u(\tau_k))]. \qquad (3.12)$$

Each Dirac delta-function preserves its cutting information (2.17, 2.17a,b).

The information contribution by finale interval $o_m$ at its inner ending moment $\tau_{m+o}$, according to (2.7b), satisfies

$$\min_{\tau_{m+o}} \Delta I_m[\tilde{x}_t(\delta u(\tau_m))] \to 0, \qquad (3.13)$$

which limits sum (3.3) at $m = n \to \infty$.

6. Since EF functional $S[\tilde{x}_t / \varsigma_t]_s^T$ limits growth of $S_{mo}|_s^T$ in (3.8), it limits the IPF in (3.5a),(3.13), hence, the IPF approaches the EF functional during time (3.10) at unlimited increase of the process dimensions.

7. Because upper time $T$ in both the EF integral and IPF functional is limited by (3.8a) and (3.10), the entropy integral converges in the path functional, and both of them are restricted at the unlimited dimension number.

8. For any of these limitations, EF measure, taken along the process trajectories for time $(T-s)$, limits maximum of total process information, while IPF extracts maximum of the process hidden information during the same time and brings more information than Shannon traditional information measure for multiple states of the process.

9. Cutting all process correlations on $(T-s)$ transforms the initial random process to a limited sequence of independent states.

10. Discrete analog of Dirac delta function is Kronicker impulse, which takes values 0 and 1.

Probability measure of impulse distributions (2.6.7) is multiplicative functional (2.8) extended on $R^n$.

Conditional probabilities of distribution $F_\delta(x) = P_\delta(-\infty, x)$ **at** $F_\delta(-\infty) = 0, F_\delta(+\infty) = 1$ satisfies Kolmogorov's 1-0 law [37, p.117] for function $f(x)|\xi$**,** $\xi, x$ is infinite sequence of independent random variables:

$$P_\delta(f(x)|\xi) = \begin{cases} 1, f(x)|\xi) \geq 0 \\ 0, f(x)|\xi) < 0 \end{cases} \ . \qquad (3.14)$$

This probability measure has applied in [44,47] for the impulse probing of an observable random process, which holds opposite Yes-No probabilities – as the unit of probability step-function.

**Part 2. Evaluation of integral information, cutting from entropy *regular* integral, information path functional, and information micro- and macroprocess**

**1. The entropy regular integral functional**

Let us have a single dimensional Eq. (1.6) with drift function $a = c\tilde{x}(t)$ at given nonrandom function $c = c(t)$ and diffusion $\sigma = \sigma(t)$.

Then, entropy functional (1.10) acquires form

$$S[\tilde{x}_t / \varsigma_t] = 1/2 \int_s^T E[c^2(t)\tilde{x}^2(t)\sigma^{-2}(t)]dt\,, \quad (1.1)$$

from which, at $\sigma(t)$ and nonrandom function $c(t)$, we get

$$S[\tilde{x}_t / \varsigma_t] = 1/2 \int_s^T [c^2(t)\sigma^{-2}(t)E_{s,x}[x^2(t)]dt = 1/2 \int_s^T c^2[2b(t)]^{-1} r_s dt, \quad (1.2)$$

where for the diffusion process, the following relations hold true:

$$2b(t) = \sigma(t)^2 = dr/dt = \dot{r}_t\,, E_{s,x}[x^2(t)] = r_s\,. \quad (1.3)$$

This allows *identifying* the entropy functional on observed Markov process $\tilde{x}_t = \tilde{x}(t)$ by measuring above correlation functions, applying positive function $u(t) = c^2(t)$ and *representing* functional (1.10) through a regular integral of non-random functions

$A(t,s) = [2b(t)]^{-1} r_s$ (1.4a) and $u(t) = c^2(t)$ (1.4b)

in form

$$S[\tilde{x}_t / \varsigma_t] = 1/2 \int_s^T u(t)A(t,s)dt\,, \quad (1.4)$$

The $n$-dimensional functional integrant (1.4) follows directly from related $n$-dimensional covariations (1.3), dispersion matrix, and applying $n$-dimensional function $u(t)$. At given nonrandom function $u(t)$, (1.4) is regular integral which measures the entropy functional of Markov process at the probability transformation (1.1.2) with additive functional (1.1.7), where the integrant averages the additive functional.

**2. Application of variation condition (1.1.5a)**

Proposition 2.1.

Integral (1.4), satisfying variation condition (1.1.5a) at linear function $c^2(t) = u(t) = c^2 t$, has form

$$S[\tilde{x}_t / \varsigma_t] = 1/2 \int_s^T u(t)o(t)dt\,, \quad (2.1)$$

where the extreme of function (1.4a) holds minimum

$$A(t, s_k^{+o}) = o(s)b_k(s_k^{+o}) / b_k(t) = o(t)\,, \quad (2.2)$$

which decreases with growing time $t = s_k^{+o} + o(t)$ at $t \to T$ and fixed both $b_k(s_k^{+o})$ and

$$o(s) = A(s,s)\,. \quad (2.3)$$

Since satisfaction of this variation condition includes transitive transformation of a current distribution to that of Feller kernel, $b_k(t)$ is transition dispersion at this transformation, which is growing with the time of the transformation.

Proof. Applying Euler's equation to variation condition (1.1.5a) of this integral, at simple linear function $c^2(t) = u(t) = c^2 t$ and fixed $c^2$ on starting moment $s_k^{+o}$, we get

$$\partial A(t, s_k^{+o})t / \partial t = \dot{A}(t, s_k^{+o})t + A(t, s_k^{+o}) = 0\,, \quad (2.4)$$

and

$$t\partial A(t,s_k^{+o})/\partial t + A(t,s_k^{+o}) = 0, \partial A(t,s_k^{+o})/A(t,s_k^{+o}) + \partial t/t = 0, t \neq 0, A(t,s_k^{+o}) \neq 0, \partial t/t \neq 0. \quad (2.4a)$$

From these it follows

$$\ln A(t,s_k^{+o}) + \ln Ct = \ln[A(t,s_k^{+o})Ct] = 0, A(t,s_k^{+o})Ct = 1,$$

$$A(t,s_k^{+o})^{-1} = Ct, \quad (2.5a)$$

$$A(s_k^{+o}, t = s_k^{+o})^{-1} = Cs_k^{+o}, C = A(s_k^{+o}, s_k^{+o})^{-1}/s \quad (2.5b)$$

On a current time interval $\Delta_t = t - s_k^{+o}$, relations (2.5a,b) determine functions

$$A(s_k^{+o}, s_k^{+o})/A(t,s_k^{+o}) = t/s_k^{+o} \ . \quad (2.6)$$

Relations (1.3) allow representations

$$A(s_k^{+o}, s_k^{+o}) = r_k\,(s_k^{+o})/2b_k\,(s_k^{+o}) \ , \quad (2.7a)$$

$$A(s_k^{+o}, t) = r_k\,(s_k^{+o})/2b_k\,(t), \quad (2.7b)$$

where correlation function

$$r(s) = \int_s^{s+o(s)} 2b(t)dt = 2b(s)o(s) \quad (2.8)$$

leads to

$$A(s,s) = [2b(s)]^{-1} 2b(s)o(s) = o(s). \quad (2.9)$$

Substitution (2.9) and (2.7a,b) in (2.6) brings

$$t = s_k^{+o} b_k\,(t)/b_k\,(s_k^{+o}). \quad (2.10)$$

This shows that with growing time $t = s_k^{+o} + o(t)$ dispersion $b_k\,(t)$ has tendency to grow.
Substituting (2.10) and (2.9) in (2.5a) determines function

$$A(t,s_k^{+o}) = o(s) b_k\,(s_k^{+o})/b_k\,(t) = o(t), \quad \Delta_t = o(t) \quad (2.11)$$

which decreases with growing time $t \to T$.
The extreme holds minimum since second derivative of function (2.5a) is positive. •

<u>Comments 2.1.</u> On interval $\Delta_t$, moment $t = t_k$ in equation (2.10) in form

$$t = s_k^{+o} \dot{r}(t)/\dot{r}(s_k), \quad (2.12)$$

at $b(t) = 1/2\dot{r}(t), b(s_k) = 1/2\dot{r}(s_k)$, (2.12a)

can be identified by controlling correlation:

$$t_k \cong s_k r_k\,(t_k)/r_k\,(s_k) \ . \quad • \quad (2.12b)$$

Thus, EF (2.10) integrates time of correlation which function $u(t) = u_t$ cuts.

## 3. Impulse action on the entropy integral

Let define $u(t) = u_t$ on space $KC(\Delta, U)$ of a piece-wise continuous step-functions $u_t(u_-^t, u_+^t)$ at $t \in \Delta$, applying to integrant of (2.1) the difference of step-down $u_-^t = u_-(t)/\delta_o$ and step-up $u_+^t = u_+(t+\delta_o)/\delta_o$ functions at fixed interval $\delta_o : u_t^{\delta_o} = [u_-(t) - u_+(t+\delta_o)]/\delta_o = u_t^{\delta_o}$ (3.1) which forms Dirac's delta-function at

$$\lim_{\delta_o \to 0} = \delta u_t \,. \qquad (3.2)$$

Proposition 3.1.

Entropy integral (2.1) under impulse control (3.2) takes the following information values:

(a)-at a switching impulse middle locality $t = \tau_k$:

$$S[\tilde{x}_t / \varsigma_t]_{t=\tau_k} = 1/2 \text{ Nats}, \qquad (3.3)$$

(b)- at switching impulse left locality $t = \tau_k^{-o}$ :

$$S[\tilde{x}_t / \varsigma_t]_{t=\tau_k^{-o}} = 1/4 Nats \qquad (3.3a)$$

(c)-and at switching impulse right locality $t = \tau_k^{+o}$:

$$S[\tilde{x}_t / \varsigma_t]_{t=\tau_k^{+}} = 1/4 Nats \,. \qquad (3.3b)$$

Proof. Applying delta-function $c^2(t,\tau_k) = \delta u_t(t-\tau_k)$ to integral

$$\Delta S[\tilde{x}_t / \varsigma_t]|_{\tau_k^{-o}}^{\tau_k^{+o}} = 1/2 \int_{\tau_k^{-}}^{\tau_k^{+}} \delta u_t(t-\tau_k) o(t) dt \ , \ \tau_k^{-o} < \tau_k < \tau_k^{+o} \qquad (3.4)$$

determines functions [36,p.678-681]

$$\Delta S[\tilde{x}_t / \varsigma_t]|_{t=\tau_k^{-o}}^{t=\tau_k^{+o}} = \begin{cases} 0, t < \tau_k^{-o} \\ 1/4 o(\tau_k^{-o}), t = \tau_k^{-o} \\ 1/4 o(\tau_k^{+o}), t = \tau_k^{+o} \\ 1/2 o(\tau_k), t = \tau_k, \tau_k^{-o} < \tau_k < \tau_k^{+o} \end{cases} \,. \qquad (3.5)$$

Or such cutoff brings amount of entropy integral $S[\tilde{x}_t / \varsigma_t]_{t=\tau_k} = 1/2 o(\tau_k) = 1/2 \text{ Nats}$, while on borders of interval $o(\tau_k)$ the integral amounts are $S[\tilde{x}_t / \varsigma_t]_{t=\tau_k^{-o}} = 1/4 o(\tau_k^{-o}) Nats$ and $S[\tilde{x}_t / \varsigma_t]_{t=\tau_k^{+o}} = 1/4 o(\tau_k^{+o}) Nats$ accordingly. •

The results concur with Sec.1.1.2.

## 4. Discrete control action on the entropy functional

Let us find a class of step-down $u_-^t = u_-(\tau_k^{-o})$ and step-up $u_+^t = u_+(\tau_k^{+o})$ functions acting on discrete interval $o(\tau_k) = \tau_k^{+o} - \tau_k^{-o}$, which will preserve the Markov diffusion process' additive and multiplicative functions within the cutting process of the impulse.

Lemma 4.1.

1.Opposite discrete functions $u_-^t$ and $u_+^t$ in form

$$u_-(\tau_k^{-o}) = \downarrow_{\tau_k^{-o}} \bar{u}_-, u_+(\tau_k^{+o}) = \uparrow_{\tau_k^{+o}} \bar{u}_+ \qquad (4.1)$$

satisfy conditions of additivity

$$[u_+^t - u_-^t] = U_a \text{ (a) or } [u_+^t + u_-^t] = U_a \text{ (b)} \qquad (4.1A)$$

and multiplicativity

$$[u_+^t \times u_-^t] = U_m \qquad (4.1B)$$

at

$\mathrm{U}_a = U_m = \mathrm{U}_{am} = c^2 > 0$, **(4.1C)**

where instance-jump $\downarrow_{\tau_k^{-o}}$ has time interval $\bar{u}_-$ and opposite instance jump $\uparrow_{\tau_k^{+o}}$ has high $\bar{u}_+$ for relation (4.1A)(a) at real values

$\bar{u}_- = 0.5$, $\bar{u}_+ = 1, \bar{u}_+ = 2\bar{u}_-$. (4.2a)

and for relation (4.1A) (b) at real values

$\bar{u}_-^o = \bar{u}_+^o = 2$. (4.2b)

2. Complex functions

$u_t(u_\pm^{t1}, u_\pm^{t2}), u_\pm^{t1} = [u_+ = (j-1), u_- = (j+1)]$ , $j = \sqrt{-1}$ (4.2c)

satisfy conditions (4.1aA), (4.B) in forms

$u_+ - u_- = (j-1) - (j+1) = -2$, $u_+ \times u_- = (j-1)\times(j+1) = (j^2 - 1) = -2$,

which however do not preserve positive (4.1C).

Opposite complex functions

$u_t(-u_\pm^{t1}) = u_t(u_\pm^{t2}), u_\pm^{t2} = [u_+ = (j+1), u_- = (j-1)]$, (4.2d)

satisfy (4.1bA)-(4.C).

<u>Proofs</u> are straight forward.

Assuming both opposite function apply on borders of interval $o(\tau_k) = (\tau_k^{+o}, \tau_k^{-o})$ in forms $u_-^{t1} = u_-(\tau_k^{-o})$, $u_+^{t1} = u_+(\tau_k^{-o})$ and $u_-^{t2} = u_-(\tau_k^{+o})$, $u_+^{t2} = u_+(\tau_k^{+o})$, (4.2e)

at $u_-^{t1}u_+^{t1} = c^2(\tau_k^{-o})$, $u_-^{t2}u_+^{t2} = c^2(\tau_k^{+o})$, $t = \tau_k^{+o}$, (4.2f)

it follows that only by end of the interval at $t = \tau_k^{+o}$ the both Markov properties satisfy, while at beginning $t = \tau_k^{-o}$ the starting process does not possess yet these properties. •

<u>Corollary 4.1</u>.

1.Conditions 4.1A-4.1C imply that $c^2(\tau_k^{-o}), c^2(\tau_k^{+o})$ are discrete functions (4.1a),(4.2f) switching on interval $\Delta_\tau = \tau_k^{+o} - \tau_k^{-o}$.

Requiring $\Delta_\tau = \delta_o$ leads to discrete delta-function $\delta^o u_t$ (3.1) which for $\delta_o = (\tau_k^{+o} - \tau_k^{-o})$ holds

$\delta^o u_{t=\tau_k} = [u_-(\tau_k^{-o}) - u_+(\tau_k^{+o})]/(\tau_k^{+o} - \tau_k^{-o})$, that at $\Delta = (\tau_k^{-o} - s_k^{+o})$ brings

$u_-(\tau_k^{-o}) = -1_{\tau_k^{-o}}\bar{u}_-, u_+(\tau_k^{+o}) = +1_{\tau_k^{+o}}\bar{u}_+$, $\bar{u}_- = 0.5$ , $\bar{u}_+ = 2$, (4.3)

when positivity of $c^2 > 0$ implies

$\delta^o u_{t=\tau_k} = [u_+(\tau_k^{+o}) - u_-(\tau_k^{-o})]/(\tau_k^{+o} - \tau_k^{-o}) > 0$. (4.3a)

2.Discrete functions $u_+(s_k^{+o}) = +1_{s_k^{+o}}\bar{u}_+, u_-(\tau_k^{o}) = -1_{\tau k}^{+o}\bar{u}_-$ (4.3b)

on $\Delta$ are multiplicative:

$(u_-(\tau_k^{-o}) - u_+(s_k^{+o}))\times(u_-(\tau_k^{-o}) - u_+(s_k^{+o})) = [u_-(\tau_k^{-o}) - u_+(s_k^{+o})]^2$.

2a. Discrete functions (4.4e) in form

$\bar{u}_+ = j\bar{u}, \bar{u}_- = -j\bar{u}, \bar{u} \neq 0$ (4.3c)

satisfy only condition (4.1A) which for functions (4.3b) holds

$[u_-(\tau_k^{-o}) - u_+(s_k^{+o})]^2 = -(j\bar{u})^2[-1_{\tau_k^{-o}} - 1_{s_k^{+o}}]^2 > 0$. • (4.3d)

Let us find discrete analog of the integral increments under *discrete* delta-function (4.3),(4.3a):

$\delta^o u_{t=\tau} = (u_-(\tau_k^{-o}) - u_+(\tau_k^{+o}))(\tau_k^{-o} - \tau_k^{+o})^{-1}$.

Proposition 4.2.

A. Applying discrete delta-function (4.3) to integral (2.1) leads to

$$\Delta S[\tilde{x}_t/\varsigma_t]|_{t=\tau_k^{-o}}^{t=\tau_k^{+o}} = \begin{cases} 0, t < \tau_k^{-o} \\ 1/4u_-(\tau_k^{-o})o(\tau_k^{-o})/\tau_k^{-o}, t=\tau_k^{-o}, 1/4\downarrow 1_{\tau_k^{-o}}\bar{u}_{ko} \\ 1/2(u_-(\tau_k^{-o}) - u_+(\tau_k^{+o}))o(\tau_k)/(\tau_k^{+o} - \tau_k^{-o}), t=\tau_k, \tau_k^{-o} < \tau_k < \tau_k^{+o}, 1/2(\downarrow 1_{\tau_k^{-o}} - \uparrow 1_{\tau_k^{+o}})\bar{u}_{km} \\ 1/4u_+(\tau_k^{+o})o(\tau_k^{+o})/\tau_k^{+o}, t=\tau_k^{+o}, 1/4\uparrow 1_{\tau_k^{+o}}\bar{u}_{k1} \end{cases} \quad (4.4)$$

which is a discrete analog of (3.4b), where

$$\bar{u}_{ko} = \bar{u}_- \times o(\tau_k^{-o})/\tau_k^{-o}, \bar{u}_{km} = (\bar{u}_+ - \bar{u}_-)\times o(\tau_k)/(\tau_k^{+o} - \tau_k^{-o}), \bar{u}_{k1} = \bar{u}_+ \times o(\tau_k^{+o})/\tau_k^{+o},$$
$$\bar{u}_{km} = 1/2(\bar{u}_+ - \bar{u}_-) = 0.75, o(\tau_k) = \tau_k^{+o} - \tau_k^{-o}, o(\tau_k^{-o})/\tau_k^{-o} = 0.5, o(\tau_k^{+o})/\tau_k^{+o} = 0.1875 \quad , \quad (4.5)$$

and $|\bar{u}_- \times \bar{u}_+| = |1/2\times 2| = |\bar{u}_k| = |1|_k$ is multiplicative measure of impulse $(\downarrow 1_{\tau_k^{-o}} - \uparrow 1_{\tau_k^{+o}})\bar{u}_k$.

All entropy increments (4.4) and below are measured in Nats.

Let middle interval $\bar{u}_{km}$ in (4.4) measures a single unit impulse $\bar{u}_k = |1|_k$ information, then relative interval $\bar{u}_{ko} = 1/3\bar{u}_{km}$ follows $\bar{u}_{km} = 1$, and then $\bar{u}_{kio} = 1/3\bar{u}_{km} = \bar{u}_{ko}$.

Functions (4.4) determine finite size the impulse parameters $\bar{u}_{ko}, \bar{u}_k, \bar{u}_{k1}$ which estimate $\bar{u}_{km}$:

$\bar{u}_{ko} = 0.25 = 1/3\bar{u}_{km}, \bar{u}_{k1} = 2\times 0.1875 = 0.375 = 0.5\bar{u}_{km}$ . • (4.6)

Proofs follows from Proposition 4.3.

Let's consider entropy unit impulse $\bar{u}_s = |1|_s$ with moments $(s_k^{-o}, s_k^o, s_k^{+o})$ prior to impulse $\bar{u}_k = |1|_k$, which measures middle interval of impulse entropy $\bar{u}_{sm}$, to find increment of $\Delta S[\tilde{x}_t/\varsigma_t]|_{s_k^+}^{\tau_k^{-o}}$ on border of impulse $\bar{u}_k$ at priory $\Delta_{\tau s+} = \delta_{sk\pm} = (s_k^{+o} - \delta_k^{\tau-})$ and posterior $\Delta_{\tau s-} = \delta_{sk\mp} = (\delta_k^{\tau-} - \delta_k^{\tau+})$ moments under impulse functions of impulse $\bar{u}_s = |1|_s$:

$$\delta^o u_{\tau=(s_k^{+o} - \delta_k^{\tau-})} = (u_+(s_k^{+o}) - u_-(\delta_k^{\tau-}))(s_k^{+o} - \delta_k^{\tau-})^{-1} = \uparrow 1_{s_k^{+o}}\bar{u} - \downarrow 1_{\delta_k^{\tau-}}\bar{u} = [\uparrow 1_{s_k^{+o}} - \downarrow 1_{\delta_k^{\tau-}}]\bar{u}, \quad (4.7)$$
$$\delta^o u_{\tau=(\delta_k^{\tau-} - \delta_k^{\tau+})} = (u_-(\delta_k^{\tau-}) - u_+(\delta_k^{\tau+}))(\delta_k^{\tau-} - \delta_k^{\tau+})^{-1} = \downarrow 1_{\delta_k^{\tau-}}\bar{u} - \uparrow 1_{\delta_k^{\tau+}}\bar{u} = [\downarrow 1_{\delta_k^{\tau-}} - \uparrow 1_{\delta_k^{\tau+}}]\bar{u}, \quad (4.8)$$
$$\delta^o u_{\tau=(\delta_k^{\tau+} - \tau_k^{-o})} = (u_+(\delta_k^{\tau+}) - u_-(\tau_k^{-o}))(\delta_k^{\tau+} - \tau_k^{-o})^{-1} = \uparrow 1_{\delta_k^{\tau+}}\bar{u} - \downarrow 1_{\tau_k^{-o}}\bar{u} = [\uparrow 1_{\delta_k^{\tau+}} - \downarrow 1_{\tau_k^{-o}}]\bar{u}. \quad (4.9)$$

Here $\bar{u}$ evaluates each impulse interval, which according to the optimal principle is an invariant.

Functions (4.7)-(4.9) apply to additive sum of each increment of the entropy functional:

$$\Delta S[\tilde{x}_t/\varsigma_t]|_{s_k^+}^{\tau_k^{-o}} = \Delta S[\tilde{x}_t/\varsigma_t]|_{s_k^+}^{\delta_k^{\tau-}} + \Delta S[\tilde{x}_t/\varsigma_t]|_{\delta_k^{\tau-}}^{\delta_k^{\tau+}} + \Delta S[\tilde{x}_t/\varsigma_t]|_{\delta_k^{\tau+}}^{\tau_k^{-o}} \quad (4.10)$$

along time interval

$$\Delta_{\tau sk\pm} = s_k^{+o} - \delta_k^{\tau-} + \delta_k^{\tau-} - \delta_k^{\tau+} + \delta_k^{\tau+} - \tau_k^{-o} = s_k^{+o} - \tau_k^{-o} = \Delta_{\tau s}. \quad (4.10a)$$

Proposition 4.3.

A. The increments of entropy functional (4.10) collected on intervals (4.10a) hold

$$\Delta S[\tilde{x}_t/\varsigma_t]|_{s_k^+}^{\delta_k^{\tau-}} = 1/2(u_+(s_k^{+o}) - u_-(\delta_k^{\tau-}))o(s_k^{+o} - \delta_k^{\tau-}))(s_k^{+o} - \delta_k^{\tau-})^{-1} = 1/2[\uparrow 1_{s_k^{+o}} - \downarrow 1_{\delta_k^{\tau-}}]\bar{u}_{ks} \quad (4.11)$$

at $\bar{u}_{ks} = \bar{u}(o(s_k^{+o} - \delta_k^{\tau-})(s_k^{+o} - \delta_k^{\tau-})^{-1}$; (4.11a)

$$\Delta S[\tilde{x}_t/\varsigma_t]|_{\delta_k^{\tau-}}^{\delta_k^{\tau+}} = 1/2(u_-(\delta_k^{\tau-}) - u_+(\delta_k^{\tau+}))o(\delta_k^{\tau-} - \delta_k^{\tau+}))(\delta_k^{\tau-} - \delta_k^{\tau+})^{-1} = 1/2[\downarrow 1_{\delta_k^{\tau-}} - \uparrow 1_{\delta_k^{\tau+}}]\bar{u}_{k\delta s}, \quad (4.12)$$

at $\bar{u}_{k\delta s} = \bar{u} \times (o(\delta_k^{\tau -} - \delta_k^{\tau +}))(\delta_k^{\tau -} - \delta_k^{\tau +})^{-1}$ (4.12a)

and

$$\Delta S[\tilde{x}_t / \varsigma_t]|_{\delta_k^{\tau +}}^{\tau_k^{-o}} = 1/2(u_+(\delta_k^{\tau +}) - u_-(\tau_k^{-o}))o(\delta_k^{\tau +} - \tau_k^{-o})(\delta_k^{\tau +} - \tau_k^{-o})^{-1} = 1`/2[\uparrow 1_{\delta_k^{\tau +}} - \downarrow 1_{\tau_k^{-o}}]\bar{u}_{k\delta}, \quad (4.13)$$

at $[\uparrow 1_{\delta_k^{\tau +}} - \downarrow 1_{\tau_k^{-o}}]\bar{u}_{k\delta} = [\uparrow 1_{\delta_k^{\tau +}} + \uparrow 1_{\tau_k^{-o}}]\bar{u}_{k\delta}$ . (4.13a)

Here each impulse interval acquires specific entropy measure:

$$\bar{u}_{k\delta} = \bar{u} \times (o(\delta_k^{\tau +} - \tau_k^{-o}))(\delta_k^{\tau +} - \tau_k^{-o})^{-1} =$$
$$\bar{u} \times (o(\delta_k^{\tau +})(\delta_k^{\tau +} - \tau_k^{-o})^{-1}) + \bar{u} \times (o(\tau_k^{-o})(\tau_k^{-o})^{-1}(\delta_k^{\tau +} - \tau_k^{-o})^{-1}\tau_k^{-o} \quad (4.14)$$

at the impulse invariant interval $\bar{u}$.

Relation (4.14) leads to impulse interval

$$\bar{u}_{k\delta} = \bar{u}_{k\delta o} + \bar{u}_{k\delta 1} \quad (4.14a)$$

with parts

$$\bar{u}_{k\delta o} = \bar{u} \times (o(\delta_k^{\tau +}))(\delta_k^{\tau +} - \tau_k^{-o})^{-1}), \ \bar{u}_{k\delta 1} = \bar{u}_{ko1} \times \bar{u}_{ko2}, \quad (4.14b)$$

$$\bar{u}_{ko1} = \bar{u} \times (o(\tau_k^{-o}))(\tau_k^{-o})^{-1}, \ \bar{u}_{ko2} = \bar{u}^{-1} \times \tau_k^{-o}(\delta_k^{\tau +} - \tau_k^{-o})^{-1}. \quad (4.14c)$$

B. Intervals $\bar{u}_{ko1}$ and $\bar{u}_{ko2}$ are multiplicative parts of impulse step-control's interval $\bar{u}_{k\delta 1}$, which satisfies relations

$$\bar{u}_{k\delta o} = \bar{u}_{k\delta 1} = 1/2\bar{u}_{k\delta}, \ \ \bar{u}_{k\delta 1} = \bar{u}_{ko1} \quad (4.15)$$

where invariant impulse $|\bar{u}_{k\delta}| = |1|_s$, acting on time interval $\delta_k^{\tau +} = 2\tau_k^{-o}$, measures

$\bar{u}_{k\delta} = \bar{u}_{ks}$ at $|\bar{u}_{k\delta 1}| = 1/2\bar{u}_{sm}$, (4.15a)

and relative time intervals of $\bar{u}_{ko}$ and $\bar{u}_{k1}$ accordingly are

$$o(\tau_k^{-o})(\tau_k^{-o})^{-1} = 0.5, \ o(\tau_k^{+o})/\tau_k^{+o}) = 0.1875. \quad (4.15b)$$

Step-control impulse $\bar{u}_{k\delta}$ applies on two equal time intervals:

$(\delta_k^{\tau +} - \tau_k^{-o}) = \delta_k^{\tau +}/2$ (4.16a) and $\tau_k^{-o} = \delta_k^{\tau +}/2$. (4.16b)

On first (4.16a), its step-up part $[\uparrow 1_{\delta_k^{\tau +}}]$ captures entropy increment

$$\Delta S[\tilde{x}_t / \varsigma_t]|_{\delta_k^{\tau +}}^{\tau_k^{-o}} = 1`/2[\uparrow 1_{\delta_k^{\tau +}}]\bar{u}_- = 1`/8[\uparrow 1_{\delta_k^{\tau +}}], \quad (4.16)$$

on second (4.16b), its step-down multiplicative part (4.14b) at $\bar{u}_{ko2} = \bar{u}^{-1}$ transfers entropy (4.16) to control part $[\downarrow 1_{\tau_k^{-o}}]$ which cuts external entropy in impulse (4.4) at $\bar{u}_{ko1} = 1/2\bar{u}_{ko}$.

C.The applied *extremal* solution (Prop.2.1.), decreasing time intervals (2.11), brings (a)-persistence continuation sequence of the process impulses; (b)-the balance condition for entropy contributions; (c)-each impulse invariant unit $\bar{u}_k = |1|_k$, supplied by entropy unit $\bar{u}_s = |1|_s$, triples information, increases density of information each following information unit. •

Proofs.

The additive sum of entropy increments under impulses(4.7-4.9)satisfies balance condition:

$$\Delta S[\tilde{x}_t / \varsigma_t]|_{s_k^+}^{\tau_k^{-o}} = \Delta S[\tilde{x}_t / \varsigma_t]|_{s_k^+}^{\delta_k^{\tau -}} + \Delta S[\tilde{x}_t / \varsigma_t]|_{\delta_k^{\tau -}}^{\delta_k^{\tau +}} + \Delta S[\tilde{x}_t / \varsigma_t]|_{\delta_k^{\tau +}}^{\tau_k^{-o}} =$$
$$1/2[\uparrow 1_{s_k^{+o}} - \downarrow 1_{\delta_k^{\tau -}}]\bar{u}_{ks} + 1/2[\uparrow 1_{\delta_k^{\tau -}} - \uparrow 1_{\delta_k^{\tau +}}]\bar{u}_{k\delta s} + 1/2 \uparrow 1_{\delta_k^{\tau +}}\bar{u}_{k\delta o} - 1/2 \downarrow 1_{|\tau|_k^{-o}}\bar{u}_{k\delta 1} = 0 \quad (4.17)$$

where action $1/2\downarrow 1_{|\tau|_k^{-o}}\bar{u}_{k\delta 1} \Rightarrow 1/4\downarrow 1_{|\tau|_k^{-o}}\bar{u}_{ko}$ transfers related entropy increment to information $\Delta I[\tilde{x}_t / \varsigma_t](\tau_k^{-o}) = 1/4\downarrow 1_{|\tau|_k^{-o}}\bar{u}_{ko}$ on discrete locality $|\tau|_k^{-o}$ of action $\downarrow 1_{|\tau|_k^{-o}}\bar{u}_{k\delta 1}$.

Fulfilment relations

$$[\uparrow 1_{s_k^{+o}}\bar{u}_{ks} - \downarrow 1_{\delta_k^{\tau-}}\bar{u}_{ks} + \uparrow 1_{\delta_k^{\tau-}}\bar{u}_{k\delta s} - \uparrow 1_{\delta_k^{\tau+}}\bar{u}_{k\delta s} + \uparrow 1_{\delta_k^{\tau+}}\bar{u}_{k\delta o} - \downarrow 1_{|\tau|_k^{-o}}\bar{u}_{k\delta 1}] = 0$$

$$[\uparrow 1_{s_k^{+o}}\bar{u}_{ks} + \uparrow 1_{\delta_k^{\tau-}}[\bar{u}_{k\delta s} - \bar{u}_{ks}] + \uparrow 1_{\delta_k^{\tau+}}[\bar{u}_{k\delta o} - \bar{u}_{k\delta s}] = \downarrow 1_{|\tau|_k^{-o}}\bar{u}_{k\delta 1}, \downarrow 1_{|\tau|_k^{-o}}\bar{u}_{k\delta 1} = -1/2\downarrow 1_{|\tau|_k^{-o}}\bar{u}_{ko},$$

leads to sum of the control intervals:

$\bar{u}_{ks} - \bar{u}_{ks} + \bar{u}_{k\delta s} + \bar{u}_{k\delta s} - \bar{u}_{k\delta s} + \bar{u}_{k\delta o} - \bar{u}_{k\delta 1} = 0$, and $\bar{u}_{k\delta 1} = -1/2\bar{u}_{ko}$

Or to $\bar{u}_{k\delta o} = \bar{u}_{k\delta 1}$ (4.17a)

Impulse $[\uparrow 1_{\delta_k^{\tau+}}\bar{u}_{k\delta o} - \downarrow 1_{|\tau|_k^{-o}}\bar{u}_{k\delta 1}] = [\uparrow 1_{\delta_k^{\tau+}} + \uparrow 1_{|\tau|_k^{-o}}]\bar{u}_{k\delta}$ contains intervals $\bar{u}_{k\delta} = \bar{u}_{k\delta o} + \bar{u}_{k\delta 1}$,

where from (4.9), (4.13a) follows $\bar{u}_{k\delta} = \bar{u}$ , and (4.17a) leads to

$\bar{u}_{k\delta o} = \bar{u}_{k\delta 1} = 1/2\bar{u}$ . (4.17b)

Interval $\bar{u}_{k\delta} = \bar{u} \times [(o(\delta_k^{\tau+}))(\delta_k^{\tau+} - \tau_k^{-o})^{-1}) + (o(\tau_k^{-o}))(\tau_k^{-o})^{-1}(\delta_k^{\tau+} - \tau_k^{-o})^{-1}\tau_k^{-o}]$ (4.17c)

consists of $\bar{u}_{k\delta}$ components:

$\bar{u}_{k\delta o} = \bar{u} \times (o(\delta_k^{\tau+}))(\delta_k^{\tau+} - \tau_k^{-o})^{-1})$ and $\bar{u}_{k\delta 1} = \bar{u}_{ko1} \times \bar{u}_{ko2} / \bar{u}$ , (4.17d)

where $\bar{u}_{ko1} = \bar{u} \times (o(\tau_k^{-o}))(\tau_k^{-o})^{-1}$, $\bar{u}_{ko2} = \bar{u}^{-1} \times \tau_k^{-o}(\delta_k^{\tau+} - \tau_k^{-o})^{-1}$.

Intervals $\bar{u}_{ko1}$ and [ $\bar{u}_{ko2} / \bar{u}$ ] are multiplicative parts of impulse interval $\bar{u}_{k\delta 1}$ covered by $|\tau|_k^{-o}$.

From (4.17b) and relations (4.17d) it follows

$\bar{u}_{k\delta o} = \bar{u} \times (o(\delta_k^{\tau+}))(\delta_k^{\tau+} - \tau_k^{-o})^{-1}) = 1/2\bar{u}$,

$(o(\delta_k^{\tau+}))(\delta_k^{\tau+} - \tau_k^{-o})^{-1}) = 1/2$ (4.18)

and $\bar{u}_{ko1} = \bar{u} \times (o(\tau_k^{-o}))(\tau_k^{-o})^{-1} = 1/2\bar{u}$ . (4.18a)

That leads to

$(o(\tau_k^{-o}))(\tau_k^{-o})^{-1} = 1/2$, (4.18b)

and from (4.18) to

$(\delta_k^{\tau+} - \tau_k^{-o})^{-1}) = (\tau_k^{-o})^{-1}, \delta_k^{\tau+} - \tau_k^{-o} = \tau_k^{-o}$, $\tau_k^{-o}(\delta_k^{\tau+} - \tau_k^{-o})^{-1} = 1$ (4.18c)

and

$\tau_k^{-o} = 1/2\delta_k^{\tau+}$. (4.18d)

From (4.18c) it follows

$\bar{u}_{ko2} = \bar{u}^{-1}$. (4.18f)

Applying sequence of Eqs (4.7-4.9), at fixed invariant control $\bar{u}$, leads to

$\bar{u} = u_+(s_k^{+o}) - u_-(\tau_k^{-o})$, (4.19)

$\bar{u} = u_+(\delta_k^{\tau+}) - u_-(\tau_k^{-o})$ at $u_+(\delta_k^{\tau+}) - u_-(\tau_k^{-o}) = \bar{u}_{kb}$ , (4.19a)

which brings invariant $|\bar{u}_s| = |1|_s$ for both impulses (4.19) and (4.19a).

Relation

$u_+(s_k^{+o}) + u_-(\tau_k^{-o}) = 2[u_-(\delta_k^{\tau-}) + (u_+(\delta_k^{\tau+})] = 0$

following from the sequence (4.7-4.9) lead to

$u_-(\delta_k^{\tau-}) = -u_+(\delta_k^{\tau+})$, (4.19b)

or to reversing (mutual neutralizing) these actions on related moments $\delta_k^{\tau-} \cong \delta_k^{\tau+}$ .

Impulse interval $\bar{u}_{k\delta}$, with $\bar{u}_{k\delta o}$ and $\bar{u}_{k\delta 1}$, starts interval of applying step-down control $o(\tau_k^{-o})(\tau_k^{-o})^{-1} = 0.5$ in (4.4) at $\bar{u}_{k\delta 1} = \bar{u}_{k\delta o} = 1/2\bar{u}_{k\delta}$.

Invariant impulse $|\bar{u}_s| = |1|_s$ consisting of $[\uparrow 1_{\delta_k^{\tau+}} \downarrow 1_{|\tau|_k^{-o}}]\bar{u}_{k\delta}$, measures its interval

$\bar{u}_{kb} = \bar{u}_{sm} = \bar{u}_s$ at $\bar{u}_{k\delta 1} = 1/2\bar{u}_{sm}$. (4.19c)

At conditions (4.18c,d) limiting time-jump in (4.13), step-control impulse $\bar{u}_{k\delta}$ applies on two equal time intervals following from (4.19c).

On the first

$(\delta_k^{\tau+} - \tau_k^{-o}) = \delta_k^{\tau+}/2$

step-up part of $\bar{u}_{k\delta}$ -action $[\uparrow 1_{\delta_k^{\tau+}}]$ captures entropy increment

$\Delta S[\tilde{x}_t / \varsigma_t]|_{\delta_k^{\tau+}}^{\tau_k^{-o}} = 1`/2[\uparrow 1_{\delta_k^{\tau+}}]\bar{u}_- = 1`/8[\uparrow 1_{\delta_k^{\tau+}}]$. (4.20)

On the second interval

$\tau_k^{-o} = \delta_k^{\tau+}/2$ ,

the captured entropy (4.20) through the step-down multiplicative part (4.17c,d) delivers to control $\bar{u}_{ko} = \bar{u}_- \times o(\tau_k^{-o})(\tau_k^{-o})^{-1}$ equal information

$\Delta I[\tilde{x}_t / \varsigma_t]|_{\delta_k^{\tau+}}^{\tau_k^{-o}} = 1/4[\downarrow 1_{\tau_k^{-o}}]\bar{u}_- = 1/8[\downarrow 1_{\tau_k^{-o}}]$. (4.20a)

The control action $[\downarrow 1_{\tau_k^{-o}}]$ at $\bar{u}_- = 0.5$ cuts external entropy of correlation in impulse (4.4) at

$\bar{u}_{ko1} = 1/2\bar{u}_{ko}$

<u>Comment</u>. Action $[\uparrow 1_{\delta_k^{\tau+}}]$ cuts the captured entropy from impulse $\bar{u}_s = |1|_s$ , while multiplicative step-down part (4.17b) transforms the detained (captured) entropy to control information, thereafter converting the entropy to information at $\bar{u}_{ko2} = \bar{u}^{-1}$. •

Analogously to (4.20a), at the end of $k$ impulse, $\bar{u}_+$ captures information from interval $\bar{u}_{kio} = \bar{u}_- \times (o(\tau_k^{+o})/\tau_k^{+o})$:

$\Delta I[\tilde{x}_t / \varsigma_t]|_{\delta_{k+}^{\tau+}}^{\tau_k^{+o}} = 1/4[\uparrow 1_{\tau_k^{-o}}]\bar{u}_{kio}\bar{u}_+ = 1/4 \times (-2\bar{u}_{kio})[\uparrow 1_{\tau_k^{-o}}]$ (4.21)

and supplies it to $k+1$ impulse.

That leads to balance equation for information contributions of $k$-impulse:

$\Delta I[\tilde{x}_t / \varsigma_t]|_{\delta_k^{\tau+}}^{\tau_k^{-o}} + \Delta I[\tilde{x}_t / \varsigma_t]|_{\tau_k^{-o}}^{\tau_k} + \Delta I[\tilde{x}_t / \varsigma_t]|_{\tau_k}^{\tau_k^{+o}} = \Delta I[\tilde{x}_t / \varsigma_t]|_{\delta_{k+}^{\tau+}}^{\tau_k^{+o}}$, (4.21a)

where interval $\bar{u}_{kio}$ holds information contribution $\Delta I[\tilde{x}_t / \varsigma_t]|_{\tau_k}^{\tau_k^{+o}} = 1/4\bar{u}_{km}$ satisfied (4.4) at $\bar{u}_+ = -2$, which is measured by $\bar{u}_{km} = 0.75$ (4.5). That leads to relations

$$0.125 + 0.75 + \bar{u}_{kio} = -2\bar{u}_{kio}, 0.125 + 0.75 + 3\bar{u}_{kio} = 0,$$
$$\bar{u}_{k1} = 3\bar{u}_{kio} = 0.375 = \bar{u}_- \times o(\tau_k^{+o})/\tau_k^{+o}) \quad (4.22)$$

and

$o(\tau_k^{+o})/\tau_k^{+o} = 0.1875,$ (4.22a)

$\bar{u}_{ko} + \bar{u}_{km} + \bar{u}_{k1} = 1.25 = 5/3\bar{u}_{km}$ (4.22b)

from which and (4.21a) it follows

$$\Delta I[\tilde{x}_t / \varsigma_t]|_{\tau_k}^{\tau_k^{+o}} = 3\Delta I[\tilde{x}_t / \varsigma_t]|_{\delta_k^{\tau+}}^{\tau_k^{-o}} . \quad (4.23)$$

Ratio $\bar{u}_{k1} / 2\bar{u}_{kio} = 3/2$ at $2\bar{u}_{kio} = 0.25$ evaluates part of $k$ impulse information transferred to $k+1$ impulse. Relations (4.17b), (4.18b,d,f), (4.19c), and (4.22a) Prove the Proposition parts A-B.

Since $\bar{u}_{_} = 0.5$ is cutting interval of impulse $\bar{u}_k$, it allows evaluate additive sum of the discrete cutoff entropy contributions (4.4) during entire impulse $(\downarrow 1_{\tau_k^{-o}} - \uparrow 1_{\tau_k^{+o}}) = \delta_k$ using $\bar{u}_{_} = \bar{u}_k$:

$$\Delta S[\tilde{x}_t / \varsigma_t]|_{\tau_k^{-o}}^{\tau_k^{+o}} = 1/4\bar{u}_k / 2 + 1/2\bar{u}_k + 1/4 \times 3/2\bar{u}_k = \bar{u}_k , \quad (4.24)$$

That determines the impulse cutoff information measure

$$\Delta S[\tilde{x}_t / \varsigma_t]_{\delta_k} = \Delta I[\tilde{x}_t / \varsigma_t]_{\delta_k} = (\downarrow 1_{\tau_k^{-o}} - \uparrow 1_{\tau_k^{+o}})\bar{u}_k = |1|\bar{u}_k , \bar{u}_k = |1|_k \ Nat \quad (4.24a)$$

equals to $\cong 1.44$ Bit, which the cutting entropy functional that random process generates.

Thus, single unit impulse $\bar{u}_k = |1|_k$ measures the relative information intervals

$$\bar{u}_{ko} = 1/3\bar{u}_{km} , \ \bar{u}_{km} = 1 , \ \bar{u}_{kio} = 1/3\bar{u}_{km} = \bar{u}_{ko}, \text{ and } \tau_k^{+o} / \tau_k^{-o} = 3 . \quad (4.24b)$$

From $\bar{u}_{ko1} = 1/2\bar{u}_{sm}$ and $\bar{u}_{ko1} = 1/2\bar{u}_{ko} = 1/6\bar{u}_{km}$ it follows

$$\bar{u}_{km} = 3\bar{u}_{sm} , \quad (4.25)$$

which shows that impulse unit $\bar{u}_k = |1|_k$ triples information supplied by entropy unit $\bar{u}_s = |1|_s$ or interval $\bar{u}_k$ compresses three intervals $\bar{u}_s$.

At the extremal principle, each impulse holds invariant interval size $|\bar{u}_k| = |1|_k$ proportional to middle impulse interval $o(\tau)$ with information $\bar{u}_{km}$ which measures $o(\tau)$.

Condition of decreasing $t - s_k^{+o} = o(t) \to 0$ with growing $t \to T$ and squeezing sequence $s_k^{+o} \to \tau_{m-1}^{+o}, k = 1,2....m$ leads to persistence continuation of the impulse sequence with transforming each previous impulse entropy to information of following impulse $\bar{u}_s = |1|_s \to \bar{u}_k = |1|_k$.

The sequence of growing and compressed information increases at

$$\bar{u}_{k+1} = |3\bar{u}_k| = |1|_{k+1} . \quad (4.25a)$$

The persistence continuation of the impulse sequence links intervals between sequential impulses $(\bar{u}_{ks}, \bar{u}_{k\delta s}, \bar{u}_{k\delta o})$ whose imaginary (virtual) function $[\uparrow 1_{s_k^{+o}} - \downarrow 1_{\delta_k^{\tau-}} + \uparrow 1_{\delta_k^{\tau+}}]u$ prognosis entropies increments (4.11), (4.12), (4.10).

Information contributions at each cut $\delta_{k-1}, \delta_k$, $k, k+1, ..., m$: $\Delta I[\tilde{x}_t / \varsigma_t]_{\delta_{k-1}}, \Delta I[\tilde{x}_t / \varsigma_t]_{\delta_k}, ....$ determine time distance interval $\tau_k^{-o} - \tau_{k-1}^{+o} = o_s(\tau_k)$, when each entropy increment $\Delta S[\tilde{x}_t / \varsigma_t]|_{\tau_{k-1}^{+o}}^{t \to \tau_k^{-o}} = 1/2uo(\tau_k) = \bar{u}_s \times o_s(\tau_k)$ supplies each $\Delta I[\tilde{x}_t / \varsigma_t]_{\delta_k}$ satisfying

$\bar{u} \times o(\tau_k) = \Delta I[\tilde{x}_t / \varsigma_t]_{\delta_k}$ at $\bar{u} \times o(\tau_k) = \bar{u}_k(\tau_k^{+o} - \tau_k^{-o})$.

Hence, impulse interval

$$\bar{u}_k = \Delta I[\tilde{x}_t / \varsigma_t]_{\delta_k} / (\tau_k^{+o} - \tau_k^{-o}) \quad (4.26)$$

measures density of information at each $\delta_k = \tau_k^{+o} - \tau_k^{-o}$, which is sequentially increases in each following Bit. Relation (4.25a,b), (4.26) confirm part C of Proposition 4.3. •

Such Bit includes three parts:
1-the cutting step-down control's information delivered by capturing external entropy of random process;
2-the cutoff information, which this control cuts from random process;
3-the information delivered by the impulse step-up control, which being transferred to the nearest impulse, keeps information connection between the impulses, providing persistence continuation of the impulse sequence during the process time $T$.

<u>Corollaries 4.2</u>.

A.The additive sum of discrete functions (4.4) during the impulse intervals determines the impulse information measure equals to Bit, generated from the cutting entropy functional of random process.

The step-down function generates $1/8+0.75=0.875Nat$ from which it spends $1/8$ $Nat$ for cutting correlation while getting $0.75$ $Nat$ from the cut. Step-up function holds $1/8$ $Nat$ while $0.675$ $Nat$ it gets from cutting $0.75$ $Nat$, from which $0.5Nat$ it transfers to next impulse leaving $0.125$ $Nat$ within $k$ impulse.

The impulse has $1/8+0.75+1/8=1Nat$ of total $1.25Nat$ from which $1/8$ $Nat$ is the captured entropy increment from a previous impulse. The impulse actually generates $0.75Nat \cong 1Bit$, while the step-up control, using $1/8Nat$ ,transfers $2/8Nat$ information to next $k$ impulse, capturing $1/8Nat$ from the entropy impulse between $k$ and $k+1$ information impulses.

B.From total maximum $0.875$, the impulse cuts minimum of that maximum $0.75$ $Nat$ implementing minimax principle. By transferring overall $0.375Nat$ to next $k+1$ impulse that $k$ impulse supplies it with its maximum of $1/3\times 0.75Nat$ from the cutting information, thereafter implementing principle maximum of minimal cut.

C.Thus, each cutting Bit is *active information unit* delivering information from previous impulse and supplying information to following impulse, which transfers information between impulses.

Such Bit includes: the cutting step-down control's information delivered through capturing external entropy of random process; the cutoff information, which the above control cuts from random process; the information delivered by the impulse step-up control, which being transferred to the nearest impulse, keeps information connection between the impulses, providing persistence continuation of the impulse sequence.

D. The amount of information that each second Bit of the cutoff sequence condenses grows in three times, which sequentially increases the Bit information density. At the invariant increments of impulse (4.4), every $\overline{u}_k$ compresses three previous intervals $\overline{u}_{k-1}$, thereafter sequentially increases both density of interval $\overline{u}_k$ and density of these increments for each following $k+1$ impulse. •

**5. Information path functional in $n$-dimensional Markov process under $n$-cutoff discrete impulses**

The IPF unites information contributions extracting along $n$ dimensional Markov process:

$$I[\tilde{x}_t / \varsigma_t]|_{s_k^-}^{\tau_n^{+o}\to T} = \lim_{k=n\to\infty} \sum_{k=1}^{k=n} \Delta I[\tilde{x}_t / \varsigma_t]_{\delta_k}\ , \qquad (5.1)$$

where each dimensional information contribution $\Delta I[\tilde{x}_t / \varsigma_t]_{\delta_k} = |1|\,\overline{u}_k Nat$, satisfying ratios of intervals

$$\overline{u}_{k+1} = |3\overline{u}_k| = |1|_{k+1}, \overline{u}_k = |3\overline{u}_{k-1}|, \qquad (5.2)$$

increases in each third interval in $\overline{u}_{(k+1)} / \overline{u}_{(k-1)} = 9$ times, concentrating in impulse $|1|_{k+1}$, at

$$\overline{u}_k = \Delta I[\tilde{x}_t / \varsigma_t]_{\delta_k} / (\tau_k^{+o} - \tau_k^{-o}) \qquad (5.3)$$

which measures density of information at each $\delta_k = \tau_k^{+o} - \tau_k^{-o}$.

Since sequence of $\bar{u}_k, k=1,...,\infty$ is limited by

$$\lim_{k\to\infty} |1| \bar{u}_k \le |1|_{k\to\infty} \ ,\qquad (5.3a)$$

information contributions of the sequence in (5.1) converges to this finite integral.

Each Bit $|1|_k$ distinguishes from other $|1|_{k+1}$ by the impulse space-time geometry[46] depending on $\bar{u}_k$ density.

Each dimensional cutoff enables converting entropy increment $\Delta S[\tilde{x}_t / \varsigma_t]_{\tau_k}$ to kernel information contribution $\Delta I[\tilde{x}_t / \varsigma_t]_{\tau_k}$ with optimal density (5.3) satisfying (5.2).

The following Propositions detail the IPF specifics.

Proposition 5.1.

A. Optimal distance between nearest $k-1,k$ information impulses, measured by difference between each previous ending finite intervals $\bar{u}_{k-1}$ and following starting finite interval $\bar{u}_k$ relatively to following interval $\bar{u}_k$: $\Delta^*_{uk} = (\bar{u}_{k-1} - \bar{u}_k)/\bar{u}_k$ , decreases twice for each fixed $k-1,k$:

$$\Delta^*_{uk} = 1/2^k \, .\qquad (5.4)$$

Indeed,

$$\Delta^*_{uk} = (\bar{u}_{k-1} - \bar{u}_k)/\bar{u}_k = (3/2\bar{u}_{k-1m} - 1/3\bar{u}_{km})/1/3\bar{u}_{km} = 1/2 \, .$$

Here $3/2\bar{u}_{k-1m}$ measures information of $k-1$ impulse's ending interval, $1/3\bar{u}_{km}$ measures information of $k$ impulse's starting interval.(Intervals $o_s(\tau_k)$ are along the EF, $\delta_k$ measures impulse intervals along the IPF).

B. With growing $k \to n$, each previous impulse decreases the between impulses distance by $\Delta^*_{un} = 1/2^n$, which at very high process dimension $n$, in limit:

$$\lim_{n\to\infty} [\Delta^*_{un} = 1/2^n] \to 0 \, .\qquad (5.4a)$$

The finite impulses entropy increment located between the information impulses:

$$\Delta S[\tilde{x}_t / \varsigma_t]|_{\tau^{+o}_{k-1}}^{t \to \tau^{-o}_k} = 1/2uo(\tau_k) = \bar{u}_s \times o_s(\tau_k)\qquad (5.4b)$$

at finite impulses $o_s(\tau_k)$ determines density measure for the impulse of invariant size of $\bar{u}_s = |1|_s$:

$$\Delta S[\tilde{x}_t / \varsigma_t]|_{\tau^{+o}_{k-1}}^{t \to \tau^{-o}_k} / o_s(\tau_k) = \bar{u}_s \ .\qquad (5.5)$$

With growing $k \to n$, decreasing interval $o_s(\tau_k) \to 0$ is limited by minimal physical time interval.

Eq (2.2.1) shows that a source of entropy increment (5.4b) between impulses is *time course*

$$\Delta_k = (\tau^{-o}_k - \tau^{+o}_{k-1}) \to o_s(\tau_k) \, ,\qquad (5.5a)$$

moving the nearest impulses closer. Moreover, each moment $\delta^{\tau+}_k$ of this time course $\Delta_k$ pushes for automatic conversion its entropy density to information density $\bar{u}_k = \Delta I[\tilde{x}_t / \varsigma_t]_{\delta_k} / (\tau^{+o}_k - \tau^{-o}_k)$ in information impulse $\bar{u}_k = |1|_k$, where the relative time intervals between impulses (5.4) measures also information density (4.23).

Distance between nearest information impulses (5.5a) evaluates interval of forming entropy increments

$$\Delta_{ks} = 2\tau^{-o}_{k-1} \to o_s \ .\qquad (5.5b)$$

Finite (5.5),(5.5a,b) limit both information density and equivalence of the entropy and information functionals. Time course intervals (5.5a,b) also runs to convert entropy increment (5.5) in kernel information contribution $\Delta I[\tilde{x}_t / \varsigma_t]_{\delta_k}$ for each cutoff dimension and drives the sequential *integration* for all contributions.

C. With decreasing $\Delta_t = t - s_k^{+o} = o(t)$ at $t \to T$, both $\Delta_k$ and $\delta_k$ are reduced to zero in limit:

$$\lim_{k\to\infty}\Delta_k = 1/2\lim_{k\to\infty}\delta_k \to 0 \ , \qquad (5.6)$$

which follows from

$$\Delta_t = t - s_k^{+o} = o(t) \to \Delta_k = o(\tau_k) \qquad (5.6a)$$

at $t \to \tau_k^{-o}, \tau_{k-1}^{+o} \to s_k^{+o}$ and reduced $o(t)$ at $t \to T$.

C. Total sum of the descending time distances at satisfaction (5.4-5.6):

$$\lim_{n\to\infty}\sum_{k=1}^{k=n}\Delta_k = 1/2\lim_{n\to\infty}\sum_{k=1}^{k=n}\delta_k = T - s \qquad (5.7)$$

is finite, converging to total interval of integrating entropy functional (2. 2.1).

D. Sum of information contributions $\Delta I[\tilde{x}_t / \varsigma_t]_{\delta_k}$ on whole $(T-s)$ is converging to both path functional integral and the entropy increments of the initial entropy functional:

$$\lim_{k=n\to\infty}\sum_{k=1}^{k\to n}\Delta I[\tilde{x}_t / \varsigma_t]_{\delta_k} \ \to I[\tilde{x}_t / \varsigma_t]_s^T = S[\tilde{x}_t / \varsigma_t]_s^T, \qquad (5.8)$$

limiting the converging integrals at the finite time interval (5.7).

The integrals contributions' time course runs integration of the impulse contributions in (2.2.1).

Information density $\overline{u}_k$ of each dimensional information contribution $\Delta I[\tilde{x}_t / \varsigma_t]_{\delta_k}$ grows according to (5.3), approaching infinity at the limit (5.6). •

Comments

1.1. Sequence of integrant of (2.2.1):

$$\delta s_k[\tilde{x}_t / \varsigma_t] = 1/2 u_k o(t_k), \mathrm{k} = 1,....,\infty \qquad (5.8a)$$

at limited $\mathrm{u}_{k\to\infty} = c^2 > 0$ leads to

$$\lim_{k\to\infty}\delta s_k[\tilde{x}_t / \varsigma_t] = 1/2 u_k o(t_k) = 0, \qquad (5.8b)$$

where each integrant (5.8a) is an entropy density, which impulse control $u_k$ converts to information density. Hence, information density at infinite dimensions is finite.

1.2. Sum of the invariant information contributions on the discrete intervals increases, whereas (5.8) integrates the previous contributions. This allows integrates any number of the process' connected information Bits, providing total process information including both random inter-states' and inter-Bits connections. •

The increments of correlation functions within optimal interval $\Delta_k = \tau_k^{-o} - s_k^{+o}$ and on the cutoff time borders $\tau_k^{-o}, \tau_k^{+o}$ determine

Proposition 5.2.

A. Correlation function on discreet interval $\Delta_t = t - s_k^{+o}$ for the extremal process holds

$$r_k^-(t) = 1/2 r_k\,(s_k^{+o})[t^2 / (s_k^{+o})^2 + 1]\,|_{s_k^{+o}}^{t\to\tau_k^{-o}}, \qquad (5.9)$$

ending with correlation on the cutoff left border $\tau_k^{-o}$:

$$r_k^-(\tau_k^{-o}) = 1/2 r_k\,(s_k^{+o})[(\tau_k^{-o} / s_k^{+o})^2 + 1] \ . \qquad (5.9a)$$

After the cutoff, correlation function on following time interval $(\tau_{k+1}^{-o}-\tau_k^{+o})$ holds
$r_k^+(t)=1/2r_k(\tau_k^{+o})[t^2/(\tau_k^{+o})^2+1]|_{\tau_k^{+o}}^{t\to\tau_{k+1}^{-o}}$. (5.10)

B. Correlation on right border $\tau_k^{+o}$ of the finite cutoff at $\tau_k^{+o}/\tau_k^{-o}=3$ holds:

$r_k^+(\tau_k^{+o})=1/2r_k(\tau_k^{-o})[(\tau_k^{+o}/\tau_k^{-o})^2+1]|_{\tau_k^{-o}}^{\tau_k^{+o}}=5r_k(\tau_k^{-o})$. (5.11)

C. Difference of these correlations, according to (5.4), at $\delta_k^r=\tau_k^{+o}-\tau_k^{-o}=1/2o(\tau_k)$ is

$r_{ko}^+(\tau_k^{+o})-r_{ko}^-(\tau_k^{-o})=\Delta r_{ko}(\delta_k),\ \Delta r_{ko}(\delta_k)=5r_k(\tau_k^{-o})-r_k(\tau_k^{-o})=4r_k(\tau_k^{-o})$ (5.12)

and its relative value during that finite cutoff holds

$\Delta r_{ko}(\delta_k)/r_k(\tau_k^{-o})=4$. (5.13)

Correlation within cutoff moment $\tau_k=1/2\delta_k^r=1/4o(\tau_k)$ evaluates

$r_k^+(\tau_k)\to 0$ at $o(\tau_k)\to 0$. (5.13a)

<u>Proof A, B,C.</u> Relation $t=s_k^{+o}b_k(t)/b_k(s_k^{+o})$, at $b_k(t)=1/2\dot r_k(t)$, determines functions

$\dot r_k(t)=2b_k(s_k^{+o})t/s_k^{+o}$ at $b_k(s_k^{+o})s_k^{+o}=1/2r_k(s_k^{+o})$ and solution

$$r_k(t)=\int_{s\to s_k^{+o}}^{t\to\tau_k^{-o}}2b_k(s_k^{+o})t/s_k^{+o}=b_k(s_k^{+o})t^2/s_k^{+o}+C_1, C_1=1/2r_k(s_k^{+o})\ . \quad (5.14)$$

From (5.14) follows correlation function on this interval (5.9) and its end (5.9a) for the extremal process.
After the cutoff, correlation function on the next time interval $(\tau_{k+1}^{-o}-\tau_k^{+o})$ holds (5.10).

The correlation, preceding the current cut on its left border $\tau_k^{-o}$:

$r_{ko}^-(\tau_k^{-o})=1/2r_k(s_k^{+o})[3^2+1]=5r_k(s_k^{+o})$, (5.15)

grows in five time of the optimal correlation for previous cutoff at $s_k^{+o}$.

Correlation on right border $\tau_k^{+o}$ of finite cutoff (5.11) allows finding both difference of these correlations on $\delta_k=\tau_k^{+o}-\tau_k^{-o}=1/2o(\tau_k)$ in (5.12) and its relative value during the finite cutoff in (5.13). •

Let us find the entropy increments under control $u_-(\tau_k^{-o}), u_+(\tau_k^{-o}-\delta_k^{\tau+})$ near a left border of cut $t=\tau_k^{-o}-\delta_k^{\tau+}$.
Applying (4.10), (4.11) at $t=\tau_k^{-o}-\delta_k^{\tau+}$ leads to

$\Delta S[\tilde x_t/\varsigma_t]|_{\tau_k^{-o}-\delta_k^{\tau+}}^{t\to\tau_k^{-o}}=-1/2(u_-(\tau_k^{-o})-u_+(\tau_k^{-o}-\delta_k^{\tau+}))(\tau_k^{-o}-\delta_k^{\tau+})(\delta_k^{\tau+})^{-1}(\tau_k^{-o}-\delta_k^{\tau+})=$

$-1/2u_-(\tau_k^{-o})(\tau_k^{-o}-\delta_k^{\tau+})-u_+(\tau_k^{-o}-\delta_k^{\tau+})(\tau_k^{-o}-\delta_k^{\tau+})(\tau_k^{-o}-\delta_k^{\tau+}))(\delta_k^{\tau+})^{-1}$ . (5.16)

If both entropy measure of these controls:

$\Delta S_u=1/2[u_-(\tau_k^{-o})-u_+(\tau_k^{-o}-\delta_k^{\tau+})](\tau_k^{-o}-\delta_k^{\tau+})$ (5.17)

and interval $(\tau_k^{-o}-\delta_k^{\tau+})$ are finite, then entropy increment near the border is infinite:

$\Delta S[\tilde x_t/\varsigma_t]|_{\tau_k^{-o}-\delta_k^{\tau+}}^{t\to\tau_k^{-o}}=\Delta S_u(\tau_k^{-o}-\delta_k^{\tau+})(\delta_k^{\tau+})^{-1}\to\infty$, at $\delta_k^{\tau+}\to 0$. (5.18)

Entropy of control (5.17):

$\Delta S_u=1/2[u_-(\tau_k^{-o})-u_+(\tau_k^{-o}-\delta_k^{\tau+})](\tau_k^{-o}-\delta_k^{\tau+})=1/2(2j[\uparrow 1_{\delta_k^{\tau+}}+\downarrow 1_{\tau_k^{-o}}])\,(\tau_k^{-o}-\delta_k^{\tau+})=j[\uparrow 1_{\delta_k^{\tau+}}+\downarrow 1_{\tau_k^{-o}}](\tau_k^{-o}-\delta_k^{\tau+})$

at $(\tau_k^{-o}-\delta_k^{\tau+})=\delta_k^{\tau+}$, (5.19)

compensates for the infinity in (5.18), when imaginary Bit of potential control $j[\uparrow 1_{\delta_k^{\tau+}}+\downarrow 1_{\tau_k^{-o}}]$ applied on interval $(\tau_k^{-o}-\delta_k^{\tau+})=\delta_k^{\tau+}$ compensates for relative interval $(\delta_k^{\tau+})^{-1}(\tau_k^{-o}-\delta_k^{\tau+})$.

Both real controls $u_-(\tau_k^{-o}),u_+(\tau_k^{-o}-\delta_k^{\tau+})$ generates real Bit $(-1_{\tau_k^{-o}}+1|_{\delta_k^{\tau+}/2})$ which actually compensates for this infinite increment.

The opposite actions of functions $u_+(\delta_k^{\tau+}/4)$ and $u_-(t=\delta_k^{\tau+}/2)\rightarrow u_-(\tau_k^{-o})$ model an interaction on $\delta_k^{\tau+}/2$ of a random process with applied control $u_-(\tau_k^{-o})$, which provides external influx entropy that this control captures.

If interactive action $u_+(\delta_k^{\tau+}/4)$ precedes $u_-(\tau_k^{-o})$, then this control is a reaction on $u_+(\delta_k^{\tau+}/4)$, while the control information covers the influx of entropy within interval

$$\delta_k^{\tau+}/2=\tau_k^{-o}. \quad (5.20)$$

Opposite symmetric actions $u_+(\delta_k^{\tau+}/4)=j(+1_{\delta_k^{\tau+}/4})$ and $u_-(t=\delta_k^{\tau+}/2)=j(-1_{\delta_k^{\tau+}/2})$, at

$$(t-\delta_k^{\tau+}/4)^{-1}(\delta_k^{\tau+}/4)^2, t=\delta_k^{\tau+}/2, (t-\delta_k^{\tau+}/4)^{-1}(\delta_k^{\tau+}/4)^2=\delta_k^{\tau+}/4, \quad (5.21)$$

bring total imaginary entropy (potential) influx:

$$\Delta S[\tilde{x}_t/\varsigma_t]|_{\delta_k^{\tau+}/4}^{\delta_k^{\tau+}/2}=-1/2[j(-1_{\delta_k^{\tau+}/2})-j(+1_{\delta_k^{\tau+}/4})](\delta_k^{\tau+}/4)=$$
$$1/4j[+1_{\delta_k^{\tau+}/4})-1_{\delta_k^{\tau+}/2}]\overline{u}_k\delta_k^{\tau+}=1/4j[1_{\delta_k^{\tau+}/4}^{\delta_k^{\tau+}/2}]\overline{u}_k\delta_k^{\tau+} \quad (5.22)$$

with two opposite imaginary entropies:

$$S_+[\tilde{x}_t/\varsigma_t]|_{\delta_k^{\tau+}/2}^{t\rightarrow\tau_k^{-o}}=1/8j[1_{\delta_k^{\tau+}/4}^{\delta_k^{\tau+}/2}]\overline{u}_k\delta_k^{\tau+}, \quad (5.23a)$$

$$S_-[\tilde{x}_t/\varsigma_t]|_{s_k^+}^{t\rightarrow\tau_k^{-o}}=-1/8j[1_{\delta_k^{\tau+}/2}^{\tau_k^{-o}}]\overline{u}_k\delta_k^{\tau+}. \quad (5.23b)$$

Action $u_-(t=\delta_k^{\tau+}/2)=j(+1_{\delta_k^{\tau+}/2})$ coincides with start of real control $u_-(\tau_k^{-o})$, while entropy (5.16) with $S_+[\tilde{x}_t/\varsigma_t]|_{\delta_k^{\tau+}/2}^{t\rightarrow\tau_k^{-o}}=1/8$ Nat evaluates difference between interactive action $u_+(\delta_k^{\tau+}/4)$ and potential reaction $u_-(t=\delta_k^{\tau+}/2)$.

Equivalent parts of entropy of interaction between these actions evaluates multiplicative relation $S_+\times S_-=1/2S_-^+$ which for (5.23a, b) leads to

$$S_-^+=-1/32. \quad (5.24)$$

Information of control, starting at $\delta_k^{\tau+}/2=\tau_k^{-o}$ with its impulse wide $\overline{u}_k\delta_k^{\tau+}$, actually implements this interaction spending part of its information

$$\Delta I_-^+=0.25\times 1/32=0.0078\cong 0.008Nat \quad (5.25)$$

on compensating the interaction, while capturing (5.22) in a move to the cut.

Thus, information covering (5.23b), includes $\Delta I_-^+$ with total contribution

$$\Delta I_-^o=\Delta I_-+\Delta I_-^+=0.125+0.008=0.133Nat. \quad (5.26)$$

The conjugated components $S_+,S_-$ start not simultaneously but with equal values (5.23a,b), acquiring by moment $\delta_k^{\tau+}/2=\tau_k^{-o}$ *dissimilarity*:

$S_-^o = 0.117 Nat$ and $\Delta I_-^o = 0.133 Nat$ (5.27)
which satisfies minimal difference between direct action and its reaction [43] at time shift
$\delta_k^{\tau+} / 4$. (5.28)

Real control $u_-(\tau_k^{-o})$, applied instead of imaginary action $u_-(\delta_k^{\tau+}/2)$, converts total entropy (5.16) on interval (5.28) to the equal control information and compensates for (5.22). That includes (5.23b) and (5.24).
Entropy gap (5.23a,b) between the anti-symmetric actions is imaginable, as well as time interval $\Delta_t = (t - s_k^{+o})$, compared with $t = \tau_k^{-o}$ when control $u_-(\tau_k^{-o})$ applies and covers the gap.
At satisfaction (5.24), the delivered information compensates for entropy
$\Delta S[\tilde{x}_t / \varsigma_t]|_{t \to \delta_k^{\tau+}/4}^{\tau_k^{-o}} \to -1/4 Nats$. (5.29)
Results (5.16)-(5.29) *extend Prop.4.3 specifying information process of capturing external entropy influx.*

## 6. Microprocess within the impulse

### 6.1.The entropy increments in the microprocess under step-functions (4.3a,b).

Within the discrete impulse (3.1) cutting Markov diffusion process arises an inner process $\tilde{x}_{otk} = \tilde{x}(t \in o(\tau_k)))$ called a microprocess under action of function $u_t(u_-^t, u_+^t) = c^2(t \in o(\tau_k))$ at each fixed impulse interval $o(\tau_k)$.
Step-down $u_-^t = u_-(\tau_k^{-o})$ and step-up $u_+^t = u_+(\tau_k^{+o})$ functions acting on discrete interval $o(\tau_k) = \tau_k^{+o} - \tau_k^{-o}$, which satisfy (4.1A-4.1C) and (4.2a-4.2d), generates the EF (1.1.10) increments:

$$\delta S_- = \delta S_-[u_-^t], \delta S_+ = \delta S_+[u_+^t] , \quad (6.1)$$

preserving the additive and multiplicative properties within the microprocess.
Starting step functions $u_\pm^{t1}$ (4.2c) at locality of beginning impulse $\tau_k^{-o}$ initiates increments of the entropies on interval $o(\tau_k - 0)$ by moment $t = \tau_k - 0$:

$$\delta S_+[u_+^{t1}] = \delta S_+^1(t = \tau_k - 0)) = \delta S_+^1(t = \tau_k^{-o}) \uparrow_{\tau_k^{+o}} (j-1),$$
$$\delta S_-[u_-^{t1}] = \delta S_-^1(t = \tau_k - 0) = S_-^1(t = \tau_k^{-o}) \downarrow_{\tau_k^{-o}} (j+1) \quad (6.2)$$

Step functions $u_\pm^{t2}$ (4.2d) starting at $t = \tau_k - 0$ contribute the entropy increments on interval $o(\tau_k)$ by moment $t = \tau_k$:

$$\delta S_+[u_+^{t2}] = \partial S_+^2(t = \tau_k) = \partial S_+^2(t = \tau_k - 0)) \uparrow_{\tau_k} (j+1),$$
$$\delta S_-[u_-^{t2}] = \partial S_-^2(t = \tau_k) = \partial S_-^2(t = \tau_k - 0)) \downarrow_{\tau_k} (-j+1) \quad (6.3)$$

Complex function $u_+^{t1}$ turns on the multiplication of functions $\delta S_+^1(t = \tau_k^{-o})$ on angle $\varphi_+^1 = -\pi/4$, and function $u_-^{t1}$ turns on the multiplication function $\partial S_-^1(t = \tau_k^{-o})$ on angle $\varphi_-^1 = \pi/4$ by moment $t = \tau_k - 0$ :

$$\delta S_+^1(t = \tau_k - 0)) = \delta S_+^1(t = \tau_k^{-o}) \times \uparrow_{\tau_k^{+o}} -\pi/4, \delta S_-^1(t = \tau_k - 0)) = \delta S_-^1(t = \tau_k^{-o}) \times \downarrow_{\tau_k^{+o}} \pi/4. \quad (6.4)$$

Analogously, step functions $u_\pm^{t2}$ starting at $t = \tau_k - 0$ turn entropy increments (6.4) on angles $\varphi_-^2 = \pi/4$ by moment $t = \tau_k$ and on angle $\varphi_+^2 = -\pi/4$ the following entropy increments by moment $t = \tau_k$:

$\delta S_{-}^{2}(t=\tau_k)=\delta S_{-}^{2}(t=\tau_k-0)\times\downarrow_{\tau_k}\pi/4, \delta S_{+}^{2}(t=\tau_k)=\delta S_{+}^{2}(t=\tau_k-0)\times\uparrow_{\tau_k}-\pi/4$ . (6.5)

The difference of angles between the functions in (6.4): $\varphi_{+}^{1}-\varphi_{-}^{1}=-\pi/2$ is overcoming on time interval $o(\tau_k-0)=\tau_k^{-o}+1/2o(\tau_k)$. After which control $u_{\pm}^{t2}$, starting with opposite increments (6.5), turns them on angle $\varphi_{-}^{2}-\varphi_{+}^{2}=\pi/2$, equalizing (6.5).

That launches entanglement of entropies increments and their angles *within* interval $o(\tau_k)$:

$\delta S_{+}^{2}(t=\tau_k)=\delta S_{-}^{2}(t=\tau_k)=\delta S_{\mp}^{2}$. (6.5a)

Turning the initial time-located vector -function $u_{-}^{t}=u_{-}(\tau_k^{-o}):\xleftarrow{\tau_k^{-o},\bar{u}_{-}=0.5,}\delta\varphi_1=0$ on angle $\delta\varphi_1=\varphi_{+}^{1}-\varphi_{-}^{1}=\pi/2$ transforms it to space vector $u_{+}(\tau_k-0)=\uparrow_{\tau_k-o}\bar{u}_{+}=1$ during a jump from moment $t=\tau_k^{-o}$ to moment $t=\tau_k-0$ on interval $o(\tau_k-0)$ in (6.4). Then vector-function $\downarrow_{\tau_k}\bar{u}_{-}^{o}=2$ , starting on time $t=\tau_k-0$ in (6.5) with space high interval $\bar{u}_{-}^{o}=2$ jumps to vector-function $\uparrow_{\tau_{k+0}}\bar{u}_{+}^{o}=2$ forming on time interval $o(\tau_k+0)=1/2o(\tau_k)+\tau_k^{+}$ the additive space- time impulse

$u_{\mp}=[\downarrow_{\tau_k+0}\bar{u}_{-}^{o}]+[\uparrow_{\tau_k^{+o}}\bar{u}_{+}^{o}]$. (6.6)

The first part of (6.6) equalizes (6.5a) within *space-time* interval $\bar{u}_{-}\times 1/2o(\tau_k)$, then joins, summing them on $\bar{u}_{-}\times o(\tau_k+0)$, which finalizes the entanglement. The last part of impulse (6.6) cuts-kills the entangled increments on interval $\bar{u}_{+}\times\tau_k^{+}$ at ending moment $\tau_k^{+}$. Section 6.4 details the time –space relation.

Relations (6.1-6.6) lead to following specifics of the microprocess.

6.1a.Step functions $u_{\pm}^{t1}$ initiate microprocess $\tilde{x}_{otk1}=\tilde{x}(t\in o(\tau_k-0))$ on beginning of the impulse discrete interval $o(\tau_k-0)$ with only additive increments (6.2).

Opposite step functions $u_{\pm}^{t2}$ continue the microprocess within interval $o(\tau_k+0)$ at $\tilde{x}_{otk2}=\tilde{x}(t\in o(\tau_k+0))$ with both additive and multiplicative increments (6.3) preserving the process Markov properties.

6.1b.Space-time impulse (6.6) within interval $o(\tau_k+0)$ processes entanglement of increments (6.5a) of microprocess $\tilde{x}_{otk2}=\tilde{x}(t\in o(\tau_k+0))$ summing these increments on $o(\tau_k)$ locality of $t=\tau_k$:

$S_{\mp}^{o}=2\delta S_{\mp}^{2}[(o(\tau_k)]$. (6.7)

Then it kills entropies (6.7) at ending moment $\tau_k^{o+}\to\tau_k^{+}$:

$S_{\mp}^{o}[\tau_k^{+}]=0$. (6.7a)

The microprocess, producing entropy increment (6.7) within the impulse interval, is reversible before killing which converts the increments in equal information contribution $S_{\mp}^{o}[\tau_k^{+}]\Rightarrow\Delta I[\tau_k^{+}]$.

The information emergence at ending impulse time interval accomplices injection of an energy with step-up control $[\uparrow_{\tau_k^{+o}}\bar{u}_{+}^{o}]$ [44, 47], which follows the impulse mutual interaction and/or with environment.

From that moment starts an irreversible information process.

6.1c. Transferring the initial time-located vector to equivalent space-vector $\uparrow_{\tau_k-o}\bar{u}_{+}$ transforms a transition impulse, concentrating within time $\tau_k^{-o}$ of interval duration $\bar{u}_{-}=0.5$, in space interval $\bar{u}_{+}=1$.

The opposite space vector $\downarrow_{\tau_k} \bar{u}_-^o = 2$ acting on relative time interval $1/2o(\tau_k)/(\tau_k^{+o} - \tau_k^{-o}) = 0.5$ forms space-time function $\downarrow_{\tau_k} \bar{u}_-^1, \bar{u}_-^1 = 2 \times 0.5 = 1$, which, as inverse equivalent of opposite function $\uparrow_{\tau_k - o} \bar{u}_+$, neutralizes it to zero cutting both $\bar{u}_+ = 1$ and time duration $\bar{u}_- = 0.5$ while concentrating them during the transition in interval $\tau_k - (\tau_k - 0) = 0_k$. Within the impulse, only step-down functions $[\downarrow_{\tau_k^{-o}} \bar{u}_-]$ on time interval $\bar{u}_- = 0.5$ and step-up function $[\uparrow_{\tau_k^{+o}} \bar{u}_+^1]$ on space-time interval $\bar{u}_+^1 = \bar{u}_+ \times \tau_k^+ = 2_{\tau_k^+}$ are left. That determines size of the discrete 1−0 impulse by multiplicative measure $U_m = |0.5 \times 2| = |1|_k = \bar{u}_k$ generating an information bit**.** This means, functions $u_+(\tau_k - 0) = \uparrow_{\tau_k - o} \bar{u}_+$ and $\downarrow_{\tau_k} \bar{u}_-^o$ are transitional during formation of that impulse and creation time-space microprocess $\tilde{x}_{otk} = \tilde{x}(t \in 1/2o(\tau_k), h_k \in 2_{\tau_k^+})$ within the impulse with final entropy increment (6.7).

**6.2. Conjugated dynamics of the microprocess within the impulse**

Opposite functions jumps $u_\pm^{t1}(t^*)$ starting at beginning of the process with relative time

$$t^* = [\mp\pi / 2\delta t^{\pm *} / \mathrm{o}(\tau_k)], \delta t^{\pm *} \in (\delta t_{ok}^\pm \to 1/2\,\mathrm{o}(\tau_k)), \quad (6.8)$$

hold directions of opposite impulses

$$u_\pm^{t1} = [u_+ = \uparrow_{\tau_k^{+o}} (j-1), u_- = \downarrow_{\tau_k^{+o}} (j+1)] \quad (6.8a)$$

on interval $\delta_o[\mathrm{t}_o^{*-}, \mathrm{t}_o^{*+}] = \delta t^* < o(\tau)$ at a locality of the impulse initial time $\tau_k^{-o}$.

The opposite jumps (6.8a) initiate relative increments of entropy:

$$\frac{\delta S}{S} / \delta t^* = u_\pm^{t1},\ u_\pm^{t1} = [u_+ = \uparrow_{\tau_k^{+o}} (j-1), u_- = \downarrow_{\tau_k^{+o}} (j+1)], \quad (6.9)$$

which in a limit leads to differential Egs

$$\dot{S}_+(t^*) = (j-1) S_+(t^*), \dot{S}_-(t^*) = (j+1) S_-(t^*). \quad (6.9a)$$

Solutions of (6.9a) describe opposite process' entropies- function of relative time $t^*$:

$$S_+(t^*) = [exp(-t^*)(\mathrm{Cos}(t^*) - jSin(t^*))]|_{t_o^{*-}}^{1/2o(\tau_k)}, S_-(t^*) = [exp(t^*)(\mathrm{Cos}(t^*) + jSin(t^*))]|_{t_o^{*+}}^{1/2o(\tau_k)} \quad (6.10)$$

with initial conditions $S_+(t_o^{*-})$, $S_-(t_o^{*+})$ at moment $t_o^{*+} = t_o^{*-} = [\mp\pi / 2\delta t_{ok}^\pm]$.

Starting moment $\delta t_{ok}^\pm = \delta t_o^\pm / \tau_k^{-o} = \pm 0.82$ determines relative wide of step-function $u_\pm^{t1}$: $\delta t_o^\pm / \mathrm{o}(\tau_k) = 0.2 + 0.005 = 0.205$ and the impulse initial relative interval of this function $\tau_k^{-o} / \mathrm{o}(\tau_k) = 0.25$.

From that follows the solutions by moment $\delta t_{ok}^\pm = \pm 0.82$:

$$S_+(t_o^+) = [exp(-\pi/2 \times 0.82)(\mathrm{Cos}(\pi/2 \times 0.82)) - jSin(\pi/2 \times -0.82))] = 0.2758 \times 1,$$
$$S_-(t_o^-) = [exp(\pi/2 \times -0.82)(\mathrm{Cos}(-\pi/2 \times -0.82) + jSin(-\pi/2 \times -0.82))] = 0.2758 \times 1. \quad (6.10a)$$

The solutions by moment $\delta t^{*\mp} = 1/2o(\tau_k)$ of time

$$t^{*-} = -\pi/2 \times 1/2\,\mathrm{o}(\tau_k)/\mathrm{o}(\tau_k) = -\pi/4, t^{*+} = \pi/2 \times 1/2\,\mathrm{o}(\tau_k)/\mathrm{o}(\tau_k) = \pi/4 \quad (6.10b)$$

are

$S_+(t^{*-}) = S_+(t_o^{*-}) \times \exp(-\pi/4)[\mathrm{Cos}(\pi/4) - jSin(\pi/4)],$
$S_-(t^{*+}) = S_-(t_o^{*+}) \times \exp(-\pi/4)[\mathrm{Cos}(-\pi/4) + jSin(-\pi/4)] =$ . (6.11)
$S_-(t_o^{*+}) \times \exp(-\pi/4)[\mathrm{Cos}(-\pi/4) - jSin(\pi/4)]$

These vector-functions at opposite moments (6.10b) hold opposite signs of their angles $\mp\pi/4$ with values:

$S_+(t^-) \cong 0.2758 \times 0.455 \cong +0.125, S_-(t^+) \cong 0.2758 \times 0.455 \cong -0.125$. (6.11a)

Function $u_{\pm}^{t2}$, starting with these opposite increments, turn them on angle $\varphi_-^2 - \varphi_+^2 = \pi/2$ that equalizes the increments and starts entangling both equal increments with their angles within interval $t = \tau_k \mp 0$:

$S_-^2(t = \tau_k + 0) = \delta S_-^1(t = \tau_k - 0) \times \downarrow_{\tau_k+0} \pi/2 =$
$S_-^1(t = \tau_k - 0) \times \exp(\pi/2 \times t_{\tau_k+0}^{*+})[\mathrm{Cos}\,\pi/2 \times t_{\tau_k+0}^{*+}) + \mathrm{jSin}(\pi/2 \times t_{\tau_k+0}^{*+}),$
$S_+^2(t = \tau_k + 0) = \delta S_+^1(t = \tau_k - 0) \times \uparrow_{\tau_k+0} \pi/2 =$ (6.12)
$S_-^1(t = \tau_k - 0) \times \exp(-\pi/2 \times t_{\tau_k+0}^{*-})[\mathrm{Cos} - \pi/2 \times t_{\tau_k+0}^{*-}) + \mathrm{jSin}(-\pi/2 \times t_{\tau_k+0}^{*-})$

at moments

$t_{\tau_k+0}^{*\pm} = [\mp\pi/2\delta t_{1k}^{\pm}], \delta t_{1k}^{\pm} = \delta t_1^{\pm}/1/2\tau_k \cong 0.4375, \delta t_1^{\pm} = \pm(0.5 - \delta t_{\pm}^{k1}),$
$\delta t_{\pm}^{k1} = \tau_k^{-o}/\tau_k + \delta t_{\pm}^{ko}/\tau_k = 0.25 + 0.03125 = 0.2895$ , (6.12a)

where $\delta t_{\pm}^{ko}/\tau_k \cong 32^{-1}$ evaluates dissimilarities (following (5.24-5.27)) between functions $u_{\pm}^{t2} = [u_+ = (j+1), u_- = (j-1)]$ at switching from $t = \tau_k - 0$ to $t = \tau_k$.

Resulting values at $t = \tau_k + 0$ are

$S_-^2(t = \tau_k + 0) = 0.125\exp(\pi/2 \times 0.4375) \times 1 \cong 0.25,$
$S_+^2(t = \tau_k + 0) = 0.125\exp(\pi/2 \times 0.4375) \times 1 \cong 0.25$ (6.13a)

which being in the same direction are summing in this locality:

$S_{\mp}^o = 2S_{\mp}^2[(\delta t_{\pm}^{ko}/\tau_k)] \cong \mp 0.5$. (6.13b)

The entanglement, starting with (6.13a), continues at (6.13b) and up to cutting all entangled entropy increments.

The $t = \tau_k \mp 0$ locality evaluates the $0_k$-vicinity of action of inverse opposite functions (6.9), whose signs imply the signs of increments in (6.13b) and in the following formulas.

The subsequent step-up function changes increment (6.13b) according to Eqs

$S_{\mp}(\tau_k^{+o}) = S_{\mp}^o(\delta t_{\pm}^{ko}/\tau_k) \times \exp(t_{\tau_k^{+o}}^{*+}), t_{\tau_k^{+o}}^{*+} = [\pi/2\delta t_k^{*o}], \delta t_k^{*o} \in (\delta t_{1k}^{*o} \to \tau_k^{+o}/\tau_k),$ (6.14)

at $\delta t_{1k}^{*o} = \delta t_{1k}^{\pm}/1/2\tau, \delta t_{1k}^{\pm} = \pm(0.5 - \delta t_{\pm}^{k1}), \delta t_{\pm}^{k1} = \delta t_{\pm}^{ko}/\tau_k + \tau_k^{+o}/\tau_k = 0.25 + 0.03125 = 0.2895,$
$\delta t_{1k}^{\pm} = \delta t_1^{\pm}/1/2\tau_k \cong 0.4375$

with resulting value

$S_{\mp}(\tau_k^{+o}) = \mp 0.5\exp(\pi/2 \times 0.4375) = \mp 0.5 \times (\cong 2) \cong \mp 1,$ (6.14a)

which measures total entropy of the impulse $\bar{u}_k = |1|_k = 1Nat$. (6.14c)

Trajectories (6.10-6.14) describe anti-symmetric conjugated dynamics of the microprocess within the impulse, which up to the cutting is reversible, generating the entangled entropy increments (6.14a).

Cutting this joint entropy at moment $\tau_k^+ \cong 0_k + \tau_k^{o+}$ coverts it to equal information contribution $S_{\mp}^o[\tau_k^+] = \Delta I[\tau_k^+] \cong 1.44\,\mathrm{bit}$ (6.14d) that each $\bar{u}_k$ impulse produces.

The cut involves an interaction which imposes irreversibility on information process with multiple cutting bits. Interacting impulse outside of the impulse microprocess delivers entropy on $0_k$-vicinity of the cutting moment:

$$S_c^*(\tau_k^+) = \exp 0_k = 1 \ . \qquad (6.14e)$$

Each current impulse requests an interaction for generating information bit from the microprocess reversible entropy, since the impulse contains requested action $[\uparrow_{\tau_k^{+o}} \bar{u}_+^o]$ (in (6.6)).

**6.3. Probabilities functions of the microprocess**

Amplitudes of the process probability functions at $S_\mp^*(\tau_k^{+o}) = |S_+^*| = |S_-^*| = 1$ are equal:

$$p_{+a} = 0.3679, p_{-a} = 0.3679 \, . \qquad (6.15)$$

That leads to

$$p_{+a} p_{-a} = p_{\pm a}^2 = 0.1353, S_{\mp a}^* = -\ln p_{a\pm}^2 = 2 \, , \qquad (6.15a)$$

or at $S_{\mp a}^* = 2, p_{a\pm} = \exp(-2) = 0.1353$, (6.15b)

where $S_{\mp a}^* = S_\mp^*(\tau_k^{+o}+) + S_c^*(\tau_k^+ +)$ (6.16)

includes the interactive components at $\tau_k^{+o}+$ .

Functions $u_+ = (j-1), u_- = (j+1)$, satisfying (4.2A), fulfill additivity at the impulse starting interval $\mathrm{o}[t_o^\mp]$, running the anti-symmetric entropy fractions, while opposite functions $u_+ = (1+j), u_- = (1-j)$, satisfying (4.2B) by the end of impulse at $\uparrow_{\tau_{k+}^{+o}} \bar{u}_\pm$, mount entanglement of these entropy fractions within impulse' $|1/2 \times 2| = |\bar{u}_k| = |1|_k$ space interval $\bar{u}_\pm = \pm 2$.

The entangling fractions hold the equal impulse probabilities (6.15), which indicates appearance of both entangled antisymmetric fractions simultaneously with starting space interval.

Probabilities $p_{\pm a}$ of interacting probability amplitudes $p_{+a}, p_{-a}$ satisfies multiplicativity $p_{\pm a} = \sqrt{p_{+a} p_{-a}}$ , but sum of the non-interacting probabilities doesn't: $p_+ + p_- = \exp(-S_+^*) + \exp(-S_-^*) = p_\pm \neq \mathrm{p}_{a\pm}$ being unequal to both interacting probability $p_{\pm a}$ and the summary probability $p_{\pm am} = 0.7358$ of the non-interacting entropy (6.15).

The interacting probabilities in transitional impulse $[\uparrow 1_{\tau|_k^-} \downarrow 1_{\tau|_k^+}] \bar{u}_k$ on $\tau_k$-locality violates their additivity, but preserves additivity of the entropy increments. The impulse microprocess on the ending interval preserves both additivity and multiplicativity only for the entropies.

These basic results are the impulse' entropy and probability equivalents for the quantum mechanics (QM) probability amplitudes relations. However the impulse cutting probabilities $p_+, p_-$, are probability of random events in the hidden correlations, while probability amplitudes $p_{+a}, p_{-a}$ are attributes of the microprocess starting within the cutting impulse.

That distinguishes the considered microprocess from the related QM equations, considered physical particles.

The entropy of multiple impulses integrate microprocess along the observing random distributions.

With minimal starting impulse entropy ½ Nat, each initial process entropy $S_\pm(t_o) = 0.25 Nat$ in such impulse self-generates entropy $S_{\mp a}^* = 0.5 Nat$ .That starts a virtual observer' [47] time–space microprocess with probability $p_{a\pm} = \exp(-0.5) = 0.6015$. Probability $p_{a\pm} = 0.1353$, relational to the impulse initial conditions, evaluates appearance of time–space real impulse that *decreases* an initial entropy on $S_{\mp a}^* = 2$ Nat (satisfying (6.16)). The impulse's invariant measure, satisfying the minimax, preserves $p_{a\pm}$ along the time-space microprocess for multiple time-space impulses. Reaching probability of appearance the time-space impulse needs $m_p = 0.6015 / 0.1353 \cong 4.4457 \approx 5$ multiplications of invariant $p_{a\pm} = 0.1353$.
Hence, each reversible microprocess within the impulse generates invariant increment of entropy, which enables sequentially minimize the starting uncertainty of the observation.
Assigning entropy (6.15b) minimal uncertainty measure $h_\alpha^o = 1/137$ - physical structural parameter of energy, which includes the Plank constant's equivalent of energy, -leads to relation:

$$S_{\mp a}^* = 2h_\alpha^o, p_{\pm a} = \exp(-2h_\alpha^o) \to 1, \qquad (6.17)$$

which evaluates probability of real impulse' physical strength of coupling independently chosen entropy fractions. The initially orthogonal non-interacting entropy fractions $S_{+a}^* = h_\alpha^o$ , $S_{-a}^* = h_\alpha^o$ , at potential mutual interactive actions, satisfy multiplicative relation

$$S_{\mp a}^* = (h_\alpha^o)^2 [\mathrm{Cos}^2(\bar{u}t) + Sin^2(\bar{u}t)]|_{t_o^\mp}^{t=1/2\tau} = (h_\alpha^o)^2 = inv \qquad (6.17a)$$

which at $S_{\mp a}^* = (h_\alpha^o)^2 \to 0$ approaches $p_{\pm a}^* = \exp[-(h_\alpha^o)^2] \to 1$.

The impulse interaction adjoins the initial orthogonal geometrical sum of entropy fractions in liner sum $2h_\alpha^o$.
Starting physical coupling with double structural $h_\alpha^o$ , according to (6.17), creates initial information triplet with probability $\to 1$, which initiates elementary information network as a unit of collective intelligence [44].
Examples. Let us find which of the entropy functional expression meets requirements (4.1A,B) within discrete intervals $\Delta_t = (t - s) \to o(t)$, particularly on $\Delta_k = (\tau_k^{-o} - s_k^{+o}) \to o(\tau_k^{-o})$ under opposite functions $u_+, u_-$:

$$u_+(s_k^{+o}) = +1_{s_k^{+o}} \bar{u} = \bar{u}(s_k^{+o}), u_- = -u_+(s_k^{+o}) = -1_{s_k^{+o}} \bar{u} . \qquad (6.18)$$

Following relations (4.11), we get entropy increments

$$S_+[\tilde{x}_t / \varsigma_t]|_{s_k^+}^{t\to\tau_k^{-o}} = -1/2[u_+(s_k^{+o})](\tau_k^{-o} - s_k^{+o})^{-1}(s_k^{+o})^2 = -1/2[u_+(s_k^{+o})(s_k^{+o})^2 / s_k^{+o}(3-1)] = -1/4[u_+(s_k^{+o})s_k^{+o}] \qquad (6.19)$$

$$S_-[\tilde{x}_t / \varsigma_t]|_{s_k^+}^{t\to\tau_k^{-o}} = 1/2[u_-(s_k^{+o})](\tau_k^{-o} - s_k^{+o})^{-1}(s_k^{+o})^2 = 1/2[u_-(s_k^{+o})(s_k^{+o})^2 / s_k^{+o}(3-1)] = 1/4[u_-(s_k^{+o})s_k^{+o}],$$

which satisfy

$$S_+[\tilde{x}_t / \varsigma_t]|_{s_k^+}^{t\to\tau_k^{-o}} = -S_-[\tilde{x}_t / \varsigma_t]|_{s_k^+}^{t\to\tau_k^{-o}}, \qquad (6.19a)$$

$$S_+[\tilde{x}_t / \varsigma_t]|_{s_k^+}^{t\to\tau_k^{-o}} - S_-[\tilde{x}_t / \varsigma_t]|_{s_k^+}^{t\to\tau_k^{-o}} = -1/2[\bar{u}(s_k^{+o})s_k^{+o}] = \Delta S[\tilde{x}_t / \varsigma_t]|_{s_k^+}^{t\to\tau_k^{-o}} . \qquad (6.19b)$$

Relations

$$4S_+[\tilde{x}_t / \varsigma_t]|_{s_k^+}^{t\to\tau_k^{-o}} / s_k^{+o} = -\bar{u}(s_k^{+o})(s_k^{+o}) = -2\times 1_{s_k^{+o}}, \quad 4S_-[\tilde{x}_t / \varsigma_t]|_{s_k^+}^{t\to\tau_k^{-o}} / s_k^{+o} = \bar{u}(s_k^{+o}) = 2\times 1_{s_k^{+o}}$$

satisfy conditions

$$4S_+[\tilde{x}_t / \varsigma_t]|_{s_k^+}^{t\to\tau_k^{-o}} / s_k^{+o} \times 4S_-[\tilde{x}_t / \varsigma_t]|_{s_k^+}^{t\to\tau_k^{-o}} / s_k^{+o} = -\bar{u}(s_k^{+o}) \times \bar{u}(s_k^{+o}) = -(2\times 1_{s_k^{+o}}) \times (2\times 1_{s_k^{+o}}) = -4\times 1_{s_k^{+o}}, \qquad (6.20a)$$

$$4S_+[\tilde{x}_t / \varsigma_t]|_{s_k^+}^{t\to\tau_k^{-o}} / s_k^{+o} - 4S_-[\tilde{x}_t / \varsigma_t]|_{s_k^+}^{t\to\tau_k^{-o}} / s_k^{+o} = -\bar{u}(s_k^{+o}) - \bar{u}(s_k^{+o}) = -(2\times 1_{s_k^{+o}}) - (2\times 1_{s_k^{+o}}) = -4\times 1_{s_k^{+o}}. \qquad (6.20b)$$

These entropy expressions at any *current* moment $t$ within $\Delta_t = (t - s_k^{+o})$ do not comply with (4.1A,B).
The same results hold true for the entropy functional increments under functions

$$u_+ = +1_{s_k^{+o}} \bar{u}, u_- = -1_{\tau_k^{-o}} \bar{u} . \qquad (6.21)$$

Indeed. For this functions on $\Delta_t = (t - s_k^{+o})$ we have

$$\Delta S[\tilde{x}_t / \varsigma_t]|_{s_k^+}^{t} = -1/2(u_-(t) - u_+(s_k^{+o}))(t - s_k^{+o})^{-1}(s_k^{+o})^2 \qquad (6.\ 22)$$

which for $t \to \tau_k^{-o}$ holds

$$\Delta S[\tilde{x}_t / \varsigma_t]|_{s_k^+}^{t\to\tau_k^{-o}} = -1/2(u_-(\tau_k^{-o}) - u_+(s_k^{+o}))(\tau_k^{-o} - s_k^{+o})^{-1}(s_k^{+o})^2,$$

and satisfies relations

$$S_+[\tilde{x}_t / \varsigma_t]|_{s_k^+}^{t\to\tau_k^{-o}} - S_-[\tilde{x}_t / \varsigma_t]|_{s_k^+}^{t\to\tau_k^{-o}} = \Delta S[\tilde{x}_t / \varsigma_t]|_{s_k^+}^{t\to\tau_k^{-o}}, \quad (6.22a)$$

$$S_+[\tilde{x}_t / \varsigma_t]|_{s_k^+}^{t\to\tau_k^{-o}} = -S_-[\tilde{x}_t / \varsigma_t]|_{s_k^+}^{t\to\tau_k^{-o}}. \quad (6.22b)$$

which determines

$$S_+[\tilde{x}_t / \varsigma_t]|_{s_k^+}^{t\to\tau_k^{-o}} = -1/4(u_-(\tau_k^{-o}) - u_+(s_k^{+o}))(\tau_k^{-o} - s_k^{+o})^{-1}(s_k^{+o})^2 \quad (6.23a)$$

$$S_-[\tilde{x}_t / \varsigma_t]|_{s_k^+}^{t\to\tau_k^{-o}} = 1/4(u_-(\tau_k^{-o}) - u_+(s_k^{+o}))(\tau_k^{-o} - s_k^{+o})^{-1}(s_k^{+o})^2. \quad (6.23b)$$

We get the entropy expressions through opposite directional discrete functions in (6.23a,b):

$$S_+[\tilde{x}_t / \varsigma_t]|_{s_k^+}^{t\to\tau_k^{-o}}\ 4(\tau_k^{-o} - s_k^{+o})^{-1}(s_k^{+o})^2 = -(u_-(\tau_k^{-o}) - u_+(s_k^{+o})),$$

$$S_-[\tilde{x}_t / \varsigma_t]|_{s_k^+}^{t\to\tau_k^{-o}}\ 4(\tau_k^{-o} - s_k^{+o})(s_k^{+o})^{-2} = u_-(\tau_k^{-o}) - u_+(s_k^{+o}),$$

which satisfy additivity at

$$-2(u_-(\tau_k^{-o}) - u_+(s_k^{+o})) = -2(-1_{\tau_k^{-o}}\bar{u} - 1_{s_k^{+o}}\bar{u})] = 2\bar{u}[1_{\tau_k^{-o}} + 1_{s_k^{+o}}] = 4[1_{\tau_k^{-o}} + 1_{s_k^{+o}}]. \quad (6.24)$$

While for each

$$S_+[\tilde{x}_t / \varsigma_t]|_{s_k^+}^{t\to\tau_k^{-o}}\ 4(\tau_k^{-o} - s_k^{+o})(s_k^{+o})^{-2} = -\bar{u}[-1_{\tau_k^{-o}} - 1_{s_k^{+o}}] \quad (6.24a)$$

$$S_-[\tilde{x}_t / \varsigma_t]|_{s_k^+}^{t\to\tau_k^{-o}}\ 4(\tau_k^{-o} - s_k^{+o})(s_k^{+o})^{-2} = \bar{u}[-1_{\tau_k^{-o}} - 1_{s_k^{+o}}] \quad (6.24b)$$

satisfaction of both 4.1A, B:

$$-(u_-(\tau_k^{-o}) - u_+(s_k^{+o})) \times (u_-(\tau_k^{-o}) - u_+(s_k^{+o})) = -[u_-(\tau_k^{-o}) - u_+(s_k^{+o})]^2,$$

requires

$$\bar{u} = -2j, \quad (6.25)$$

when $-[u_-(\tau_k^{-o}) - u_+(s_k^{+o})]^2 = (-2j)^2[-1_{\tau_k^{-o}} - 1_{s_k^{+o}}]^2.$ (6.25a)

Simultaneous satisfaction of both 4.1.A, B leads to

$$\Delta S\ [\tilde{x}_t / \varsigma_t]|_{s_k^+}^{t\to\tau_k^{-o}}\ 2(\tau_k^{-o} - s_k^{+o})(s_k^{+o})^{-2} = -2\bar{u}[-1_{\tau_k^{-o}} - 1_{s_k^{+o}}] = 4j[1_{\tau_k^{-o}} + 1_{s_k^{+o}}],$$

$$-(-1_{\tau_k^{-o}}\bar{u} - 1_{s_k^{+o}}\bar{u}) \times (-1_{\tau_k^{-o}}\bar{u} - 1_{s_k^{+o}}\bar{u}) = (-2j)^2(-1_{\tau_k^{-o}} + 1_{s_k^{+o}})^2. \quad (6.25b)$$

At $o(t) \to 0$, these admit an instant existence of both $(-1_{\tau_k^{-o}}\bar{u} + 1_{s_k^{+o}}\bar{u})$.

Thus, under function (6.25) entropy expressions are imaginary:

$$S_+[\tilde{x}_t / \varsigma_t]|_{s_k^+}^{t\to\tau_k^{-o}}\ 4(\tau_k^{-o} - s_k^{+o})(s_k^{+o})^{-2} = -2j[-1_{\tau_k^{-o}} + 1_{s_k^{+o}}] = -2j[1_{s_k^{+o}}^{\tau_k^{-o}}], \quad (6.26a)$$

$$S_-[\tilde{x}_t / \varsigma_t]|_{s_k^+}^{t\to\tau_k^{-o}}\ 4(\tau_k^{-o} - s_k^{+o})(s_k^{+o})^{-2} = 2j[-1_{\tau_k^{-o}} + 1_{s_k^{+o}}] = 2j[1_{s_k^{+o}}^{\tau_k^{-o}}], \quad (6.26b)$$

at their multiplicative and additive relations:

$$S_-[\tilde{x}_t / \varsigma_t]|_{s_k^+}^{t\to\tau_k^{-o}}\ 4(\tau_k^{-o} - s_k^{+o})(s_k^{+o})^{-2} \times S_+[\tilde{x}_t / \varsigma_t]|_{s_k^+}^{t\to\tau_k^{-o}}\ 4(\tau_k^{-o} - s_k^{+o})(s_k^{+o})^{-2} = 4, \quad (6.27a)$$

$$S_+[\tilde{x}_t / \varsigma_t]|_{s_k^+}^{t\to\tau_k^{-o}}\ 4(\tau_k^{-o} - s_k^{+o})(s_k^{+o})^{-2} - S_-[\tilde{x}_t / \varsigma_t]|_{s_k^+}^{t\to\tau_k^{-o}}\ 4(\tau_k^{-o} - s_k^{+o})(s_k^{+o})^{-2} = -j2[1_{\tau_k^{-o}} + 1_{s_k^{+o}}] = -j2[1_{s_k^{+o}}^{\tau_k^{-o}}],$$

$$S_+[\tilde{x}_t / \varsigma_t]|_{s_k^+}^{t\to\tau_k^{-o}} - S_-[\tilde{x}_t / \varsigma_t]|_{s_k^+}^{t\to\tau_k^{-o}} = 1/2j[1_{s_k^{+o}}^{\tau_k^{-o}}], \quad (6.27b)$$

Relations (6.24),(6.24a,b) satisfy additivity only at points $\tau_k^{-o}, s_k^{+o}$.

Between these points, within $\Delta_t = (t - s_k^{+o}) \to o(t)$, the entropy expressions (6.26a, b), (6.27b) are imaginary.

Time direction may go back within this interval until an interaction occurs.
Within this interval, entropy $S_t = S_+ - S_-$ holds relations

$$(S_+ - S_-)^2 = S_+^2 + S_-^2 - 2S_+ \times S_-, S_+ = -1/2\,jS_t^+, S_- = 1/2\,jS_t^-, S_+^2 = -1/4S_t^{\pm 2}, S_-^2 = 1/4S_t^{\pm 2}, \quad (6.28)$$

leading to

$-2S_+ \times S_- = -2(-1/4)\,jjS_t^{\pm 2} = -1/2S_t^{\pm 2}$, while $S_+^2 + S_-^2 = -1/2S_t^{\pm 2}$ and $(S_+ - S_-)^2 = (-jS_t^+)^2 = -(S_t^+)^2$.

At fulfillment of 4.1A,B follows relations

$$S_{t=\tau}^2 = -1/2\,jS_{t\to\tau}, S_{t=\tau}^2 = S_+ \times S_- = -1/4S_{t\to\tau}^{\pm 2}, S_{t\to\tau} = \pm 2\,jS_{t=\tau}, \quad (6.29)$$

from which also follows

$$S_{t\to\tau} = 2j. \quad (6.30)$$

These examples concur with *Secs. 6.2,6.3 and illustrate it.*

**6.4. The relation between the curved time and equivalent space length within an impulse**

Let us have a two-dimensional rectangle impulse with wide $p$ measured in time $[\tau]$ unit and high $h$ measured in space length $[l]$ unit, with the rectangle measure $M_i = p \times h$. (6.31)

The problem: Having a measure of wide part of the impulse $\mathrm{M}_p$ to *find* high $h$ at equal measures of both parts:

$$\mathrm{M}_p = \mathrm{M}_h \text{ and } \mathrm{M}_p + \mathrm{M}_h = M_i. \quad (6.32)$$

Since that, $\mathrm{M}_h = 1/2M_i = 1/2p \times h$, (6.32a), and measure $\mathrm{M}_p$ of the impulse wide $1/2p$ is equal to measure $\mathrm{M}_h$ of the rectangle with high $h$ and the same wide as measured by $\mathrm{M}_p$.

Assuming that first part of the impulse $\mathrm{M}_p$ has only wide part $1/2p$, it equals $\mathrm{M}_p = (1/2p)^2$.

Then from $\mathrm{M}_p = (1/2p)^2 = \mathrm{M}_h = 1/2p \times h$ follows

$$h/p = 1/2. \quad (6.33)$$

Let us find a length unit $[l]$ of the curved wide time unit $[\tau]$ using relations

$$2\pi h[l]/4 = 1/2p[\tau] \ (6.35a), \quad [\tau]/[l] = \pi h/p. \quad (6.35b)$$

Substitution (6.31) leads to ratio of the measured units:

$$[\tau]/[l] = \pi/2. \quad (6.36)$$

Relation (6.36) sustains orthogonality of these units in time-space coordinate system, but since initial relations (6.32) are linear, ratio (6.36) represents a linear connection of the time-space units (6.35a), received through curving the time unit in the impulse-jumps (6.9). The microprocess, built in rotation movement, curving the impulse time, adjoins the initial orthogonal axis of time and space coordinates (**Fig.1**).

The impulses, preserving multiplicative and additive measures (6.31), (6.32), have common ratio of $h/p = 1/2$, whose curving wide part $p = 1/2$ brings universal ratio (6.36), which concurs with Lemma 4.1 (4.2a).

At above assumption, measure $\mathrm{M}_h$ does not exist until the impulse-jump curves its only time wide $1/2p$ at transition of the impulse, measured only in time, to the impulse, measured both in time $1/2p$ and space coordinate $h$.

According to (6.32a), measure $\mathrm{M}_h$ emerges only on a half of that impulse' total measure $M_i$.

The transitional impulse could start on border of the virtual impulses $\downarrow\uparrow$, where transition curving time $\delta t_p = 1/2p$ under impulse-jump during $\delta t_p \to 0$ leads to

$$M_h \Rightarrow M_i = p \times h, M_p \to 0. \quad (6.37)$$

If a virtual impulses $\downarrow\uparrow$ has equal opposite functions $u_-(t), u_+(t+\Delta)$, at $\overline{u}_+ = \overline{u}_-$, additive condition for measure (4.2a): $\mathrm{U}_a(\Delta)=0$ violates, and the impulse holds only multiplicative measure $\mathrm{U}_m(\Delta)\neq 0$; its relation (4.2C): $U_m(\Delta)=\mathrm{U}_{am}$ is finite only at $\overline{u}_+ = \overline{u}_- \neq 0$. If any of $\overline{u}_+ = 0$, *or* $\overline{u}_- = 0$, both multiplicative $\mathrm{U}_m(\Delta)=0$ and additive $\mathrm{U}_a(\Delta)=0$ disappear. At $\overline{u}_- \neq 0$, $\mathrm{U}_a(\Delta)$ is a finite and positive, specifically, at $\overline{u}_- = 1$ leads to $\mathrm{U}_a(\Delta)=1$ preserving measure $\mathrm{U}_{amk} = |\mathrm{U}_a|_k$.

Existing of the transitional impulse has shown in Secs. 4 (4.19a,b), 6.1, and 6.2.

An impulse-jump at $\mathrm{o}[\mathrm{t}_o^{\mp}] \to \delta t_p \to 0$ curves a "needle pleat" space at transition to the finite form impulse.

The Bayes probabilities measure may overcome this transitive gap.

## 7. Information dynamic processes determined by EF and IPF functionals

1.The EF-IPF connection.

Since the IPF functional integrates finite information and converges with the initial entropy functional, which had expressed through the additive functional, the EF covers both cutoff information contributions and entropy increments between them. The IPF at $n \to \infty$ integrates unlimited discrete sequence of the EF cutoff fractions. The integration of the discrete fractions and solving a classical variation problem for the IPF to find continuous extreme dynamics presents a *difficult mathematical task*.

Integral (2.1.4) measures the EF and IPF maximal limit approaching EF at $o(t \to T) \to 0, n \to \infty$ and avoids the direct access to Markov random process.

Extreme of this integral provides dynamic process $x(t)$, which minimizes the distance (1.1.5a) and dynamically approximates movement $\tilde{x}_t$ to $\varsigma_t$ evaluating transition to Feller kernel.

Initial entropy functional (1.1.10) presents potential information functional of the Markov process until the applied impulse control, carrying the cutoff increments, transforms it to physical IPF (5.1).

Process $x(t)$ carries information collected by the maximal IPF at $n \to \infty$, as the IPF information dynamic macroprocess [38], when each interval $\Delta_t = (t-s) \to o(t)$.

Increment of the EF at the end of interval $o_m \to 0$ approaches zero satisfying (1.3.13):

$$\lim_{t_m = T} \Delta S_m[\tilde{x}_t(\tau_m \to t_m))] \to 0. \qquad (7.1)$$

The IPF extracts the finite amount of integral information on all cutoff intervals, approaching initial $S[\tilde{x}_t / \varsigma_t]$ (1.110) before its cutting.

The sequential cuts on $(T-s)$ lose the process correlations and the states functional connections, which transforms the initial random process to a limited sequence of independent states.

In the limit, the IPF extracts a deterministic process (with probability (1.2.13)-within each cutoff), which approaches the EF extremal trajectories. Thus, information, collected from the diffusion process by the IPF, approaches its source, measured by the EF, when intervals become infinite small $\Delta_t \to o(t)$.

However without impulse action releasing information from the EF is unfeasible.

Impulse composes Yes-No actions founding Bit and all communications.

That's why starting from Sec.1.2 we study 'an anatomy' of the impulse.

2.Estimation of extremal process.

Mathematical expectations of Ito's Eqs:

$$E[a] = \dot{\overline{x}}(t) = E[c\tilde{x}(t)] = cE[\tilde{x}(t)] = c\overline{x}(t) \qquad (7.2)$$

approximates a regular differential Eqs

$$\dot{\overline{x}}(t) = c\overline{x}(t), \qquad (7.3)$$

whose common solution averages the random movement by dynamic *macroprocess* $\bar{x}(t)$:

$\bar{x}(t)=\bar{x}(s)\exp ct, \bar{x}(s)=E[\tilde{x}(s)]$. (7.3a)

Within discrete $o(t)=\delta_o$, opposite controls $u_+, u_-$, satisfying relation

$c^2=|u_+u_-|=c_+c_-=\bar{u}^2, c_+=u_+, c_-=u_-$, $|u_+u_-|=\bar{u}^2$

are imaginable, presenting an opposite discrete complex:

$u_+=j\bar{u}, u_-=-j\bar{u}$. (7.4)

Conditions 4.1A, B fulfilled at

$\bar{u}=-2j$. (7.4a)

when

$u_+u_-=j\bar{u}(-j\bar{u})=\bar{u}^2$, $-j\bar{u}-(+j\bar{u})=-2j\bar{u}$,

$\bar{u}^2=-2j\bar{u}, \bar{u}=-2j$. (7.4b)

The controls are real when

$u_+=j(-2j)=2, u_-=-j(-2j)=-2$. (7.5)

Relations (7.3), (7.4), satisfy two differential equations

$\dot{x}_+(t)=j\bar{u}x_+(t), \dot{x}_-(t)=-j\bar{u}x_-(t)$ (7.6)

describing microprocess ($x_+(t)$, $x_-(t)$) under controls (7.4,7.5) on time interval $\Delta_t=t-s$, $\Delta_t\rightarrow o(t)$.

Solutions of (7.6) takes forms

$\ln x_+(t)=Cu_+t, \ln x_-(t)=Cu_-t, x_+(t)=C\exp(j\bar{u}t), x_-(t)=C\exp(-j\bar{u}t), \mathrm{C}=x_-(s^{+o})=x_+(s^{+o})$, (7.7a)

$x_+(t)=x_+(s^{+o})(\mathrm{Cos}\,\bar{u}t+jSin\bar{u}t), x_-(t)=x_-(s^{+o})(\mathrm{Cos}(\bar{u}t-jSin\bar{u}t)$, (7.7)

where moment $t=t^e$ of reaching minimal entropy functional identifies Eqs (2.2.12a,b).

Correlation function for microprocess (7.7):

$r(x_+(t),x_-(t))=r_s=x_+(s^{+o})\times x_-(s^{+o}), Cos^2(\bar{u}t)+\mathrm{Sin}^2(\bar{u}t)=1$ (7.7b)

depends on interaction at moment $\delta^{+o}/4, \delta^{+o}/2$ on an interactive edge of the impulse. During this fixed correlation, the conjugated anti-symmetric entropies (6.14a) interact producing entropy flow (6.16).

Applying formula [41, p.27] for the correlation between moments $\delta^{+o}/4, \delta^{+o}/2$ (Sec.2.4) leads to

$r_s=\sqrt{(\delta^{+o}/4)(/\delta^{+o}/2)}$, $r_s=\sqrt{0.5}$ . (7.7c)

During this correlation acts impulse $[1_{\delta_k^{\tau+}/4}^{\delta_k^{\tau+}/2}]\bar{u}_k$, $\bar{u}_k=|\pm 1|_k$, which includes both opposite controls instantaneously.

If real control, cutting the entropy influx, does not cover it, the states' correlation will not dissolve, and the state, carrying both opposite controls, will hold during the correlation. It corresponds the states' entanglement, which may exists between the entropy gap (5.26a,b) balancing entropy (5.29) at the anti-symmetric actions.

Process (7.7) is microprocess on $o(t)\rightarrow o(\tau_n^{-o})$ compared to macroprocess (7.3a).

The microprocess becomes an inner part of the dynamics process, minimizing distance (1.1.5a), when its time intervals satisfy optimal time between the impulse cutoff information at

$\tau_k^{-o}/\tau_k^{+o}=3, \delta_k=\tau_k^{+o}-\tau_k^{-o}=2\tau_k^{-o}, \tau_k^{-o}=3\delta_k/2$. (7.8)

It implies that imaginary time interval (5.5b) triples the cutoff discrete intervals:

$\Delta_k=3\delta_k$, (7.8a)

while microprocess (7.7) is within these cutoff discrete intervals.

To find dynamic process $x(t)$, we consider solution of (7.3) under real control (7.5) starting at moment $t=t^e$:

$x_\pm(t^e)=x(s^+)\exp(u_\pm t^e)$, $t^e=s_k^{+o}b_k(t)/b_k(s_k^{+o})$, $u_\pm=\pm 2$ (7.9)

approximating an extreme of the entropy functional within each $\Delta_t = t - s$ following from (2.2.10).
Solutions in form

$$dx(t)/x(t) = cdt, \ln x(t) = ct, t = s_k^{+o} b_k(t)/b_k(s_k^{+o}), \ln x(t) = cs_k^{+o} b_k(t)/b_k(s_k^{+o}) \quad (7.10)$$

starting on time $t = t^e$ integrate minimal distance of $\Delta_t = t - s$:

$$x(t) = \exp[cs_k^{+o} b_k(t)/b_k(s_k^{+o})], x(s_k^{+o}) = \exp(cs_k^{+o}), c = \ln(x(s_k^{+o}))/s_k^{+o}, x(s_k^{+o}) = \overline{x}(s_k^{+o}),$$

$$x(t) = \exp[\ln(x(s_k^{+o})) b_k(t)/b_k(s_k^{+o})] = \exp[\ln(x(s_k^{+o}))t/s)], \ln x(t) = \ln(x(s_k^{+o}))t/s, \quad (7.10a)$$

which at $t \to T$ approaches

$$\ln x(T) = \lim_{t \to T} [\ln(x(s_k^{+o}))T/s], x(T) \to x(s_k^{+o}))T/s. \quad (7.10b)$$

Process $x(t)$ (7.3a), (7.10) is the extremal solution of *macroprocess* $\overline{x}(t)$, which averages solution of Ito Eq. under the optimal controls' multiple cutoffs of the EF for $n$-dimensional Markov process within $\Delta_t = (t - s) \to o(t)$.
Process $x(t)$ carries the EF increments, while the information dynamic macroprocess collects the maximal IPF contributions at $o(t) \to 0$, $n \to \infty$, and each $\Delta_t \to o(t)$.
Information, collected from the diffusion process by the IPF, approaches the EF entropy functional.
Finding process $x(t)$ which the EF generates requires solution of the EF variation problem.

## 8. The solution of variation problem for the entropy functional

Applying the variation principle to the entropy functional, we consider an integral functional

$$S = \int_s^T L(t, x, \dot{x}) dt = S[x_t] \quad , \quad (8.1)$$

which minimizes the entropy functional (1.1.10) of the diffusion process in the form

$$\min_{u_t \in KC(\Delta, U)} \tilde{S}[\tilde{x}_t(u)] = S[x_t], \; Q \in KC(\Delta, R^n). \quad (8.1a)$$

Specifically, for integral (2.1.1), it leads to optimal solution of variation problem

$$\operatorname{extr} S[\tilde{x}_t / \varsigma_t] = \operatorname*{extr}_{c^2(t)} 1/2 \int_s^T c^2(t) A(t,s) dt, \; c^2(t) = \dot{x}(t). \quad (8.1b)$$

Proposition 8.1.
1.An *extremal solution* of variation problem (8.1a, 8.1) for entropy functional (1.1.10), brings the following equations of extremals for vector $x$ and conjugate vector $X$ accordingly:

$$\dot{x} = a^u, \; a^u = a(u, t, x) \; (t, x) \in Q \; , \quad (8.2)$$

$$\dot{X} = -\partial P / \partial x - \partial V / \partial x, \quad (8.3)$$

where

$$P = (a^u)^T \frac{\partial S}{\partial x} + b^T \frac{\partial^2 S}{\partial x^2} \; , \quad (8.4)$$

$S(t, x)$ is function of action on extremals (8.2,8.3); $V(t, x)$ is integrant of the additive functional (1.7), which defines the probability function (1.3).
Proof. Using the Jacobi-Hamilton (JH) equations for function of action $S = S(t, x)$, defined on the extremals $x_t = x(t), (t, x) \in Q$ of functional (8.1), leads to

$$-\frac{\partial S}{\partial t} = H, H = \dot{x}^T X - L, \quad (8.5)$$

where $X$ is a conjugate vector for $x$ and $H$ is a Hamiltonian for this functional.

(All derivations here and below have vector form).
From (8.1a) it follows

$$\frac{\partial S}{\partial t}=\frac{\partial \tilde{S}}{\partial t},\frac{\partial \tilde{S}}{\partial x}=\frac{\partial S}{\partial x}, \qquad (8.6)$$

where for the JH we have

$$\frac{\partial S}{\partial x}=X,-\frac{\partial S}{\partial t}=H\ . \qquad (8.6a)$$

The Kolmogorov Eq. for functional (1.1.10) on division process allows joining it with Eq. (8.6a) in the form

$$-\frac{\partial \tilde{S}}{\partial t}=(a^u)^T X+b\frac{\partial X}{\partial x}+1/2a^u(2b)^{-1}a^u=-\frac{\partial S}{\partial t}=H\ , \qquad (8.7)$$

where dynamic Hamiltonian $H=V+P$ includes function $V=d\varphi/ds$ from (1.1.7) and potential function

$$P(t,x)=(a^u)^T X+b^T\frac{\partial X}{\partial x}. \qquad (8.8)$$

Applying Hamilton equations $\frac{\partial H}{\partial X}=\dot{x}$ and $\frac{\partial H}{\partial x}=-\dot{X}$ to (8.7) we get the extremals for vector $x$ and $X$ in the forms (8.2) and (8.3) accordingly. • More details in [38].

<u>Proposition 8.2.</u>

A *minimal solution* of variation problem (8.1a,8.1) for the entropy functional (1.1.10) brings the following equations of extremals for $x$ and $X$ accordingly:

$$\dot{x}=2bX_o\ , \qquad (8.9)$$

satisfying condition

$$\min_{x(t)} P=P[x(\tau)]=0\ . \qquad (8.10)$$

Condition (8.10) is a dynamic constraint, which is imposed on the solutions (8.2), (8.3) at some set of the functional's field $Q\in KC(\Delta,R^n)$, where the following relations hold:

$$Q^o\subset Q,\ Q^o=R^n\times\Delta^o,\Delta^o=[0,\tau],\tau=\{\tau_k\},k=1,...,m \qquad (8.11)$$

for cutting process $x(t)_{t=\tau}=x(\tau)$ at $\tau$ -localities.
Hamiltonian

$$H_o=-\frac{\partial S_o}{\partial t} \qquad (8.12)$$

is defined for function of action $S_o(t,x)$, which on extremals Eq.(8.9) satisfies condition

$$\min(-\partial\tilde{S}/\partial t)=-\partial\tilde{S}_o/\partial t\ . \qquad (8.13)$$

Hamiltonian (8.12) and Eq. (8.9) determine a second order differential Eq. of extremals:

$$d^2x/dt^2=dx/dt[\dot{b}b^{-1}-2H_o]. \qquad (8.14)$$

<u>Proof.</u> Using (8.4) and (8.6), we find the equation for Lagrangian in (8.1) in the form

$$L=-b\frac{\partial X}{\partial x}-1/2\dot{x}^T(2b)^{-1}\dot{x}\ . \qquad (8.15)$$

On extremals (8.2,8.3), both functions drift and diffusion (in 1.1.10) are nonrandom.
After substitution the extremal Eqs to (8.1), the integral functional $\tilde{S}$ on the extremals holds:

$$\tilde{S}[x(t)]=\int_s^T 1/2(a^u)^T(2b)^{-1}a^u dt\ , \qquad (8.15a)$$

which should satisfy variation conditions (8.1a), or

$\tilde{S}[x(t)] = S_o[x(t)]$, (8.15b)

where both integrals are determined on the same extremals.
From (8.15), (8.15a,b) it follows

$L_o = 1/2(a^u)^T(2b)^{-1}a^u$, or $L_o = \dot{x}^T(2b)^{-1}\dot{x}$ . (8.16)

Both expressions for Lagrangian (8.15) and (8.16) coincide on the extremals, where potential (8.7) satisfies condition (8.10) in the form

$$P_o = P[x(t)] = (a^u)^T(2b)^{-1}a^u + b^T\frac{\partial X_o}{\partial x} = 0, \quad (8.17)$$

while Hamiltonian (8.12), and function of action $S_o(t,x)$ satisfies (8.13).
From (8.15b) it also follows

$$E\{\tilde{S}[x(t)]\} = \tilde{S}[x(t)] = S_o[x(t)]. \quad (8.17a)$$

Applying Lagrangian (8.16) to Lagrange's equation

$$\frac{\partial L_o}{\partial \dot{x}} = X_o, \quad (8.17b)$$

leads to equations for vector

$$X_o = (2b)^{-1}\dot{x} \quad (8.17c)$$

and extremals (8.8).
Both Lagrangian and Hamiltonian here are *information forms* of JH solution for the EF.
Lagrangian (8.16) satisfies the maximum principle for functional (8.1.7.1a), from which also follows (8.17a).
Functional (8.1) reaches its minimum on extremals (8.8), while it is maximal on extremals (8.2,8.3) of (8.6).
Hamiltonian (8.7), at satisfaction of (8.17), reaches minimum:

$$\min H = \min[V + P] = 1/2(a^u)^T(2b)^{-1}a^u = H_o, \quad (8.18)$$

from which it follows (8.10) at

$$\min_{x(t)} P = P[x(\tau)] = 0 \quad . \quad (8.19)$$

Function $(-\partial\tilde{S}(t,x)/\partial t) = H$ in (8.6) on extremals (8.2,8.3) reaches a *maximum* when constraint (8.10) is not imposed. Both the minimum and maximum are conditional with respect to the constraint imposition.
Variation conditions (8.18), imposing constraint (8.10), selects Hamiltonian

$$H_o = -\frac{\partial S_o}{\partial t} = 1/2(a^u)^T(2b)^{-1}a^u \quad (8.20)$$

on the extremals (8.2,8.3) at discrete moments $(\tau_k)$ (8.11).

The variation principle identifies two Hamiltonians: $H$-satisfying (8.6) with function of action $S(t,x)$, and $H_o$ (8.20), whose function action $S_o(t,x)$ reaches absolute minimum at moments $(\tau_k)$ (8.11) of imposing constraint $P_o = P_o[x(\tau)]$.
Substituting (8.2) and (8.17b) in both (8.16) and (8.20), leads to Lagrangian and Hamiltonian on the extremals:

$$L_o(x, X_o) = 1/2\dot{x}^T X_o = H_o. \quad (8.21)$$

Using $\dot{X}_o = -\partial H_o/\partial x$ brings $\dot{X}_o = -\partial H_o/\partial x = -1/2\dot{x}^T\partial X_o/\partial x$,
and from constraint (8.10) it follows

$$\partial X_o/\partial x = -b^{-1}\dot{x}^T X_o, \text{ and } \partial H_o/\partial x = 1/2\dot{x}^T b^{-1}\dot{x}^T X_o = 2H_o X_o, \quad (8.22)$$

which after substituting (8.17b) leads to extremals (8.9).
Using Eq. for the conjugate vector (8.3), constraint (8.10) in form

$$\frac{\partial X_o}{\partial x} = -2X_o X_o^T \qquad (8.23)$$

follows from (8.7), (8.8) and (8.17c).
Differentiation of (8.9) leads to a second order differential Eqs on the extremals:

$$\ddot{x} = 2b\dot{X}_o + 2\dot{b}X_o, \qquad (8.24)$$

which after substituting (8.22) leads to (8.14). •

**8a. About initial conditions for the entropy functional and its extremals**

1.According to definition (1.1.1), the *initial conditions for the EF* determines ratio of a primary a priori- a posteriory probabilities $p(\mathrm{o}_s^p) = \frac{P_{s,x}^a}{P_{s,x}^p}(\mathrm{o}_s^p)$ beginning the probabilistic observation.

Start of the observation evaluates $p(\mathrm{o}_s^p) \cong 1.65 \times 10^{-4}$ [47] and $\Delta s_{ap}(\mathrm{o}_s^p) = -\ln p(\mathrm{o}_s^p) = 0.5 \times 10^{-4}$ with minimal posterior probability $P_{poo} \approx 1 \times 10^{-4}$.That gives estimation of an average initial entropy of the observation:

$$S(\mathrm{o}_s^p) = [-\ln p(\mathrm{o}_s^p) \times P_{poo}] \cong 0.5 \times 10^{-8}\, Nat. \qquad (8.25)$$

Based on physical coupling parameter $h_\alpha^o = 1/137$, physical observation theoretically starts with entropy(6.17):

$$S(\mathrm{o}_{rs}^p) = 2/137 \cong 0.0146 Nat, \qquad (8.26)$$

while entropy of the first real 'half-impulse' probing action starts at moment $t^{oe}$:

$$S_{ko}(t^{oe}) = 0.358834 Nat \qquad (8.27)$$

with a priori-a posteriory probabilities $P_{ako} = 0.601, P_{pko} = 0.86$.

In this approach, involving no material entities, physical process begins with the real probing action, and physical coupling may start with minimizing this entropy beginning the information process.

At real cut of posteriory probability $P_{po} \to 1$, ratio of their priory-a posteriori probabilities $P_{ao}/P_{po} \cong 0.8437$ determines minimal entropy shift between interacting probabilities $P_{ao} \to P_{po}$ during real cut: $\Delta s_{apo} = -\ln(0.8437) \cong 0.117 Nat$,which after averaging at $P_{po} = 1$ leads to

$$S(\mathrm{o}_r^p) = 0.117 Nat. \qquad (8.28)$$

Minimal entropy cost on covering a gap during it conversion to information is $s_{ev} \cong 0.0636 Nat$ [47].

Theoretical start of observation (8.26) is a potential until an Observer gets information from that entropy.

If we *define* the launch of real Information Observer by minimal converting entropy (8.28), then opening real observation *defines* entropy of first real 'half-impulse' probing actions (8.27), which could be multiple for a many dimensional process.

Since the potential observing physical process with structural entities is possible with entropy (8.26), this observation we call *virtual* until real observation generating information Observer starts and confirms it.

The term 'virtual' associates with physically possibility, until this physical objectivity becomes information-physical reality for the information Observer.

Specifically, if start of virtual observation associates with entropy binding primary a priori-a posteriory probability (8.25), then Virtual Observer identifies entropy of appearing potential structures (8.26).

Here, we have *identified* both beginning of virtual and physical observations and the virtual and real information Observers, based on the actual quantitative parameters, which are independent on each particular Observer.

But it may impose specific limitations after the Observer forms [47, Sec.2.5.2].

2.The initial conditions for the EF *extremals* determine function $x_{\pm}(t^e)=x(s^+)\exp(u_{\pm}t^e)$ which at moment $t^e=s_k^{+o}b_k(t^e)/b_k(s_k^{+o})$ starts virtual or real observations, depending on required minimal entropy of related observations (8.25-8.28). It brings functions

$x_{\pm}(t^e)=x(s^+)\exp(\pm 2t^e)$ (8.29a), $t^e=s_k^{+o}b_k(t^e)/b_k(s_k^{+o})$, (8.29b)

where (8.29b) at known dispersion function $b_k(t^e,s_k^{+o})$ identifies dependency $t^e=t^e(s_k^{+o},b_k)$, while (8.29a) identifies initial conditions for the EF extreme conjugated process:

$$x_{\pm}(t^e)=x(s^+)\exp(\pm 2t^e(s_k^{+o},b_k)). \quad (8.29)$$

Applying the EF solutions (6.10, 6.10a) at opposite relative time $t_-^*=-t_+^*$ lead to entropy functions $S_+(t_+^*)=[exp(-t_+^*)(\mathrm{Cos}(t_+^*)-jSin(t_+^*))]|, S_-(t_-^*)=[exp(-t_+^*)(\mathrm{Cos}(-t_+^*)+jSin(-t_+^*))]$ at

$$S_{\pm}(t_{\pm})=1/2S_+(t_+^*)\times S_-(t_-^*)=1/2[\exp(-2t_+^*)(\mathrm{Cos}^2(t_+^*)+Sin^2(t_+^*)-2Sin^2(t_+^*))]=$$
$$1/2[\exp(-2t_+^*)((+1-2(1/2-Cos(2t_+^*))))]=1/2\exp(-2t_+^*)Cos(2t_+^*). \quad (8.30)$$

This interactive entropy $S_{\pm}(t_{\pm})$ becomes minimal interactive threshold (8.28) at $t_+^*=t_*^e$, which starts the information Observer. From (8.28) it follows:

$$S(t_+^*=t_*^e)=1/2\exp(-2t_*^e)Cos(2t_*^e)=0.117 \quad (8.31)$$

with relative time $t_*^e=\pm\pi/2t^e$.

Solution (8.31) will bring real $t^e$, which after substitution in (8.29b) determines $s_k^{+o}=s_k^{+o}(t^e,b_k(t^e,s_k^{+o}))$ if dispersion functions $b_k(t^e,s_k^{+o})$ are known. Substituting relative $s_{k*}^{+o}=s_k^{+o}(t_*^e,b_k)$ to

$$S(s_{k*}^{+o})=1/2\exp(-2s_{k*}^{+o})Cos(2s_{k*}^{+o}) \quad (8.32)$$

would allow finding unknown initial posterior entropy $S_{\pm}(s_k^{+o})$ starting virtual Observer at $s_k^{+o}=s_{k*}^{+o}/(\pi/2)$.

To find the moment of time starting virtual Observer at maximal uncertainty measure (8.26), when dispersion functions unknown, only joint pre-requirement (8.32)+(8.26) is available.

Applying (7.7) for initial conditions holds $\ln x(s_k^{+o})=x(s_k^{+o})u_{\pm}s_k^{+o}$ which after integration leads to $1/2[\ln x(s_k^{+o})]^2=u_{\pm}(s_k^{+o})^2, \ln x(s_k^{+o})=\sqrt{2u_{\pm}}(s_k^{+o})$ and to

$x_{\pm}(s_k^{+o})=\exp(\pm\sqrt{2u_{\pm}})(s_k^{+o}), x_+(s_k^{+o})=\exp(\pm\sqrt{2\times 2})s_k^{+o}, x_-(s_k^{+o})=\exp(\pm \mathrm{j}\sqrt{2\times 2})s_k^{+o}$.

It bring both real and imaginary initial conditions:

$$x_+^r(s_k^{+o})=\exp(\pm 2s_k^{+o}), x_-^{im}(s_k^{+o})=\exp(\pm 2js_k^{+o})=Cos(2s_k^{+o})\pm jSin(2s_k^{+o}), \quad (8.33)$$

which corresponds related extremal trajectories starting from (8.33) that average micro- and macroprocesses.

Finally the initial conditions starting extreme process (8.29a) at $t^e$ take form

$$x_{\pm}^r(t^e)=\exp(\pm 2s_k^{+o})\exp(\pm 2t^e), x_{\pm}^{im}(t^e)=[Cos(2s_k^{+o})\pm jSin(2s_k^{+o})]\exp(\pm 2t^e). \quad (8.34)$$

Let's numerically validate the results (8.31-8.34). Solution of (8.31):

$$\ln 1/2-2t_*^e+\ln[Cos(2t_*^e)]=\ln 0.117, -0.693+2.1456=2t_*^e-\ln[Cos(2t_*^e)], 1.4526=2t_*^e+\ln[Cos(2t_*^e)]$$

leads to $2t_*^e\approx 1.45, t^e=1.45/\pi\approx 0.46$- as on one of possible answer.

Applying condition (8.26) to (8.32): $S(s_{k*}^{+o})=1/2\exp(-2s_{k*}^{+o})Cos(2s_{k*}^{+o})=2/137$ leads to solution

$-0.693+4.22683=2s_{k*}^{+o}-\ln[\mathrm{Cos}(2s_{k*}^{+o})], 3.534=2s_{k*}^{+o}-\ln[\mathrm{Cos}(2s_{k*}^{+o})]$ with a result $s_{k*}^{+o}\approx 1.767, s_k^{+o}\approx 1.12$.

Applying condition of beginning virtual observation (8.25) at relative time $o_{s*}^p$ to

$$S(o_{s*}^p)=1/2\exp(2o_{s*}^p)Cos(2o_{s*}^p)=0.5\times 10^{-8} \quad (8.34a)$$

leads to $-0.693+12.8=2\,o_{s*}^{p}-\ln[Cos(2\,o_{s*}^{p})]$ with solutions $2\,o_{s*}^{p}\approx 12$ and $o_{s}^{p}\approx 3.85$.

Applying condition of starting information observation (8.27) with (8.30) at relative time $t_{*}^{oe}$, leads to solution $2t_{*}^{oe}\approx 0.33, t^{oe}\approx 0.1$.

These times are counting from the real Observer after overcoming the threshold (8.28). That means that virtual observation evaluates time interval $o_{s}^{p}\approx 3.85$, virtual observer starts on time interval $s_{k}^{+o}\approx 1.12$, and real observer starts on $t^{e}\approx 0.46$ while observing the first real 'half-impulse' probing action takes *part* of this time: $t^{oe}\approx 0.1$.

The real initial conditions starting extemal process (8.34) evaluate pair of states for the conjugated process:

$$x_{+}^{r}(t^{e})=3.65\times 1.58\cong 5.78, x_{-}^{r}(t^{e})=0.326\times 0.0.6313\cong 0.206\,. \quad (8.35)$$

Imaginary initial conditions evaluate four options:

$$x_{-}^{im1}(t^{e})=1.58[Cos(2\times 1.12)\pm j\,\mathrm{Sin}(2\times 1.12)]=1.58[0.999\pm j0.039]=1.578\pm j0.0617 \quad (8.35a)$$

$$x_{-}^{im2}(t^{e})=0.206[Cos(2\times 1.12)\pm j\,\mathrm{Sin}(2\times 1.12)]=0.206[0.999\pm j0.039]=0.2057\pm j0.008\,. \quad (8.35b)$$

**8b. Applying equation of extremals $\dot{x}=a^{u}$ to a dynamic model's traditional form** [48]:

$$\dot{x}=Ax+u, u=Av, \dot{x}=A(x+v)\,, \quad (8.36)$$

where $v$ is a control reduced to the state vector $x$, allows finding optimal control $v$ that solves the initial variation problem (VP) and identifies matrix $A$ under this control's action.

Proposition 8.3.

The reduced control is formed by a feedback function of macrostates $x(\tau)=\{x(\tau_{k})\}, k=1,...,m$ in form**:**

$$v(\tau)=-2x(\tau)\,, \quad (8.36a)$$

or applied to macroprocess' speed:

$$u(\tau)=-2Ax(\tau)=-2\dot{x}(\tau), \quad (8.36b)$$

at the localities of moments $\tau=(\tau_{k})$ (8.11), when matrix $A$ identifies equation

$$A(\tau)=-b(\tau)r_{v}^{-1}(\tau), r_{v}=E[(x+v)(x+v)^{T}], b=1/2\dot{r}, r=E[\tilde{x}\tilde{x}^{T}] \quad (8.37)$$

through the correlation function and it derivative, or directly, via the dispersion matrix $b$ from (2.1):

$$|A(\tau)|=b(\tau)(2\int_{\tau-o}^{\tau}b(t)dt)^{-1}>0,\ \tau-o=(\tau_{k}-o), k=1...,m\,. \quad \bullet \quad (8.37a)$$

Proof. Using Eq. for the conjugate vector (8.3) allows writing constraint (8.10) in the form

$$\frac{\partial X}{\partial x}(\tau)=-2XX^{T}(\tau)\,, \quad (8.38)$$

where for model (8.36) it leads to

$$X=(2b)^{-1}A(x+v), X^{T}=(x+v)^{T}A^{T}(2b)^{-1}, \frac{\partial X}{\partial x}=(2b)^{-1}A,\ b\neq 0\,, \quad (8.38a)$$

and (8.38a) acquires form

$$(2b)^{-1}A=-2E[(2b)^{-1}A(x+v)(x+v)^{T}A^{T}(2b)^{-1}], \quad (8.38b)$$

from which, at a nonrandom $A$ and $E[b]=b$, the identification equations (8.37) follows strait.

Completion of both (8.38a,b)is reached with aid of the control's action, which is found using (8.38b) in form

$A(\tau)E[(x(\tau)+v(\tau))(x(\tau)+v(\tau))^{T}]=-E[\dot{x}(\tau)x(\tau)^{T}]$, at $\dot{r}=2E[\dot{x}(\tau)x(\tau)^{T}]$**.**

This relation after substituting (8.36) leads to

$A(\tau)E[(x(\tau)+v(\tau))(x(\tau)+v(\tau))^{T}]=-A(\tau)E[(x(\tau)+v(\tau))x(\tau)^{T}]$, and then to

$$E[(x(\tau)+v(\tau))(x(\tau)+v(\tau))^{T}+(x(\tau)+v(\tau))x(\tau)^{T}]=0\,,$$

which is satisfied at applying the control (8.36a).
Since $x(\tau)$ is a discrete set of states, satisfying (8.11), (8.13), the control has a discrete form.
Each stepwise control (8.36a) with its inverse value, doubling controlled state $x(\tau)$, applies to both (8.38a,b) implementing (8.38) which follows from variation conditions (8.1a), and, therefore, it fulfills this condition. This control, applied to both Ito Eqs of random process and the extremal segments, also imposes the constraint that transforms the process to extremals. During the transformation, this control also initiates the identification of matrix $A(\tau)$ following (8.37, 8.37a). Applying the control step- down and step-up actions to satisfy conditions (8.7) and (8.10), the control sequentially starts and terminates the constraint on each segment, while extracting the cut off hidden information on the $x(\tau)$-localities. •
Obtaining this control here *simplifies* some results of Theorems [38].
Corollary 8.3. Control that turns the constraint on creates a Hamilton dynamic model with complex conjugated eigenvalues of matrix $A$. After the constraint's termination, the control transforms this matrix to its *real* form (on the diffusion process' boundary point [43]), which is identified by diffusion matrix in (8.37a). •
Proposition 8.4.
Let us consider controllable dynamics of a closed system, described by operator $A^v(t,\tau)$ with eigenfunctions $\lambda_i^v(t_i,\tau_k)_{i,k=1}^{n,m}$, whose matrix equation:

$$\dot{x}(t) = A^v x(t), \qquad (8.39)$$

includes the feedback control (8.36a). The drift vector for both models (8.36) and (8.39) has same form:

$$a^u(\tau, x(\tau,t)) = A(\tau,t)(x(\tau,t)+v(\tau)); A(\tau)(x(\tau)+v(\tau)) = A^v(\tau)x(\tau) \,. \qquad (8.39a)$$

Then the followings hold true:
(1)-Matrix $A(t,\tau)$ under control $v(\tau_k^o) = -2x(\tau_k^o)$, applied during time interval $t_k = \tau_k^1 - \tau_k^o$ : $A(t_k,\tau_k^1)$ depends on initial matrix $A(\tau_k^o)$ at the moment $\tau_k^o$ according to the Eq

$$A(t_k,\tau_k^1) = A(\tau_k^o)\exp(A(\tau_k^o)t_k)[2-\exp(A(\tau_k^o)t_k)]^{-1} \,. \qquad (8.39b)$$

(2)- The identification Eq.(8.37) at $\tau_k^1 = \tau$ holds

$$A^v(\tau) = -A(\tau) = 1/2b(\tau)r_v^{-1}(\tau), b(\tau) = 1/2\dot{r}_v(\tau) \,, \qquad (8.39c)$$

whose covariation function $r_v(\tau_k^o)$, starting at the moment $\tau_k^o$, by the end of this time interval acquires form

$$r_v(\tau_k^1) = [2-\exp(A(\tau_k^o)t_k)]r(\tau_k^o)[2-\exp(A^T(\tau_k^o)t_k)]. \qquad (8.39d)$$

(3a)-At the moment $\tau_k^o + o$ following $\tau_k^o$ at applying control $v(\tau_k^o) = -2x(\tau_k^o)$, the controllable matrix gets form

$$A^v(\tau_k^1)_{t_k \to 0} = A^v(\tau_k^o + o) = -A(\tau_k^o), \qquad (8.39e)$$

changing its sign from the initial matrix.
(3b)- When this control, applied at the moment $\tau_k^1$, ends the dynamic process on extremals in the following moment at $x(\tau_k^1 + o) \to 0$, function $a^u = A^v x(t)$ in (8.36) turns to

$$a^u(x(\tau_k^1 + o)) \to 0 \,; \qquad (8.39f)$$

which brings (8.39a) to its dynamic form $a^u = A(\tau_k^1 + o)v(\tau_k^1 + o) \to 0$ that requires turning the control off, at $v(\tau_k^1 + o) \to 0$.

(3c)- At fulfilment of (8.39f), initial stochastic Eq. (1.1.1) includes only its diffusion component, which allows identifying dynamic matrix $A(\tau_k^1+o)$, being transformed in the following moment $\tau_{k+1}^1$:

$A(\tau_k^1+o)\to A(\tau_{k+1}^1)$ via correlation matrix $r(\tau_{k+1}^1)$ using relation

$A(\tau_{k+1}^1)=1/2\dot{r}(\tau_{k+1}^1)r^{-1}(\tau_{k+1}^1)$ at $r(\tau_{k+1}^1)=E[\tilde{x}(t)\tilde{x}(t+\tau_{k+1}^1)^T]=r^v(\tau_{k+1}^1)_{v(\tau_k^1+o)\to 0}$.

(3d)-Dispersion matrix on extremals with covariation matrix (8.39d) acquires forms

$$\partial r_v(\tau_k^1)/\partial t_k=-A(\tau_k^o)\exp(A(\tau_k^o)t_k)r(\tau_k^o)[2-\exp(A^T(\tau_k^o)t_k)]$$
$$+[2-\exp(A(\tau_k^o)t_k)]r(\tau_k^o)[-A^T(\tau_k^o)\exp(A^T(\tau_k^o)t_k)], \quad (8.40)$$

where for symmetric matrix $A(\tau_k^o)$ it holds relations for dispersion matrixes

$\partial r_v(\tau_k^1)/\partial t_k\,|_{t_k=\tau_k^1}=-2A(\tau_k^o)\exp(A(\tau_k^o)\tau_k^1)r(\tau_k^o), b(\tau_k^1)=-A(\tau_k^o)\exp(A(\tau_k^o)t_k)r(\tau_k^o)$,

$b(\tau_k^1=\tau_k^o)=-A(\tau_k^o)\exp(A(\tau_k^o)0_k)r(\tau_k^o)=-A(\tau_k^o)r(\tau_k^o)$ and ratio

$$b(\tau_k^1)/b(\tau_k^o)=\exp(A(\tau_k^o)\tau_k^1)A(\tau_k^o)^{-1}\ . \quad (8.40a)$$

Relation (8.40a) for a single dimension, at $A(\tau_k^o)=\alpha_1(\tau_{k1}^o)$, leads to

$$b(\tau_{k1}^1)/b(\tau_{k1}^o)=\exp\alpha_1(\tau_{k1}^o)/\alpha_1(\tau_{k1}^o) \quad (8.40b)$$

which after applying Brownian relation (2.1.3) $b(\tau_{k1}^1)/b(\tau_{k1}^o)=\tau_k^1/\tau_{k1}^o$ leads to

$$\tau_k^1=\tau_{k1}^o\exp\alpha_1(\tau_{k1}^o)/\alpha_1(\tau_{k1}^o)\ , \quad (8.40c)$$

connecting (8.40c) with information speed on beginning of interval $t_k=\tau_k^1-\tau_k^o:\tau_{k1}^1=\tau_{k1}^o[\alpha_1(\tau_{k1}^o)]$.

(3e)- Equation for conjugated vector (8.17) for each extremal segments follows from relation:

$$X_o(t_k)=2b(t_k)^{-1}\dot{x}(t_k)=2A(\tau_k^o)\exp(A(\tau_k^o)t_k)x(\tau_k^o)A^T(\tau_k^o)\exp(A^T(\tau_k^o)t_k)\ . \quad (8.41)$$

(3f)- Entropy increment $\Delta S_{io}$ on optimal trajectory at

$E[\frac{\partial\tilde{S}}{\partial t}(\tau)]=1/4Tr[A(\tau)]=H(\tau)$, $A(\tau)=-1/2\sum_{i=1}^{n}\dot{r}_i(\tau)r_i^{-1}(\tau),(r_i\ )=r$,

measured on the cutting localities, determine the EF between the nearest segments' time interval below: $\Delta S_{io}=$

$$I_{x_t}^p=-1/8\int_s^T Tr[\dot{r}r^{-1}]dt=-1/8Tr[\ln(r(T)/\ln r(s)],(s=\tau_o,\tau_1,...,\tau_n=T). \quad (8.42)\ \bullet$$

<u>Proof</u> (1). Control $v(\tau_k^o)=-2x(\tau_k^o)$, imposing the constraint at $\tau_k^o$ on both (8.36) and (8.39) and terminating it at $\tau_k^1$ on time interval $t_k=\tau_k^1-\tau_k^o$, brings solutions of (8.36) by the end of this interval:

$$x(\tau_k^1)=x(\tau_k^o)[2-\exp(A(\tau_k^o)t_k)]. \quad (8.43)$$

Substituting this solution to the right side of $\dot{x}(\tau_k^1)=A^v(\tau_k^1)x(\tau_k^1)$ and the its derivative to the left side leads to

$-x(\tau_k^o)(A(\tau_k^o)t_k)\exp(A(\tau_k^o)t_k)=A^v(\tau_k^o)x(\tau_k^o)[2-\exp(A(\tau_k^o)t_k)])]$,

or to connection of both matrixes $A^v(\tau_k^{1`})$ and $A(\tau_k^1)$ (at the interval end) with the matrix $A(\tau_k^o)$ (at the interval beginning) for the closed system (8.39) in the forms:

$$A^v(t_k,\tau_k^1)=-A(\tau_k^o)\exp(A(\tau_k^o)t_k)[2-\exp(A(\tau_k^o)t_k)]\ , \quad (8.44)$$

and $A^v(\tau_k^1)=-A(\tau_k^1)$ by the moment $\tau_k^1$, from which folows (8.39e). Other Proofs are straight forward. •

The initial conditional probability measure (1.1.2) along the extremal trajectory determines the extremal EF:

$$p[x(t)]=p[x(s)]\exp(-S[x(t)]) \quad (8.44a)$$

where starting probability

$p[x(s)=\exp(-S[x(s)])$ (8.44b)

follows from formula (8.32) and numerical values (8.34a),(8.35,8.35a,b).

(3e)-Invariant relations. Using (8.36b) in form $u(\tau)=-2Ax(\tau)=-2\dot{x}(\tau)$, and

$c^2=|u_+u_-|=c_+c_-=\bar{u}^2, c_+=u_+, c_-=u_-$ leads to

$$c^2=\dot{x}(\tau)=-2A^v x(\tau), \ln x(\tau)/\ln x(s)=\mathrm{A}^v(\tau-s)=\mathrm{u}_\pm(\tau-s),$$
$$\mathrm{A}^v(\tau-s)=\mathrm{u}_\pm(\tau-s)=inv, c^2=\dot{x}(\tau)=a^u=inv \quad , \qquad (8.45)$$

where $(\tau-s)$ is equivalent of interval $t_k=\tau_k^1-\tau_k^o$ imposing constraint (8.10,8.38) between the discrete moments (8.11)on each impulse invariant interval. As result of constraint (8.10), it follows

$\mathrm{A}^v(\tau-s)=\mathrm{u}_\pm(\tau-s)=inv=\mathbf{a}_o$, (8.45a)

where function $\mathbf{a}_o=\mathbf{a}_o(\gamma_k)$ [44] depends on ratio of imaginary to real eigenvalues of operator (8.44): $\gamma_k=\beta_{ko}/\alpha_{ko}$ . From (8.45a), in particular, follows $\mathrm{A}^v=\left[\pm 2\right]_{n\times n}$. (8.45b)

That concurs with Eqs (8.33) and the impulse invariant information in Corollaries 4.2.

For optimal model with $\mathbf{a}_o(\gamma_k\to 0.5)=\ln 2$, it leads to $(\tau-s)=\ln 2/2\cong 0.346$ that is close to evaluation $\delta_k=\tau_k^{+o}-\tau_k^{-o}\cong 0.35$ [47].

Invariant $\mathbf{a}_o=\ln 2\cong 0.7$ measures information generating at each impulse cut (Secs.1.1.2,2.2.3).

Dispersion matrix (8.39b), measured by the `optimal model's invariant, is

$r_v(\tau_k^1)=[2-\exp\mathbf{a}_o\ )]r(\tau_k^o)[2-\exp(\mathbf{a}_o)]=r(\tau_k^o)\left[1.5^2\right]$, (8.46)

and the extreme states (8.40) measures vector $x(\tau_k^1)=x(\tau_k^o)\left[1.5\right]$. Conjugate vector

$X_o(\tau_k^1)=2A(\tau_k^o)\exp(\mathbf{a}_o)x(\tau_k^o)A^T(\tau_k^o)\exp(\mathbf{a}_o)$ (8.46a)

for a single dimension holds

$X_{o1}(\tau_k^1)=2\alpha_1(\tau_{k1}^o)^2\exp(2\mathbf{a}_o)x(\tau_{k1}^o)$, (8.46b)

or at $\alpha_1(\tau_{k1}^o)=2\mathbf{a}_o/t_k, \alpha_1(\tau_{k1}^o)=2, t_k=\mathbf{a}_o$ it acquires $X_{o1}(\tau_k^1)=8\times 4x(\tau_{k1}^o)\alpha_1(\tau_{k1}^o)^2=128x(\tau_{k1}^o)$.

The EF-IPF *estimate* the invariant's information measure $\mathbf{a}_o(\gamma_k)$, which allows count both the segment's and inter-segment's increments:

$\tilde{S}_{\tau m}^i=\sum_{k=1}^m(\mathbf{a}_o(\gamma_k)+\mathbf{a}_o^2(\gamma_k)), \tilde{S}_\tau=\sum_{i=1}^n\tilde{S}_{\tau m}^i$,, (8.47)

where $m$ is the number of the segments, $n$ is the model dimension (assuming each segment has a single $\tau_k$ -locality). However, to *predict* each $\tau_k$ - locality, where information should be measured, only invariant $\mathbf{a}_o(\gamma_k)$ needs. Sum of process's invariants

$\tilde{S}_{\tau m}^{io}=\sum_{k=1}^m\mathbf{a}_o(\gamma_k)$, $\tilde{S}_\tau^o=\sum_{i=1}^n\tilde{S}_{\tau m}^{io}$ (8.48)

estimates the IPF entropy with a maximal process' probability. Knowing *this* entropy allows encoding the *random process* using the Shannon formula for an average optimal code-word length:

$l_c\ge\tilde{S}_\tau^o/\ln D$, (8.49)

where $D$ is the number of letters of the code's alphabet, which encodes $\tilde{S}_\tau^o$ (8.48).

An elementary code-word to encode the process' segment is

$l_{cs} \geq \mathbf{a}_o(\gamma_k)\,[\text{bit}]/\log_2 D_o$, (8.50)

where $D_o$ is a segment's code alphabet, which implements the macrostate connections.

At $\mathbf{a}_o(\gamma_k \to 0.5) \cong 0.7$, $D_o$=2, it follows $l_{cs} \geq 1$ , or a bit per the encoding letter.

Initial segment's $x(\tau^o_{k1o})$ determines the values of initial conditions $x^r_{\pm}(t^e)$, $x^{im1}_{-}(t^e), x^{im2}_{-}(t^e)$ (8.35,8.35a,b), which depend on observations at moment $t^e$.

Finally, the extremal trajectory, with each following segment, determines $x(\tau^o_{k1o})$ which starts also conjugated process, and moment $t^e$ identifies dispersion and correlation on the trajectories that determine operator of information speed $A(t,\tau_k)$ and both EF-IPF on the optimal trajectory segments. Optimal control starts with the above initial conditions. The detail identification of information micro-macrodynamics (IMD), based on the *invariant* $\mathbf{a}_o(\gamma_k)$ *description,* is in [43,44,47], where the dynamics' scale parameter $\gamma^{\alpha}_{k,m}$ depends on a frequency spectrum of observations detecting through $\gamma_k$. The observer IMD self-scales the observation to build the observer time-space distributed information network.

These relations allow finalize not only applications of main general result but also validate them numerically.

**9. Discussing the theoretical results**

Initial entropy functional (1.1.5, 1.1.10) presents a potential information functional of the Markov process until the applied impulse control, carrying the cutoff contributions (1.2.16a,b),(1.2.17), and (2.3.5), transforms it to the informational path functional (2.5.1). Markov random process is a source of each information contribution, whose entropy increment of cutting random states delivers information, hidden between these states.

The finite restriction on the cutting function determines the discrete impulse's step-up and step-down controls, acting between impulse cutoff $\delta_k = \tau^{+o}_k - \tau^{-o}_k$,which hold the information received on $\tau^{-o}_k$ and transfers it on $\tau^{+o}_k$ while starting interval $\Delta_t \to o(t)$ following next cutoff, delivers new hidden process' information, and so on.

Cutting entropy on $\tau^{-o}_k$ produces the equivalent physical information and memorizes it.

That is why information is a physical entity, which distinguishes from entropy that can be virtual.

In a multi-dimensional diffusion process, the step-wise controls, acting on the process all dimensions, sequentially stops and starts the process, evaluating the multiple functional information. Impulses delta-function $\delta u_t$ or discrete $\delta u_{\tau_k}$ implement transitional transformations (1.1.2-1.1.3), initiating the Feller kernels along the process and extracting total kernel information for $n$ -dimensional process with $m$ cuts off.

The maximal sum measures the interstates information connections held by the process along the trajectories during its real time $(T-s)$, which are hidden by the random process correlating states.

The EF functional information measure on trajectories is not covered by traditional Shannon entropy measure.

The dissolved element of the functional's correlation matrix at the cutoff moments provides independence of the cutting off fractions, leading to orthogonality of the correlation matrix for these cut off fractions.

Intervals between the impulses do not generate information, being imaginary-potential for getting information, since no real double controls are applying within these intervals.

The minimized increments of entropy functional between the cutoffs allow prediction each following cutoff with maximal conditional probability.

A sequence of the functional a priori-posteriori probabilities provides Bayesian entropy measuring a *probabilistic causally* [44- 47], which is transforming to physical casualty in information macrodynamics.

Sum of information contributions, extracted from the EF, approaches its theoretical measure (1.1.10) which evaluates the upper limit of the sum.

Since the sum of additive fractions of the EF on the finite time intervals is less than EF, which is defined by the additive functional, the additive principle for a process' information, measured by the EF is *violated*.

Each $k$-cutoff "kills" its process dimension after moment $\tau_k^{+o}$, creating process that balances killing at the same rate [39]. Then $k=n$, and condition (1.1.38) requires infinite process dimensions, continuing the processes balance creation.

The EF measure, taken along the process trajectories during time $(T-s)$, limits maximum of total process information, extracting its hidden cutoff information (during the same time), and brings more information than Shannon traditional information measure for multiple states of the process.

The limited time integration and the last cutting correlation bind the EF last cut.

Maximum of the process cutoff information, extracting its total hidden information, approaches the EF information measure. Or total physical information, collected by IPF in the infinite dimensional Markov diffusion process, is finite. Since rising process dimension up to $n \to \infty$ increases number of the dimensional kernels, information of the cutting off kernels grows, and when the EF transforms to the IPF kernels, the IPF finally measures all kernels finite information.

Total process uncertainty-entropy, measured by the EF is also finite on the finite time of its transition to total process certainty-information, measured by the IPF. The EF integrates this time (2.1.2) along the trajectories.

The IPF formally defines the distributed actions of multi-dimensional delta-function on the EF via the multi-dimensional additive functional (1.3.7), which leads to analytical solution and representation by Furies series.

The delta-impulse, generating spectrum of multiple frequencies, is a source of experimental probabilities in formal theory [37].

Since entropy requires a direction of time course –arrow of time, cutting entropy memorizes the cutting time interval which freezes the probability of events with related dynamics of information micro- macroprocess.

The cutting off random process brings information Bit that includes: information delivered by capturing external entropy during transition to the cut; information cut from the random process; information transferring to the nearest impulse that keeps persistence continuation of the impulse sequence and the Bits.

Such persistent Bit sequentially and automatically converts entropy to information, holding the cutoff information of random process, which connects the Bits sequences in the IPF.

The connecting Bit is different from that cutoff from random process [47].

The IPF maximum, integrating unlimited number of Bits' units with finite distances, limits the total information carrying by the process' Bits.

The limited total time of collecting the cutoff information (at $n \to \infty$) decreases the time interval of each cutoff, which increases the inclusive quantity of information extracted on this interval.

At the same cutting off information for each impulse, density of this information, related to the impulse interval, grows, which integrates sum of all previous cutoffs.

The limitation on the cutting discrete function determines two process' classes for cutting EF functional: microprocess with entropy increment on each time interval $o(t)$, as a carrier of information contribution to each $\delta_k$ driving the integration, and macroprocess over real time $(T-s)$ describing the total process of transformation EF-IPF.

The microprocess specifics within each $o(t)$ are: an imaginary time compared to real time of real microprocess on information cutoff $\delta_k$; two opposite sources of entropy-information as an interactive reaction from random process at transition $\delta_k^{\tau+}$ to $\tau_k^{-o}$, which carry two symmetrical conjugated imaginary entropy increments until the control interactive capturing brings their dissimilarity (the asymmetry break between controls on $\delta_k^{\tau+}$ evaluates relations (5.25), (5.27)); the entropy increments correlate at $\delta_k^{\tau+}$-locality, before the cutting action on $\tau_k^{-o}$ transforms the adjoin increments to real information; the cutting action dissolves the correlation and generates the impulse information contribution by moment $\tau_k^{+o}$; the entropy cut memorizes the cutting information, while a gap within $\delta_k^{\tau+}$ delivers external influx of entropy (5.27), covered by real step-wise action, which carries energy for the cut.
Microprocess may exist within Markov kernel or beyond, and even prior interactions, independently on randomness.
Microprocess with imaginary time between the impulses belongs to random process, whose cutoff transfers it to information microprocess within each impulse $\delta_k$.

Time course on $\Delta_k$ is a source of the entropy increment between impulses, measured in Nats, which moves the nearest impulses closer. The relative intervals (5.4) has a measure of information density (5.3).
Between the impulse No and Yes action emerges a transitional impulse which transforms interim time to the following space interval ending with holding the cutting information.
Superimposing interaction, measured through the multiplication of conjugated entropy and probability functions (Sec.2.2.6), brings the observable values (Sec.2.2.8a, b). The conjugated and probabilistic dynamics of the microprocess within the impulse is different from Physics in Quantum Mechanics.

Shifting $\delta_k^{\tau+}$ in real time course $\Delta_k$ moves to automatically convert its entropy to information, working as Maxwell's Demon [42], which enables compensate for the transitive gap [44,p.60] from entropy to information.
The macroprocess, integrating the imaginary entropy between impulses with an imaginary microprocesses and the $\delta_k$ cutoff information of real impulses, builds information process of the collected entropy converted to physical information process.
The EF extremal trajectories describe the IPF information macroprocess which averages all the microprocess.
At $n \to \infty$, such information macroprocess is extremals of both EF and IPF.
The EF imaginary entropy, measured by logarithmic conditional probability of the observing random process, is distinct from Boltzmann physical entropy satisfying Second Thermodynamic Law.
The EF-IPF transformations provide the Information Path from Randomness and Uncertainty to Information, Thermodynamics, and Intelligence of Observer [44, 45]. This includes a virtual observer, acting with imaginary control and integrating imaginary entropy increments in EF, which is mathematically established in Secs.2.2.5-2.6 and numerically verified in Ses.2.2.8a, b. Actual information observer acts with real control cutting impulse's information, which integrates the IPF. During the process' time-space observation, the virtual and actual Observer rises, created on a Path from uncertainty to certainty-information without any priory physical law. Multiple interactive impulse actions on the above Path coverts the integrated maxmin entropy to information in real time-space, transforming the information to physical laws.
The information processes correctly describe all worlds' natural process because they logically chose and connect all observations in certain–most probable sequence of events, which include hidden information between interacting events. Primary virtual observer through probing impulses sequentially increases the observing correlations, reducing entropy fractions, and integrates them in entropy functional.
The chosen certain fractions create the information observer whose path functional integrates the fractions in information process, describing a factual events-series as physical processes.
According to Feynman [49], a physical law describes the most probable events of observation process, which implies applying a variation principle (VP) for finding the law.
However, a common form of arising physical law formulates only the informational VP whose solution brings maximal probability on the VP extremals-information processes (Sec.2.2.8).

**10.Estimation of information hidden by interstates' connection of the diffusion process, other applications**

The evaluated information effect of losing the functional's bound information at the cutoff moments, according to (1.3.5), holds amount of 0.5 Nats (~0.772 bits). These cutoff in the form of $\delta^i[\tau_k]$-function, applied $k=m$ times to each of the process dimension $i=1,...,n$, bring total information

$I_c \cong 0.772 m \times n$ bits.

Thus, the process functional's information measures (1.10),(2.3.5,6) encloses $I_c$ bits more, compared to the information measure, applied separately to each of the process $m \times n$ states (during the same time).
This result is applicable to a comparative information evaluation of the divided and undivided fractions of an information process, measured by corresponding EF, where each two bits of undivided process' pair contains 0.772 bits more hidden information, measured by the functional. That means that information *process* holds more information than any divided number of its fractions, and the considered entropy functional, measured this process, evaluates the quantity of information that *connect* these fractions.
Moreover, by knowing this initially hidden information, one could determine which information is necessary to connect the data, being measured separately, toward composing a unit of some information process.
According to the evaluation of an upper bound entropy per an English character (token) [50], its minimum is estimated by 1.75 bits, with the average amount between 4.66-7 bits per character.
The evaluation includes the inner information bound by a character. At minimal entropy per symbol in 1 bit, a minimal symbol's bound information is 0.75, which is close to our evaluation at the cutoff.
The impulse cut-off method was *applied* in different solidification processes with impulse controls' automatic system [51], which reveal some unidentified phenomena (such as a compulsive appearance centers of crystallization-indicators of generation information, integrated in the IPF).
Possibility of natural increase of correlations illustrate experimental results [52],[53], more are cited in [46].
Retinal Ganglion Cells are Eyes discrete receptors interacting with observations and generating information which transmission integrates [54]. Applications in different artificial and cognitive systems [44], [45], others model cognitive information dynamics.

**Conclusive Remarks**

1. Numerous different interactive observations possess randomness whose probability is a source of its entropy-uncertainty. Uncertainty measures conditional entropy of relative a priori—a posteriory probabilities. Since theoretical uncertainty with probability zero has infinite entropy measures, such conditional entropy does not exist. Therefore, the meaningful measured notion of uncertainty is *finite and conditional (relative)*, while closeness to the infinity holds a posteriory probability. We measure uncertainty by non-correlating a priori-a posteriory probabilities, when their connection is absent approaching zero correlation.

2. Information, as notion of certainty–opposite to uncertainty, measures reduction of uncertainty to posterior probability 1(evaluating a probabilistic fact-truth), which however related negative entropy cannot determines.

Since standard unit of information a Bit is a *discrete* entity- an impulse composing of Yes-NO actions, information originates from the impulse *discrete* actions directed on reductions of the entropy.
The interactive observations as source of randomness, which model series of random impulses- a random process possessing entropy, whose cutting by the impulses may produce information.
Moreover, each cut brings memory of the entropy being cut, which provides both reduction of the process entropy and discrete unit of the cutting entropy – a Bit. That allows us to define information:
*'Information is memorized entropy cutting in random observations which process interactions.'*
3. The origin of information, thus, associates with an "anatomy of creation of impulse" enables both cut and stipulate random process, generating information under the cut, where memory is the impulse' cut time interval.
4. Since information composes multiple bits, the measurement of the information manifolds requires integral measure applied to an observable process, which for the discrete units is mathematically problematic.
Integral measuring entropy along trajectories of random process, as source of information, is also challenging. A view of natural origin of information from general time –space continuum does not lead to discrete nature of information, while, according to Second Law, entropy increases with time course. Cutting entropy memorizes the cutting time interval which freezes the probability of events with related dynamics of information process.
5. The origin of information may initiate the probing impulses-virtual from beginning and actual with rising information Observer, which can reverse natural time course and decrease entropy in the Observer processes. Virtual (observation and/or observer) means a potential real, which confirms obtaining its information (if not, it remains virtual). Observation under random impulses with opposite Yes-No probability events reveals hidden correlation, which connects Bayesian probabilities increasing each posterior correlation. That sequentially reduces relational entropy along the process, conveying probabilistic casualty with temporal memory collecting correlations which interactive impulse *innately* cuts. Uncertainty losses the probabilistic casualty, which holds *objective measure*. Definition and measuring information are concepts subsequent to information is a *phenomenon* of interactions from which all Universe composes.
6. Each impulse No action cuts maximum of impulse minimal information while following Yes action transfers this maxim between the impulses performing that dual principle of converting the process entropy to information. Multiple maxmin-minimax leads to the dual variation principle, arising in natural interactions. Cutting entropy of correlation reveals information hidden within correlation.
7. Within the hidden correlation emerges reversible time–space *microprocess* with Yes-No conjugated entangled entropy flows, which the impulse dynamically cuts and memorizes the cutting entropy as information-certainty.
Sequential interactive cuts along the process integrates the cutoff information in *information macroprocess* with irreversible time course. Each memorized information binds the reversible microprocess within impulse with the irreversible information macroprocess along the multi-dimensional process, which consecutively and automatically converts entropy to information and holds it.
8. The impulse (0-1) actions convert maximal entropy of the cutting correlation to equivalent information Bit of information microprocess, which memorizes *logic* of the observed entropy probes' prehistory.
Superimposed structuring information starts with quantum entangled Bit of the information microprocess, formed by the (0-1) actions, which holds Yes-No logic of a primary information observer as Wheeler's Bit-Participator built without any priory physical law.
The impulse, bringing the information Bit and information microprocess, carries the information logical cost of getting the Bit.
The sequential impulse cuts along the process convert its entropy to information, connect the Bits sequences, and *integrate* the cutting microprocesses in information macroprocess.
9. Integrating process entropy in an entropy functional and Bits information in information path integral' measures embraces the variation problem with the minimax law which determines all processes regularities.

Solving the problem, mathematically describes the micro-macro processes, information network (IN) that the processes build, and invariant conditions of Observer's IN self-scaling, self-organization and self-replication. The integral maximum, integrating unlimited number of Bits' units, limits total information enclosed in process. The process regularities include constraints, thresholds, and gaps the conversion information overcoming. The considered random observations under interactive impulse is a constituent of any natural events in physical interactive processes, which originates information processes including brain logic, human thoughts, artificial intelligence.
The paper first time completes *Mathematical Foundation of Information Dynamics*, confirmed in many Applications.

**Fig.1. Illustration of origin the impulse space coordinate measure $h[l]$ at curving time coordinate measure $1/2p[\tau]$ in transional movement.**

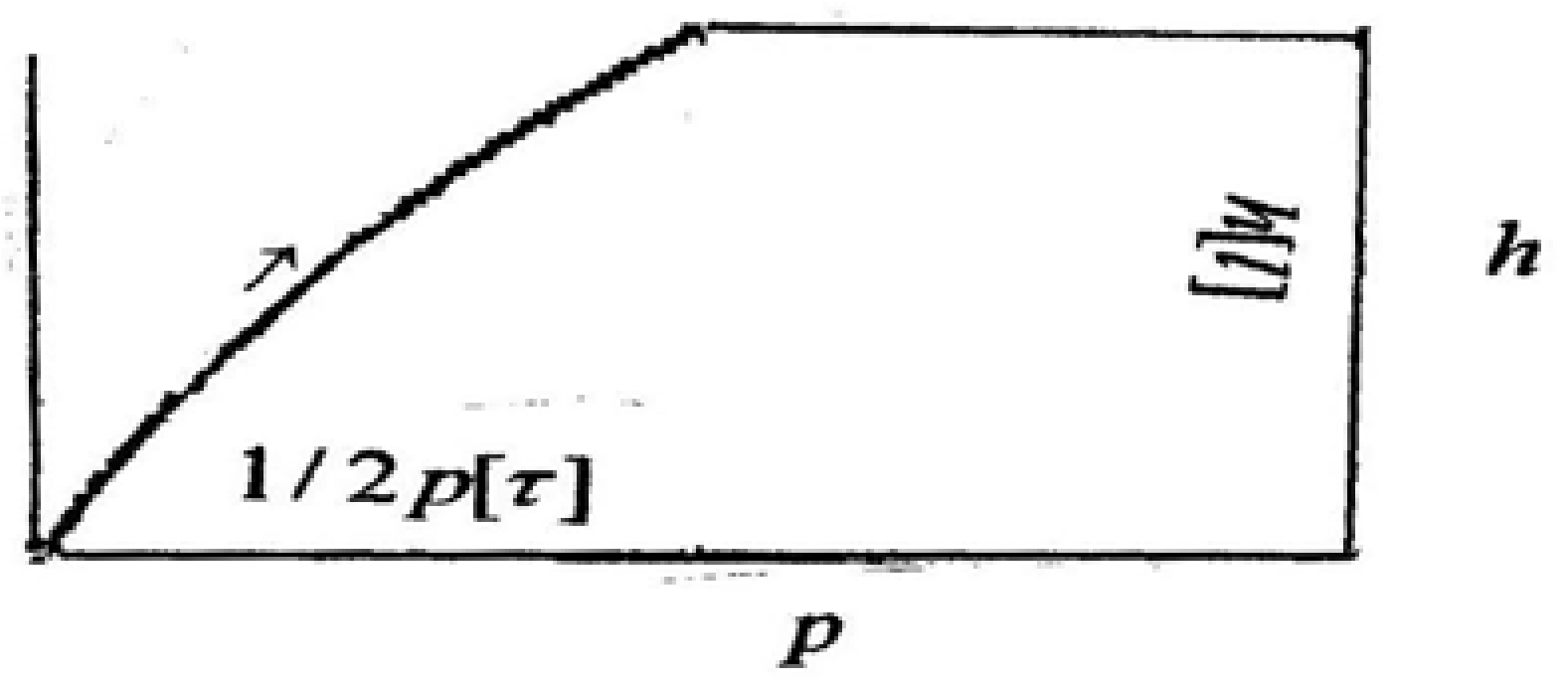